\documentclass[single, twocolumn]{aastex631}
\usepackage{eqnarray, amsmath}				
\usepackage{graphicx}
\usepackage{verbatim}
\usepackage{grffile}					
\usepackage{mathrsfs}
\usepackage[mathscr]{euscript}

\usepackage{jmc_merge_uniq}

\newcommand{\Figures}{./}           

\newcommand{\NewFigures}{\Figures}

\newcommand{\Cloudlet}{\Figures}
\newcommand{\Otherfigures}{\Figures}

\newcommand{\figmused}{0.85}
\newcommand{\figmusederr}{0.05}

\newcommand{\epsoonepercent}{40}

\begin{document}

\newif\iffrbFAST		
\frbFASTtrue			
\frbFASTfalse			

\newif\ifextrafigs		
\extrafigstrue
\extrafigsfalse

\parindent 10pt

\shorttitle{FRB Propagation through Interstellar and Intergalactic Media}
\title{Redshift Estimation and Constraints on Intergalactic and Interstellar Media from Dispersion and Scattering  of Fast Radio Bursts}

\author[0000-0002-4049-1882]{J. M. Cordes}
\author[0000-0002-4941-5333]{Stella Koch Ocker}
\author[0000-0002-2878-1502]{Shami Chatterjee}
\affiliation{Department of Astronomy and Cornell Center for Astrophysics and Planetary Science, Cornell University, Ithaca, NY, 14853, USA}
\correspondingauthor{JMC}
\email{jmc33@cornell.edu}
 
\begin{abstract}
\noindent
A sample of 14 FRBs with measured redshifts and scattering times  is used to assess contributions to dispersion and scattering from the intergalactic medium (IGM), galaxy halos, and the disks of host galaxies.  The IGM and galaxy halos contribute significantly to  dispersion measures but evidently not to scattering, which is then dominated by host galaxies. This enables  usage of  scattering times  for estimating DM contributions from   host galaxies and also for   a combined scattering-dispersion redshift estimator.  
Redshift estimation is calibrated using scattering of Galactic pulsars  after taking into account different scattering geometries for Galactic  and intergalactic lines of sight.     
The DM-only estimator  has a bias $\sim 0.1$  and RMS error $\sim 0.15$ in the redshift estimate for an  assumed ad-hoc  value  of 50~$\DMunits$ for the host galaxy's DM contribution.  The combined redshift estimator shows less bias by a factor of four to ten and a 20 to 40\% smaller   RMS error. 
\added{We find that }values for the  baryonic fraction of the ionized IGM $\figm \simeq \figmused\pm \figmusederr$ optimize redshift estimation using dispersion and scattering.   
Our study suggests that two of the 14 candidate galaxy associations (FRB~20190523A and FRB~20190611B) should be reconsidered. 
\end{abstract}


\section{Introduction}

Of the hundreds of distinct sources of fast radio bursts (FRBs) that have been recognized to date, there are only 14 published cases with  associated galaxies and redshifts.   By contrast, the dispersion measure (DM), the path integrated electron density, is necessarily measured  concomitantly with burst detections and, in many cases,  measurements or upper limits are also obtained on characteristic scattering times from multipath propagation.   

Because FRB distances are necessary for understanding both the energetics and the size of the FRB source population, significant efforts now aim to make sub-arcsecond localizations that aid  subsequent spectroscopic observations to determine redshifts.    It will be some time before these efforts yield large numbers of redshifts.
In the meantime,  coarser redshift estimates can be made using DM values combined with electron-density models for the Milky Way and the intergalactic medium (IGM) and assumptions about contributions from host galaxies.    Indeed, a recent study has shown a trend for larger DM with increasing redshifts, as expected, but with significant scatter \cite[][]{2020Natur.581..391M}.   Some of this scatter is from cosmic variance in the  electron density  in the IGM but the sizable range of  DM contributions from host galaxies also contributes. 
In this paper we develop and assess a redshift estimator that uses scattering measurements in tandem with DM values to better constrain host-galaxy DMs and thus tighten constraints on redshifts.    Balmer-line measurements also contribute to this analysis \citep[][]{tbc+17, bta+17}
and will play an increasingly important role in the future as more FRB sources are localized \cite[e.g.][]{2021arXiv210711334S}. 

\explain{This paragraph moved from earlier in Introduction}
While nearing completion of this paper,  FRB~20190520B  with large total DM was found to be associated with a low-redshift galaxy \citep[][]{2021arXiv211007418N}.  This object corroborates the main results of this paper but a detailed analysis is deferred to another paper \citep[][]{2022arXiv220213458O}.

Section 2 discusses contributions to DM and presents posterior PDFs for host-galaxy DMs for the fourteen objects with redshifts of associated galaxies (which are tentative in a few cases). 

Section 3 assesses contributions to scattering  in the context of  a parameterized cloudlet model that is calibrated against  scattering of Galactic pulsars.    The section presents alternative geometries for FRB scattering under the assumption that IGM and galaxy halos do not contribute significantly to measured scattering. 

Section 4 presents a combined analysis of dispersion and scattering for the nine objects with scattering and redshift measurements. It includes estimates of the $\Ftilde$ parameter that is a measure of the scattering strength in the cloudlet model. 

Section 5 considers redshift estimation  using only dispersion measures versus usage of scattering in tandem with dispersion.  It presents a criterion for when scattering can usefully constrain the redshift and presents results that also constrain the baryonic fraction of the ionized IGM.  

Section 6 presents a summary and our conclusions. 

The Appendix presents details for the cloudlet scattering model. 

\added{
Naming convention:   In  the text and in Tables  1 to 3 we use the full FRB name (e.g. FRB20121102A) given by the Transient Name Server\footnote{\url{https://www.wis-tns.org}}.    Most figure labeling  uses short labels (e.g. 121102), which is unambiguous for the sample we analyze and discuss.}

\section{Dispersion Measure Inventory}
\label{sec:DMinventory}

The  objects we analyze are listed in Table~\ref{tab:FRBproperties} with columns
(1) FRB name;
(2-3) Galactic coordinates $l,b$;
(4) DM;
(5) NE2001 estimate for the MW contribution to DM (sans a MW halo contribution);
(6) burst width; for the repeating FRBs 20121102A and 20180916B, this is a typical value;
(7) $\taud$ (measurement or limit);
(9-10)  + and - RMS errors in $\taud$;
(11) radio frequency for the $\taud$ entries;
(12)  reference for $\zh$;
and
(13) reference for $\taud$.

We note that determinations of $\taud$ need to be used with caution because some are made on bursts with low signal-to-noise ratios or that appear in  a narrow frequency band.   Another problem is  the frequency drift (`sad trombone') phenomenon seen in many bursts \citep[e.g.][]{2019ApJ...876L..23H,2019ApJ...885L..24C,2020ApJ...891L...6F,2020ApJ...891L..38C} that 
 can produce asymmetries in wide bandwidth burst profiles that are similar to those expected from scattering.  
 We assume that all  scattering measurements  in  Table~\ref{tab:FRBproperties} are unaffected by frequency-drifts or other effects that can masquerade as scattering asymmetries.
 

\begin{deluxetable*}{l R R C C R C  r C c c c c}[t]
\tabletypesize{\footnotesize}
\tablecaption{ FRB Scattering and Redshift Sample \label{tab:FRBproperties}}
\tablehead{
\colhead{FRB} &
\colhead{$l$} & 
\colhead{$b$} & 
\colhead{DM} & 
\colhead{$\DM_{\rm NE2001}$} &
\colhead{$W$} & 
\colhead{$z_{\rm h}$} & 
\colhead{$\taud$} & 
\colhead{$\sigma_-$} &
\colhead{$\sigma_+$} &
\colhead{$\nu_{\taud}$} & 
\multicolumn{2}{c}{References} 
\\
\colhead{} & \multicolumn{2}{c}{(Degrees)} & 
\colhead{($\DMunits$)} & 
\colhead{($\DMunits$)} &
\colhead{(ms)}&
 & 
\colhead{(ms)} &
\colhead{(ms)} &
\colhead{(ms)} &
\colhead{(GHz)} &
\colhead{$z_{\rm h}$\tablenotemark{\footnotesize d}} &
\colhead{$\taud$\tablenotemark{\footnotesize e}}
\\
\colhead{(1)} &
\colhead{(2)} &
\colhead{(3)} &
\colhead{(4)} &
\colhead{(5)} &
\colhead{(6)} &
\colhead{(7)} &
\colhead{(8)} &
\colhead{(9)} &
\colhead{(10)} &
\colhead{(11)} &
\colhead{(12)} &
\colhead{(13)} 
 }
\tablecolumns{13}
\startdata
20121102A          & 174.9 	&  -0.2 	&   557 	&   188 	& 3.0 	& 0.193 	& $<$   9.6 	& $\cdots$  	& $\cdots$ 	&  0.50 	& 1 &  1 \\
20180916B\tablenotemark{\footnotesize a}          & 129.7 	&   3.7 	&   349 	&   199 	& 0.87 	& 0.034 	& $<$   1.7 	& $\cdots$  	& $\cdots$ 	&  0.35 	& 2  & 2 \\
20180924A          &   0.74 		& -49.4 	&   361 	&    40 	&  1.30 	& 0.321  	&   0.68 		&  0.03 		&  0.030 	&  1.27 	& 3 &  3\\
20181112A          & -17.4 	& -47.7 	&   589 	&    42 	&  2.1 	& 0.475  	&   0.021 		&  0.001 		&  0.001 		&  1.30 	& 4 &  4\\
20190102B          & -47.4 	& -33.5 	&   364 	&    57 	&  1.7 	& 0.291  	&   0.041 		&  0.003 		&  0.002 		&  1.27 	& 5  & 3 \\
20190523A          & 117.0 	&  44.0 	&   761 	&    37 	&  0.42 	& 0.660  	&   1.4 		&  0.2 		&  0.20 		&  1.0 	& 6 &  5 \\
20190608B          &  53.2 	& -48.5 	&   339 	&    37 	&  6.0 	& 0.118  	&   3.3 		&  0.2 		&  0.20 		&  1.27 	& 5 &  3 \\
20190611B\tablenotemark{\footnotesize b}          	
		         	& -47.1 	& -33.3 	&   321 	&    58 	&  2 .0	& 0.378  	&   0.18 		&  0.02 		&  0.020 		&  1.30 	& 5,7 & 3 \\
20190711A          & -49.1 	& -33.9 	&   593 	&    56 	& 6.5 	& 0.522 	& $<$  1.12 	& $\cdots$  	& $\cdots$ 	&  1.30 	& 5 & 6 \\
20190714A           & -71.1        &   48.7      &  504     &    39       & 2.0       & 0.2365  & $<$  2            & $\cdots$        & $\cdots$         &  1.27     & 7 &  7\tablenotemark{\footnotesize f}  \\
20191001A           & -17.3         &  -44.0     &   507     &    44      &  10        & 0.234    &  3.3                & 0.2                 & 0.2                &  0.824   &  7  & 8 \\
20200430A           &  17.1         &  52.5       &  380      &   27      &  15.        & 0.16      & 10                 &    5                 &  5                   &  0.865   & 7  &  9\tablenotemark{\footnotesize f}  \\
20200120E\tablenotemark{\footnotesize c}   & 142.2 	& 41.22 	& 88	&    41 	&     0.1    & $(3.6~\rm Mpc)$ 	& $<$ 30~ns   	& $\cdots$  	& $\cdots$	&  1.40      & $\cdots$ & 10 \\
20201124A     & 177.8         & -8.52     & 414     &   140     & 3.2        & 0.098     &   5.6              &    3                 & 3                   & 0.865     & 8,9      & 11 \\
\hline
 \enddata
\tablenotetext{a}{Pulse broadening has been measured at 0.15~GHz \citep[][]{2021Natur.596..505P} from Galactic scattering; the upper bound for this object refers to any extragalactic scattering}
\tablenotetext{b}{The association of this FRB with the candidate galaxy at the redshift $z_{\rm h}$ in column 7 is stated to be
tentative \citep[][]{2020Natur.581..391M}.}
\tablenotetext{c}{FRB source is associated with a globular cluster in the M81 system \citep[][]{2022Natur.602..585K} at a distance of 3.6~Mpc with a formally negative redshift.  Measured scintillations are Galactic in origin and correspond to a scattering time $\sim 27$~ns \citep[][]{2022NatAs.tmp...43N}.   Extragalactic scattering is not evident so we take 30~ns as an upper limit.}      
\tablenotetext{d}{References for $z_{\rm h}$: 
(1) \citet[][]{tbc+17}
(2) \citet[][]{2020Natur.577..190M};
(3) \citet[][]{2019Sci...365..565B};
(4) \citet[][]{2019Sci...366..231P};
(5) \citet[][]{2020Natur.581..391M};
(6) \citet[][]{2019Natur.572..352R}; 
(7) \citet[][]{2020ApJ...903..152H};
(8) \citet[][]{2021ATel14516....1K};
(9) \citet[][]{2021arXiv210609710R}.
}
\tablenotetext{e}{References for $\taud$: 
(1) \citet[][]{2019ApJ...882L..18J}; 
(2) \citet[][]{2020ApJ...896L..41C}; 
(3) \citet[][]{2020MNRAS.497.3335D};  				
(4) \citet[][]{2020ApJ...891L..38C}; 
(5) \citet[][]{2019Natur.572..352R};      						
(6) \citet[][]{2020MNRAS.497.1382Q};
(7)  \citet[][]{2019ATel12940....1B};
(8) \citet[][]{2020ApJ...901L..20B};
(9) \citet[][]{2020ATel13694....1K};
(10) \citet[][]{2022NatAs.tmp...43N};
(11) \citet[][]{2021ATel14502....1K}.
}
\tablenotetext{f}{Estimate for $\tau$ is  based on dynamic spectrum linked to quoted reference.}

\end{deluxetable*}

The dispersion measure, $\DM = \int ds\, \nelec(s)$ expressed in  standard units of $\DMunits$,  is estimated from chromatic arrival times and receives contributions from all non-relativistic plasmas along the line of sight (\LOS).   While we exclude the small contributions originating within the solar system we include all others between the solar system and an FRB source.  

 As is usual in the FRB literature, we write the measured DM for a source at redshift $\zh$ as the sum,
\be
\DM = \DMMW +  \DMigm(\zh) + \frac{\DMig}{1+ \zig} + \frac{\DMh}{1+z_{\rm h}},
\label{eq:DMexpression}
\ee
that includes  components from the Milky Way (mw),  the intergalactic medium (igm), a possible intervening galaxy or halo (igh), and a host galaxy (h), including its halo.   
The Milky Way term includes both the non-halo (`disk') and halo components,
$\DMMW = \DMMWdisk + \DMMWhalo$ that are estimated separately because their phenomenology and characterization differ substantially.     The IGM contribution displays cosmic variance indicative of the stochastic distribution of galaxy halos and requires  a statistical dependence on redshift, $z$. 
 The last two terms involve reduction of  the rest-frame dispersion measures, $\DMig$ and $\DMh$, by  $1/(1+z)$ factors for the intervening and host galaxies.     For simplicity,  all possible contributions to $\DMh$ (galaxy disk, halo, and circumsource region) are lumped together.  
  
\subsection{Milky Way Contribution}
\label{sec:DMmw}

The `disk'   contribution to DM from the MW   is obtained by integrating the direction-dependent NE2001 model \cite[][]{cl02} through the entire Galaxy to give  $\DMMWdisk(l,b)$. 
 The NE2001 model actually  comprises  two disk components, spiral arms, and localized regions.  The differences between the NE2001 model and the alternative YMW16 model \cite[][]{2017apj...835...29y} are negligible for FRBs at Galactic latitudes $\gtrsim 20^{\circ}$ but NE2001 is more accurate for FRB~20121102A in the Galactic anticenter direction \citep[][]{Ocker_2021}.  Also, the YMW16 model does not properly estimate scattering observables and thus cannot be used in our analysis of scattering. 
  
For high-latitude lines of sight,  the spread in estimated values for $\DMMW$ is several tens of $\DMunits$, primarily from uncertainties in the contribution from the Galactic halo.  For the two low latitude cases, FRB~20121102A and 
FRB~20180916B, the uncertainty in $\DMMW$ could be substantially  larger.  However, for
FRB~20121102A,   the measured redshift and  the independent constraint on $\DMh$ from
Balmer-line measurements \citep[][]{tbc+17}  provide tighter ranges for the host-galaxy and IGM contributions.     
For FRB~20180916B,  the total DM is small enough that  a substantially larger $\DMMW$ than provided by 
the NE2001 model for $\DMMWdisk$ is not allowed, particularly for a larger estimated Galactic halo contribution, $\DMMWhalo$. 

 To include uncertainties in the disk DM estimate from NE2001,  we employ a flat probability density function
(PDF) $\pdfmwdisk(\DM)$ with a \epsoonepercent\% spread  (i.e. $\pm 20$\% deviation from the mean) centered on the  NE2001 estimate.   While larger departures from NE2001 (or YMW16) estimates are seen for some  individual {\em Galactic} pulsars due to unmodeled HII regions,  estimates at Galactic latitudes $\vert b \vert \gtrsim 20^{\circ}$ appear to have much less estimation error, gauged in part by near agreement of the NE2001 and YMW16 models and also by consistency (in the mean) with parallax distances of high-latitude pulsars \citep[][]{2019ApJ...875..100D}.   As a test, using a smaller 20\% spread  on NE2001 DM values yielded very little change in the final results. 

Estimates in the literature for the MW's halo contribution to DM  range from $25~\DMunits$
to   $\sim 80$~$\DMunits$ \citep[][]{2018apj...852l..11s, 2018mnras.474..318p,2019MNRAS.485..648P,Yamasaki_2020}, large enough to impact estimates of  extragalactic contributions.  
Though it has been argued that the MW halo could contribute as little as 10~$\DMunits$
\citep[][]{10.1093/mnrasl/slaa095},
we conservatively use a flat distribution $\pdfmwhalo(\DM)$ extending from 25 to 80~$\DMunits$. 
We note however that FRB~20200120E   in the direction of M81 \citep[][]{Bhardwaj_2021} shows a total
$\DM = 87.8~\DMunits$ in the direction $(l,b )= (142.^{\!\!\circ}19, 41.^{\!\!\circ}2)$. With estimates
of $\DMMWdisk \sim 40$ and 35~$\DMunits$ for the NE2001 and YMW16 models, respectively, 
only 48 to 53~$\DMunits$ is allowed for $\DMMWhalo + \DMigm + \DM_{\rm M81}$.    

Recent work has shown that 
the burst source is coincident with a globular cluster in the M81 system \citep[][]{2022Natur.602..585K},
so the disk of  M81 and the globular cluster  make no or little contribution to the DM.  If we take the assumed minimum MW halo contribution of $\DMMWhalo= 25~\DMunits$,   only 23 to 28~$\DMunits$ are contributed by M81's halo along with a minimal contribution from the IGM.   An alternative reckoning is to attribute $\DMigm \lesssim 1~\DMunits$ using the mean cosmic  density $\nezero$ (next subsection) and the 3.6~Mpc distance to M81, leaving a total
of $\lesssim 53~\DMunits$ for the summed contributions of the MW and M81 halos.    While   the halo of M81 may be smaller and less dense than that of the MW,  FRB~20200120E provides constraints that are not inconsistent with our adoption of $25~\DMunits$ as the minimum of the MW's halo contribution.    

In our analysis we marginalize over the total Milky Way \DM\ contribution using the  PDF
for the sum $\DMMWdisk + \DMMWhalo$ 
 that is the convolution  of the disk and halo PDFs, 
 $\pdfmw(DM) = \pdfmwdisk * \pdfmwhalo, $
which  is trapezoidal in form.

\subsection{Intergalactic Medium Contribution}
\label{sec:DMigm}

The FRBs analyzed in this paper have  redshifts   $z<1$, so it is reasonable to consider the 
IGM to be almost completely ionized.
We calculate the IGM term using  a nominal electron density  for the diffuse IGM  at $z=0$ given by 
a fraction $\figm$ of the  baryonic contribution to the closure density, $\nezero = 2.2\times10^{-7}$~cm$^{-3}\ \figm $, evaluated	
using cosmological parameters  from the Planck 2018 analysis implemented in {\tt Astropy}. 
\citet[][]{2018apj...852l..11s} specify a fiducial range   $\figm \approx 0.6\pm 0.1$  for the baryon fraction although
\citet[][]{Yamasaki_2020} adopt a range $[0.6, 0.9]$ consistent with an earlier conclusion that $\figm > 0.5$ 
\citep[][]{2012ApJ...759...23S}.   Measurements of the kinematic Sunayev-Zeldovich effect 
\cite[e.g.][]{2016PhRvL.117e1301H,2021PhRvD.104d3518K}
demonstrate consistency of the baryon fraction with big-bang nucleosynthesis by attributing the apparent deficit of baryons near galaxies  to the presence of ionized gas.    Those results suggest   $\figm \sim 0.8$ according to the baryon budget presented in \citet[][Figure 10]{2012ApJ...759...23S} in agreement with \citet[][]{2018ApJ...867L..21Z}.  In this paper we consider a range of values $0.4 \le \figm \le 1$ for most of the analysis  but adopt $\figm = 0.85$ when a specific nominal  value is needed.   We also show that a value $\sim 0.85$ minimizes the bias and minimum error for a combined dispersion-scattering redshift predictor for the sample of FRBs that have both redshift and scattering measurements. 

For a constant co-moving density the IGM makes a mean contribution,
\be
\!\!\!\!
 \DMigmbar(z) &=& \nezero \dH \int_0^z d\zp\, \frac{(1+\zp)}{E(\zp)} 
  \nonumber \\
  &\equiv& \nezero \dH \rtilde_1(z) \approx  972\,\DMunits\, \figm \rtilde_1(z),
\label{eq:DMIGMbar}
\ee
where $\dH = c / \Hzero$ is the Hubble distance and  $E(z) = [\Omega_{\rm m} (1+z)^3 + 1 - \Omega_{\rm m}]^{1/2}$ for a flat $\Lambda$CDM universe with a matter density $\Omega_{\rm m}$. 
The second equality  defines the integral  $\rtilde_1(z)$ where $\rtilde_1 \simeq z$ for $z\ll 1$. 

\begin{figure}[htbp] 
   \centering
   \includegraphics[width=\linewidth]{\NewFigures/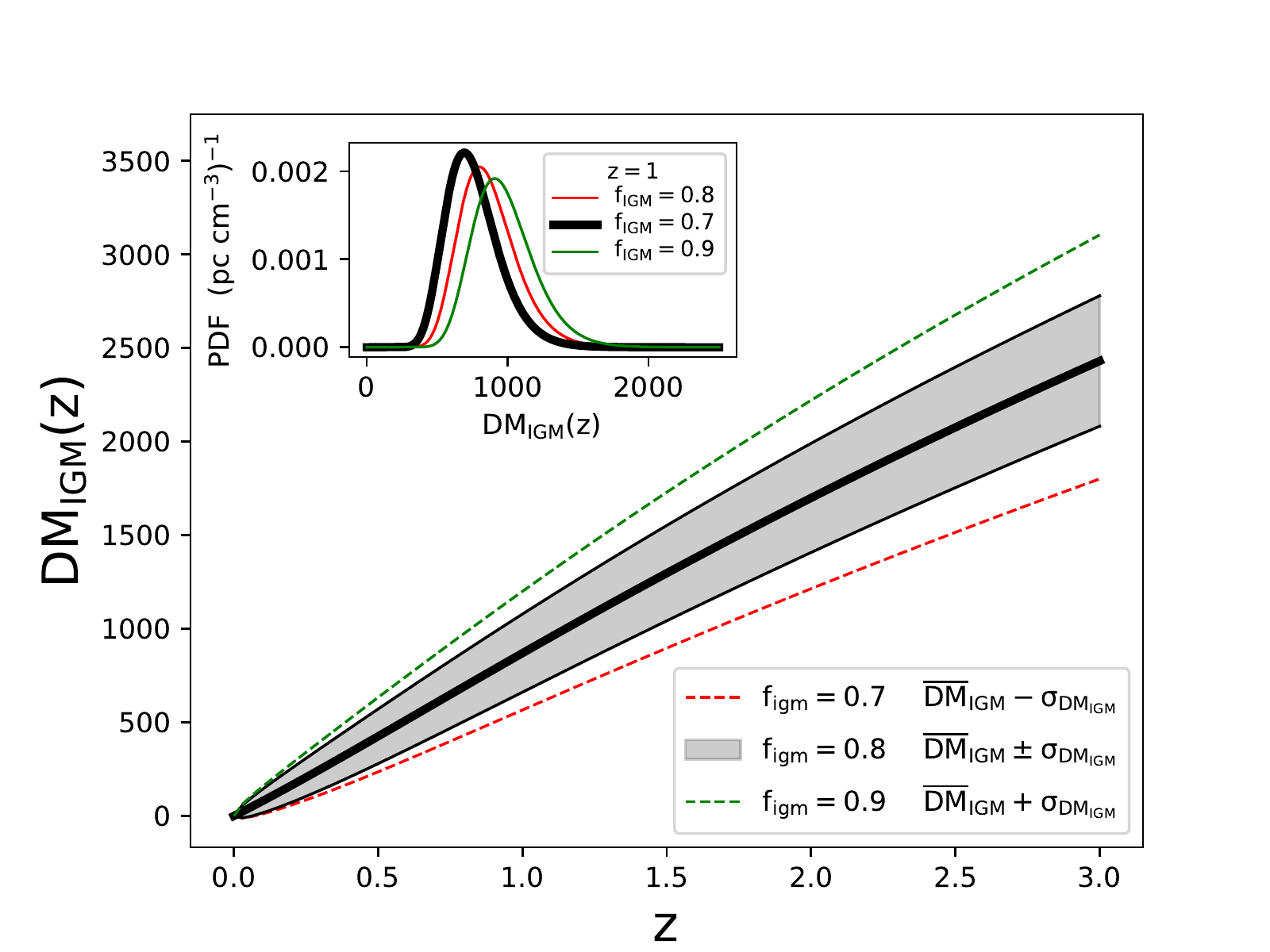}
   \caption{Modeled IGM contribution to DM as a function of redshift for three values of the 
   baryonic fraction, $\figm$.   The grey band shows  $\DMigmbar(z) \pm \sigDMigm(z)$ for
   $\figm=0.8$ while the dashed green line is the upper range $\DMigmbar(z) +\sigDMigm(z)$
   for $\figm = 0.9$ and the dashed red line is the lower range $\DMigmbar(z)+\sigDMigm(z)$
   for $\figm=0.7$.   The inset shows the PDF of $\DMigm$ at $z=1$ for the three values
   of $\figm$. 
   }
   \label{fig:DMvsz_LCDM}
\end{figure}

%


\begin{deluxetable*}{l c r    r  r   C  r  r  r   C  r r r    R R   r r r}[t]
\tabletypesize{\footnotesize}
\tablecaption{ FRB DM Inventory and Redshift Estimates from DM \label{tab:FRBinventory}}
\tablehead{
\colhead{FRB} &
\colhead{DM} & 
\colhead{$z_{\rm h}$} &
\multicolumn{2}{c}{$\DMMW^{\tablenotemark{\footnotesize a}}$} &&
\multicolumn{3}{c}{$\DMigm^{\tablenotemark{\footnotesize b}}(\figm=0.85)$} &&
\multicolumn{3}{c}{$\DMh^{\tablenotemark{\footnotesize c}}(\figm=0.85)$} &&
\multicolumn{3}{c}{$\zhat(\DM, \figm=0.85)$} 
\\
\cline{4-5}
\cline{7-9}
\cline{11-13}
\cline{15-17}
\colhead{} & 
 \colhead{($\DMunits$)} &
 \colhead{} & 
\multicolumn{2}{c}{($\DMunits$)}  & &
 \multicolumn{3}{c}{($\DMunits$)}  & &
 \multicolumn{3}{c}{($\DMunits$)}  
\\
\colhead{(1)} &
\colhead{(2)} &
\colhead{(3)} &
 \multicolumn{1}{c}{(4)}  & 
 \multicolumn{1}{c}{(5)}  & &
 \multicolumn{1}{c}{(6)}  & 
 \multicolumn{1}{c}{(7)}  &
 \colhead{(8)} &&
  \colhead{(9)} &
  \colhead{(10)} &
  \colhead{(11)} &&
  \colhead{(12)} &
  \colhead{(13)} &
  \colhead{(14)} &
 }
\tablecolumns{17}
\startdata
20121102A &  557  & 0.193 & 241 & $\pm$27 && 152 &  $-$59  & $+$97 && 215 &  $-83$  & $+$69 && 0.373 & $-$0.117 & $+$0.125  \\ 
20180916B &  349  & 0.034 & 252 & $\pm$28 && 24 &  $-$12  & $+$36 && 82 &  $-33$  & $+$33 && 0.103 & $-$0.048 & $+$0.066  \\ 
20180924A &  361  & 0.321 & 93 & $\pm$17 && 268 &  $-$87  & $+$129 && 99 &  $-52$  & $+$67 && 0.319 & $-$0.105 & $+$0.113  \\ 
20181112A &  589  & 0.475 & 94 & $\pm$17 && 411 &  $-$114  & $+$159 && 206 &  $-118$  & $+$128 && 0.571 & $-$0.153 & $+$0.158  \\ 
20190102B &  364  & 0.291 & 110 & $\pm$17 && 240 &  $-$81  & $+$122 && 100 &  $-53$  & $+$64 && 0.302 & $-$0.101 & $+$0.110  \\ 
20190523A &  761  & 0.660 & 90 & $\pm$16 && 585 &  $-$142  & $+$188 && 261 &  $-155$  & $+$178 && 0.762 & $-$0.183 & $+$0.188  \\ 
20190608B &  339  & 0.118 & 90 & $\pm$16 && 87 &  $-$39  & $+$72 && 190 &  $-68$  & $+$45 && 0.297 & $-$0.100 & $+$0.108  \\ 
20190611B &  321  & 0.378 & 110 & $\pm$17 && 320 &  $-$97  & $+$141 && 58 &  $-26$  & $+$43 && 0.253 & $-$0.089 & $+$0.099  \\ 
20190711A &  593  & 0.522 & 109 & $\pm$17 && 455 &  $-$122  & $+$167 && 171 &  $-100$  & $+$123 && 0.559 & $-$0.151 & $+$0.157  \\ 
20190714A &  504  & 0.236 & 91 & $\pm$16 && 191 &  $-$69  & $+$109 && 289 &  $-116$  & $+$83 && 0.481 & $-$0.138 & $+$0.143  \\ 
20191001A &  507  & 0.234 & 97 & $\pm$17 && 188 &  $-$68  & $+$109 && 287 &  $-115$  & $+$82 && 0.477 & $-$0.137 & $+$0.143  \\ 
20200430A &  380  & 0.160 & 80 & $\pm$16 && 123 &  $-$50  & $+$87 && 217 &  $-84$  & $+$58 && 0.356 & $-$0.113 & $+$0.120  \\ 
FRB20200120E &  88  & $\cdots$ & 93 & $\pm$17 && $< 1$ &  $\cdots$  & $\cdots$ && 13 &  $-$6  & $+$7 && $\cdots$& $\cdots$ & $\cdots$  \\ 
20201124A &  414  & 0.098 & 192 & $\pm$23 && 71 &  $-$33  & $+$64 && 172 &  $-60$  & $+$44 && 0.305 & $-$0.100 & $+$0.107  \\ 
\hline
\enddata
\tablenotetext{a}{$\DMMW$ includes contributions from the disk using the NE2001 model  and halo  described in  \S\ref{sec:DMmw}.}
\tablenotetext{b}{$\DMigm$ is calculated using the redshift $\zh$ and the log-normal model of \S\ref{sec:DMigm}.}
\tablenotetext{c}{$\DMh$ values are in the frame of the host galaxy at redshift $\zh$ and are calculated by integrating over the PDFs for $\DMMW$ and $\DMigm$ using Eq.~\ref{eq:postDMh}.}
\end{deluxetable*}

Cosmic variance  of the IGM density 
\cite[e.g.][]{mcq14}
 produces variations in \DM\ characterized as a zero-mean  process 
$\dDMigm$  with a distance-dependent RMS, $\sigDMigm(z)$.  We approximate the results of 
cosmological simulations by adopting a simple scaling law,
\be
\sigDMigm(z) = \left[\DMigmbar(z) \DM_{\rm c}\right]^{1/2},
\label{eq:DMIGMsig}
\ee
where $\DMc = 50~\DMunits$.   For $z=1$ this gives $\sigDMigm(1) = 233\figm~\DMunits$.
We obtained results by  increasing $\DMc$ to 100~$\DMunits$, i.e. a 41\% increase in $\sigDMigm$, and found little change in the net results described in the rest of the paper.

Our scaling law implies a decrease in the fractional variation of $\DMigmbar$ with increasing redshift as $\sigDMigm(z) /  \DMigmbar(z) = [\DM_{\rm c} / \DMigmbar(z)]^{1/2}$, which is consistent with simulation results reported by \citet[][]{iok03,ino04, mcq14} and \citet[][]{dgbb15}, 
although there is considerable uncertainty related to the number of halos encountered
along a \LOS\ and their sizes.  This is exemplified in  
\citet[][]{2019ApJ...886..135P},  who report substantially different DM distributions between uniform weighting and matter-weighted \LOS\ integrals.      Simulations also indicate substantial skewness of $\DMigm$ toward larger values.

Cosmic variance in $\DMigm$ is implemented using a redshift-dependent probability density function (PDF)
  $\pdfigm(\DMigm; z, \figm)$ that is  log-normal  in form,  ${\cal N}(\mu, \sigma)$ with parameters 
\be
\sigma &=& \left\{ \ln [ 1 + (\DMigmsig / \DMigmbar)^2] \right\}^{1/2} ,
\\
\mu &=& \ln \DMigmbar - \sigma^2/2.     
\ee
The  skewness of the distribution, 
$\gamma = (e^{\sigma^2} + 2) \sqrt{e^{\sigma^2} - 1}$, decreases with redshift and so is at least qualitatively consistent with  published simulations cited above. 

\Fig\ref{fig:DMvsz_LCDM} shows $\DMigm(z)$  for
 $\figm = 0.8 \pm 0.1$ using the parameterization of Eq.~\ref{eq:DMIGMbar} and \ref{eq:DMIGMsig}. 
 The inset shows the PDFs of $\DMigm$ at $z=1$ for the same values of $\figm$.   Cosmic variance in $\DMigm$ implies considerable variations in DM-derived values of redshift even if the baryonic fraction $\figm$ is known.  Likewise,   uncertainties in $\figm$ exacerbate those of DM-derived redshifts.    In a later section we demonstrate that scattering measurements can further improve   redshift estimates as well as constrain the value of $\figm$. 

\begin{figure*}[t] 
   \centering
    \includegraphics[width=0.9\linewidth]{\NewFigures/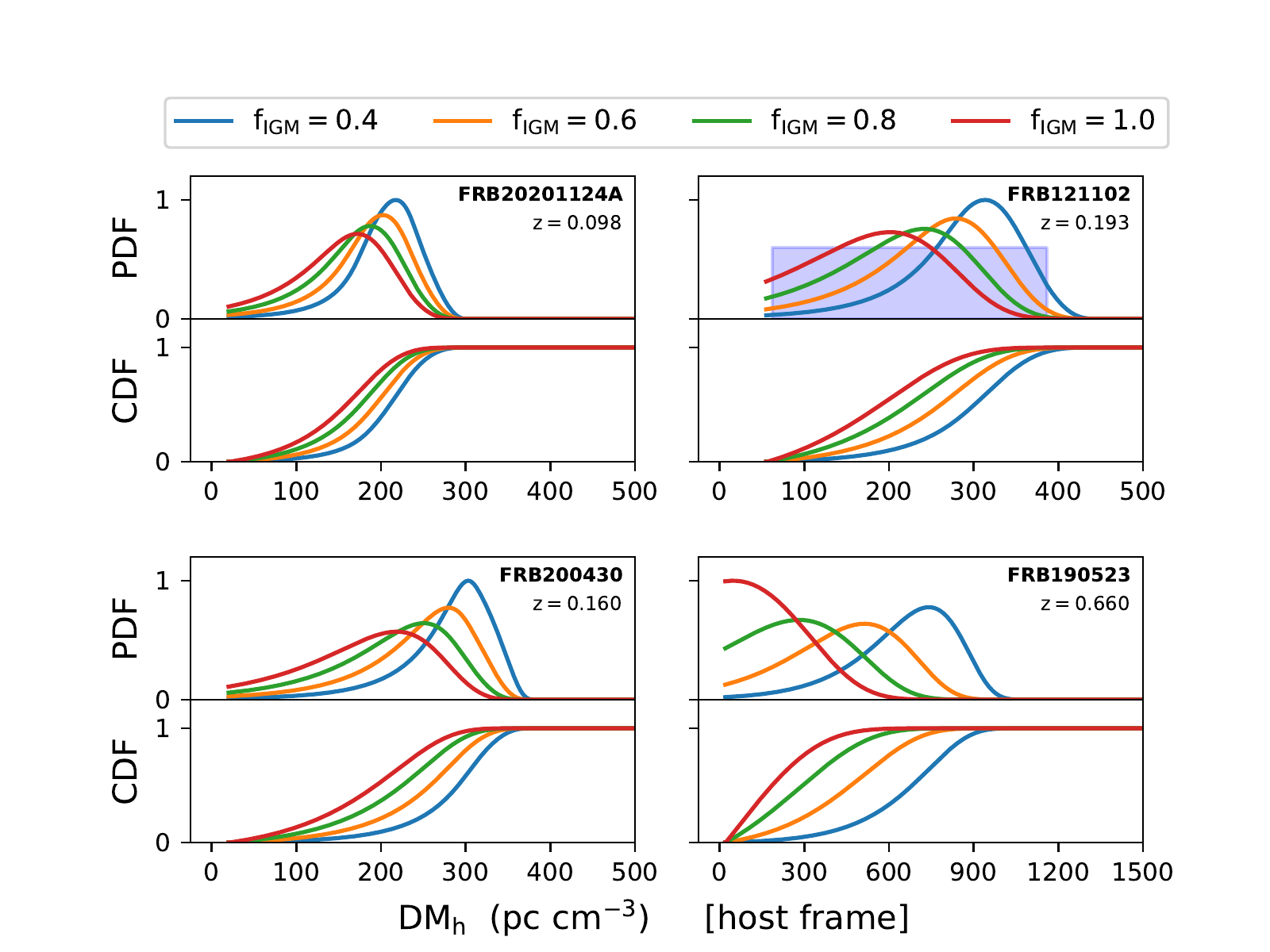}
   \caption{\explain{White space was removed between each PDF and CDF pair.}  Posterior PDF and CDF for $\DMh$ (in the host galaxy's frame) for four selected FRBs and  four values 
   of the IGM's baryonic fraction $\figm$.  PDFs are normalized to unit maximum.   A minimum value, $\DM_{\rm h, min} = 20~\DMunits$, has been imposed as a prior on $\DMh$ except for FRB~20121102A, for which 50~$\DMunits$ was used.  Results are not sensitive to this minimum except  for FRB~20200120E (not shown), which has a small extragalactic contribution to DM.   The shaded region for FRB~20121102A designates the constraint on $\DMh$ from Balmer line measurements \citep[][]{tbc+17,bta+17}.
   }
   \label{fig:DMh_PDF_CDF_page3}
\end{figure*}

\subsection{Posterior PDF for $\DMh$ for FRBs with  Redshifts}
\label{sec:postDMh}

A frequent assumption that appears in the FRB literature is a constant host-galaxy contribution to DM, often with a value
$\DMh^{\rm (assumed)} = 50~\DMunits$ \citep[e.g.][and references therein]{2021MNRAS.501.5319A} accompanied by a statement that a range of values does not matter in an analysis of mostly large FRB DMs.    We find that this is not the case for published FRBs with associated galaxy redshifts.   Indeed our conclusion is underscored by the discovery of the low-redshift FRB~190520  with a large total DM ($z = 0.241$, DM = 1202~$\DMunits$), that requires a large $\DMh$ \citep[][]{2021arXiv211007418N}.
 In this paper a necessary step is to calculate the Bayesian posterior PDF for each FRB. 

\begin{figure}[t] 
   \centering
   \includegraphics[width=\linewidth]{\Figures/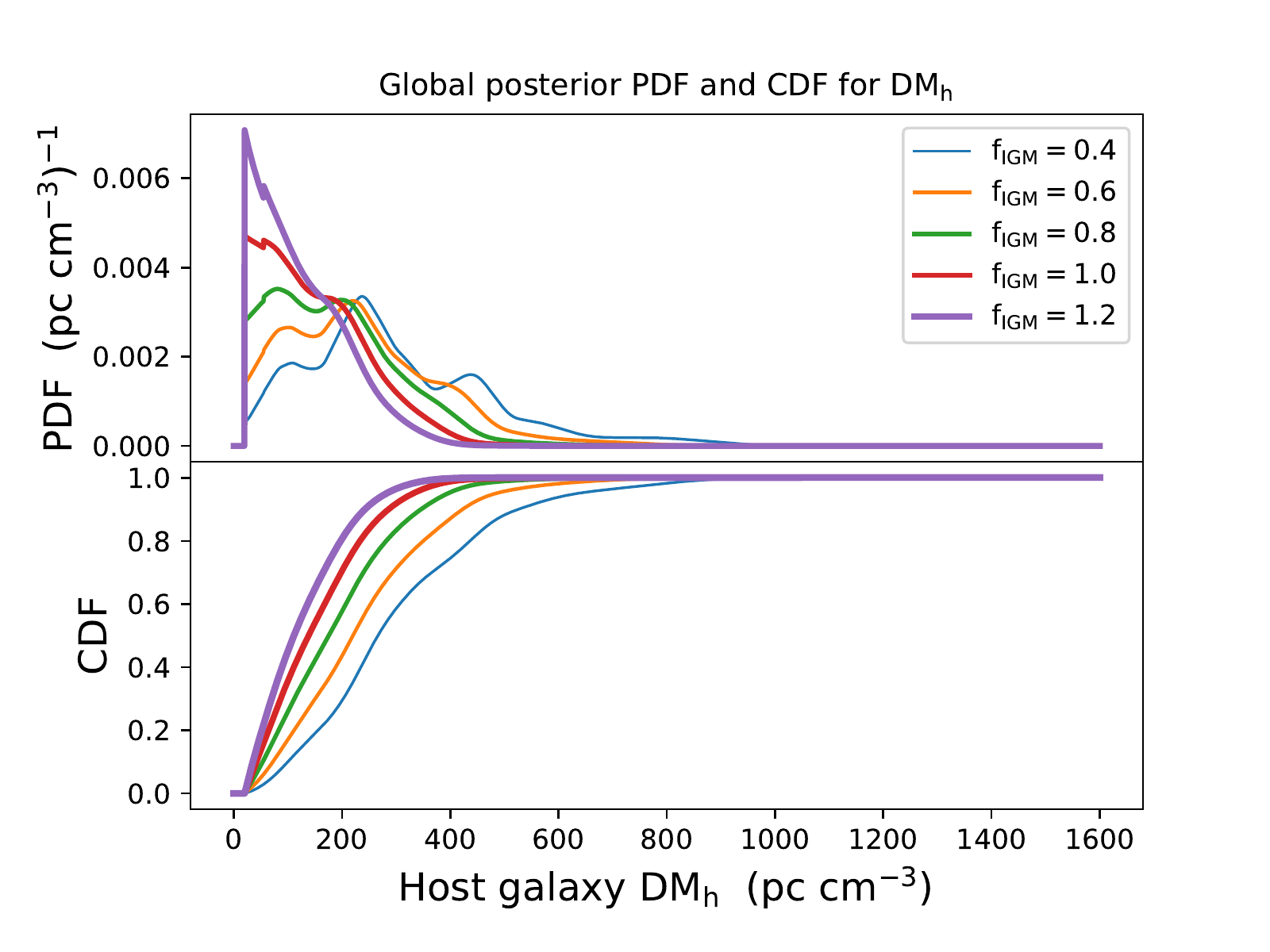}
   \caption{Composite PDF and CDF of host-galaxy dispersion measures, $\DMh$, calculated for five  values of 
   the fraction of baryons $\figm$ in the diffuse IGM.     The  composite PDF is the unweighted sum of the posterior PDFs for the FRBs in Table~\ref{tab:FRBproperties} with FRB~20200120E excluded due to its special geometry with respect to M81.
   }
   \label{fig:global_pdf_cdf_DMh}
\end{figure}

We wish to estimate the dispersion measure  $\DMh$ contributed by a host galaxy (in its rest frame) taking into account  uncertainties in the MW and IGM contributions.    We assume that measurements of \DM\ and redshift $z$ have negligible error. 
From  \Eq\ref{eq:DMexpression} the conditional PDF for $\DMh$ is
\be
\pdfh(\DMh\vert \DM, \DMMW, \zh) &=& 
\nonumber \\
\!\!\!\! \!\!\!\! 
(1+\zh)^{-1} \pdfigm(\DM - \DMMW &-& \DMh / (1+\zh)) .
\label{eq:postDMh}
\ee
Marginalization over the the PDF $\pdfmw$ for the MW contribution $\DMMW$   then gives 
$\pdfh(\DMh\vert \DM,  \zh)$.

\Fig\ref{fig:DMh_PDF_CDF_page3} shows the PDF $\pdfh$ and the corresponding CDF for four selected objects and four different values for the baryonic fraction, $\figm$.    The FRBs include   FRB~20200120E with a very small total \DM\ compared to another with a potentially large host-galaxy DM, FRB~20190523A.   
\explain{{\bf This text deleted:} Different baryonic fractions yield negligible differences for FRB~20200120E, while differences in $\DMh$ are of order 100 to 200~$\DMunits$ for FRB~20121102A and FRB~20200430A and several hundred $\DMunits$ for FRB~20190523A.}
\added{
Due to the proximity of FRB~20200120E, different assumed values for the baryonic fraction yield negligible changes in the estimates of $\DMh$ because the IGM contributes very little to DM.  However, changes in 
$\DMigm$ and thus $\DMh$ are of order 100 to 200~$\DMunits$ for FRB~20121102A and FRB~20200430A and several hundred $\DMunits$ for FRB~20190523A for different values of $\figm$. }
The range of $\DMh$ for FRB~20121102A is consistent with that found from analysis of Balmer lines  
\citep[][]{tbc+17,bta+17}, designated by the shaded band in the figure.  

Table~\ref{tab:FRBinventory} gives the DM inventory for all of the FRBs from Table~\ref{tab:FRBproperties}. 
Columns (1)-(3) give the FRB name, measured DM, and host redshift. The next eight columns give the \DM\ and credible range for the Milky Way,  IGM, and host-galaxy contributions.   The last three columns give redshift estimates using only the DM inventory, as discussed in \S\ref{sec:dm2z}.  In that section,  the DM-based redshifts are compared with those obtained using a combined scattering-DM redshift estimator.

The line of sight to FRB~20200120E,   the FRB in a globular cluster associated with M81, evidently does not sample the disk of M81.   Using the PDF for the MW contribution from the disk and halo and the negligible contribution from the diffuse IGM, 
the halo of M81 is found to contribute a DM of $\DM_{\rm M81, halo} =  13^{+ 21}_{-8}~\DMunits$. 

The posterior PDFs for host-galaxy DMs are combined in \Fig\ref{fig:global_pdf_cdf_DMh}, which shows the global PDF and CDF for 13 objects (excluding FRB~20200120E) using five values for $\figm$, including a value of 1.2 that exceeds the nominal limit for the diffuse IGM.   Note again that $\DMh$ is defined in the host-galaxy frame, not the observer's frame.    The PDF shifts to larger values of $\DMh$ for larger $\figm$. 
We find the  range $\figm \simeq 0.85\pm 0.05$  to be a good representation of our overall results (see \S\ref{sec:dmtau2z}).  For $\figm = 0.85$,   the 68\% credible interval is $\DMh = 166^{+122}_{-100}~\DMunits$,
a result that is not inconsistent with those of  \citet[][]{2022MNRAS.509.4775J}.
%
The CDF implies that about 5\% of FRBs will show  $\DMh \gtrsim 400~\DMunits$.    The discovery of FRB~20190520B with an implied $\DMh$ well in excess of $400~\DMunits$ 
\citep[][]{2021arXiv211007418N}
 will extend the tail of the global PDF further but is not overly  inconsistent with the statistics of  the sample we have analyzed in Table~\ref{tab:FRBproperties}.   A detailed analysis  of FRB~20190520B is given in 
 \citet[][]{2022arXiv220213458O}.
 
\section{Scattering Inventory}
\label{sec:ScatteringInventory}

The scattering time $\taud$ is the other propagation observable that constrains intervening plasmas. 
The scintillation bandwidth $\dnud \simeq (2\pi\taud)^{-1}$ yields the same information, though in practice it has only been measured convincingly for scintillation caused by Galactic scattering 
\citep[e.g.][]{mls+15b,2018ApJ...863....2G,2019ApJ...876L..23H,2020Natur.577..190M,2020ApJ...901L..20B}
whereas directly measured pulse broadening has been identified \added{primarily} from scattering that is extragalactic in origin \added{except for FRB~20180916B, which shows Galactic scattering with 
$\tau = 46\pm10$~ms at 0.15~GHz \cite[][]{2021Natur.596..505P} that is consistent with scintillation bandwidths measured at higher frequencies. }

\explain{This text deleted: Of course low-Galactic latitude and low-frequency FRBs will eventually show Galactic pulse broadening. }

Contributions to scattering times from different media along the \LOS\ are additive
(c.f. \Eq\ref{eq:tau}). 
Parallel to the DM inventory in \Eq\ref{eq:DMexpression},  
we expand $\tau$ into terms involving the MW (disk and halo), the IGM,
a possible intervening galaxy or halo, and a host galaxy (including its halo), 
\be
\taud(\nu) &=& \taumw(\nu) + \taudigm(\nu, z) 
\nonumber \\
&&
	+  \frac{\taudig(\nu)} {(1+\zig)^{\xtau-1}} 
	+ \frac {\taudh(\nu)} {(1+\zh)^{\xtau-1}} ,
\label{eq:tauexpression}
\ee
where we adopt a power-law  frequency  scaling, $\taud(\nu) \propto \nu^{-\xtau}$,  with
an index $\xtau \simeq 4$.
The redshift scalings in the last two terms of Eq.~\ref{eq:tauexpression} take into account that 
scattering  occurs at $\nuprime = \nu(1+z) $ in a galaxy's rest frame
for an observation frequency $\nu$ and that  dilation of the observed scattering time  is by a factor $(1+z)$   
\citep[see also][]{2013apj...776..125m}.

In the following we develop a model for scattering media and compare it against Galactic pulsar measurements.   We argue that only the disk components of galaxies contribute significantly to scattering while galaxy  halos and the IGM  contribute negligibly.  

\begin{figure}[t] 
   \centering
   \includegraphics[width=\linewidth]{\Otherfigures/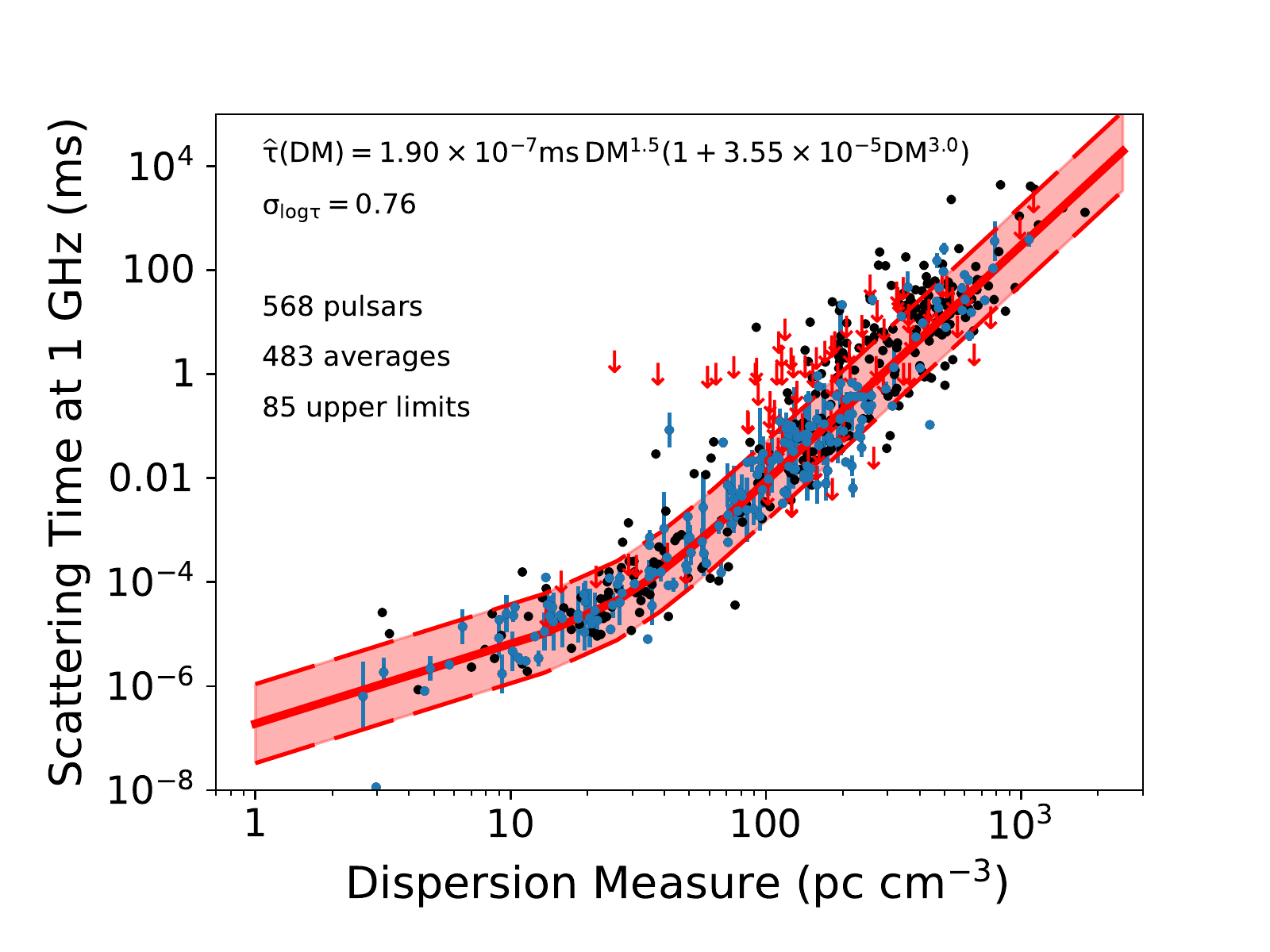}
   \caption{$\taud$ vs. DM  for Galactic pulsars and FRBs.   The fitted line (solid red) and $\pm 1\sigma$ variations (dashed red) are  based  on measurements and upper limits on $\taud$ for Galactic pulsars.  Blue points with error bars are averages over multiple measurements while black points are single measurements from the literature.  The plotted pulsar values are from numerous literature sources and are available on request from the corresponding author.
}
   \label{fig:pulsar_tauDM}
\end{figure}

\begin{figure}[t] 
   \centering
    \includegraphics[width=\linewidth]{\Cloudlet/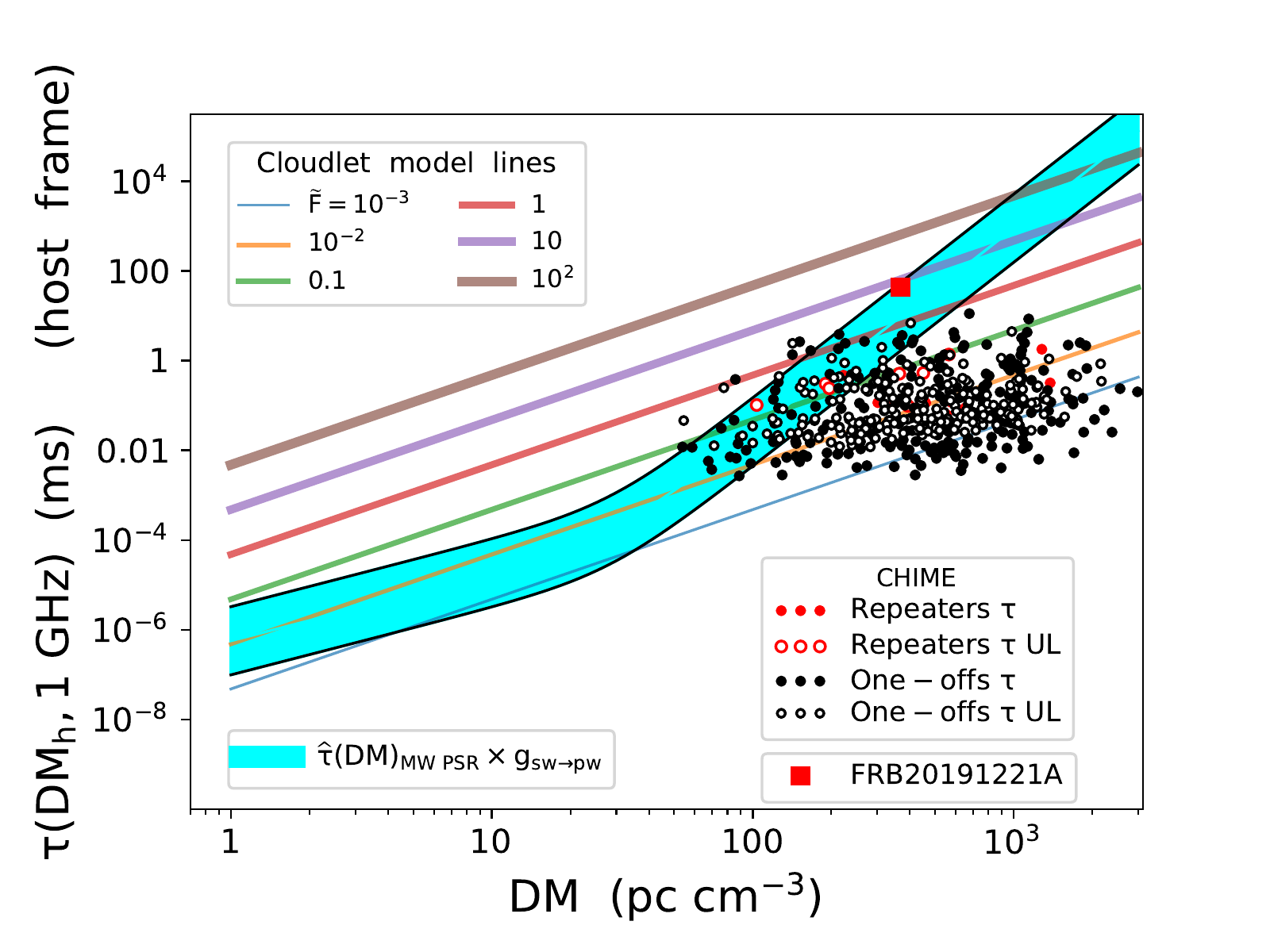}
   \caption{The scattering time  $\taudh$ vs. dispersion measure $\DM$.   
   The cyan band shows the range of scattering times seen from Galactic pulsars evaluated
   using  $\pm\sigma_{\log_10(\taud)} = 0.76$  about the mean curve.
    The $\tau$-DM relation for pulsars from \Fig\ref{fig:pulsar_tauDM}  is shown here after shifting it upward to account for 
   plane wave scattering relevant to FRBs (see text). 
    Lines show the scattering time for the cloudlet model of Eq.~\ref{eq:taucloudlet} for different values of the
   fluctuation parameter $\Ftilde$ (Eq.~\ref{eq:ftilde}).  The units of $\Ftilde$ are 
   ${\rm (pc^2~km)^{-1/3}}$.   
    Applying the relation to a host galaxy then requires that the  DM axis corresponds to  $\DMh$,  the host-galaxy's contribution,  and the $\tau$ axis is in the rest frame of the host galaxy.
    Also shown are FRBs from the CHIME FRB catalog \citep[][]{2021ApJS..257...59A}  (black points)
    The large red square is for the heavily scattered FRB~20191221A detected with CHIME 
    \citep[][]{2021arXiv210708463T}.
   For plotting the FRB points, the {\it total} DM has been used, which includes 
   Galactic and IGM contributions in addition to the host galaxy contribution.    
}
   \label{fig:taudm}
\end{figure}

\subsection{The $\taud$-\DM\ Relation for Galactic Scattering}

We incorporate much of what has been learned about temporal scattering from Galactic pulsars. 
\Fig\ref{fig:pulsar_tauDM} shows scattering times plotted against dispersion measure for 568 pulsars, including upper limits,  using data from the literature. Scattering times from different radio frequencies have been scaled to 1~GHz using a scaling law with $\xtau = 4$.  

Multifrequency observations yield a range of roughly $3 \lesssim \xtau \lesssim 4.5$ for the power-law index $\xtau$, whereas idealized models of diffraction from small-scale density fluctuations in the interstellar plasma  indicate $\xtau = 4$  \cite[e.g.][]{sch68, ric90} or $\xtau = 2\beta / (\beta-2) = 4.4$ for the simplest form of Kolmogorov fluctuations with
a wavenumber spectral slope $\beta = 11/3$.
Departures from $\xtau=4.4$ are expected if the inner 
scale for the fluctuations is larger than the diffraction scale \cite[][]{sg90,bcc+04,2009MNRAS.395.1391R}, or if
scattering is anisotropic  \cite[e.g.][]{bmg+10}. Scattering regions that are finite in size transverse to the \LOS\
also alter the scaling law \cite[][]{2001ApJ...549..997C}.    
These effects invariably {\it reduce} $\xtau$ from the simple Kolmogorov value. 
Keeping this variety of scaling  exponents in mind, we adopt $\xtau=4$ as a fiducial value.  
This value  is also consistent with the multifrequency analysis of 
\citet[][]{bcc+04} and \citet[][]{kmn15}.

\replaced{Applying  a fitting}{Fitting a}  function $\taudhat(\DM) = A\times \DM^{a} (1 + B\times\DM^{b})$ 
\cite[][]{rmd+97}
to the pulsar data 
 yields the scattering-\DM\ relation for Galactic pulsars at frequencies $\nu$ in GHz,
\be
\left[\taudhat(\DM, \nu)\right]_{\rm mw,psr}  &=& 1.90\times10^{-7}\, {\rm ms}\times
\nu^{-\xtau}\DM^{1.5} 
\nonumber \\
&&\times(1 + 3.55\times10^{-5}\,\DM^{3.0}) ,
\label{eq:hockeystick}
\ee
with  scatter  $\sigma_{\log\taud} = 0.76$ 
\citep[][]{bcc+04, 2019ARAA..57..417C}.
The fit is shown in \Fig\ref{fig:pulsar_tauDM} as a red band with a centroid line given by \Eq\ref{eq:hockeystick} and the upper and lower boundaries corresponding to $\pm 1~\sigma_{\log\taud} $. 
The band steepens significantly at large 
DMs, a feature that is due to the larger density fluctuations in the inner Galaxy, where large-DM pulsars are located, compared to those near the solar system or in the outer galaxy \citep[][]{cwf+91, cl02}. 

The measured scattering times necessarily include the fact that  Galactic pulsars are embedded in the interstellar scattering medium.   The same medium will scatter FRBs but by larger amounts because of their much larger distances.  The difference between spherical wavefronts from Galactic pulsars and plane waves from distant extragalactic FRBs amounts to an increase by a factor
 $\gswpw = 3$ in the scattering time.    The same holds true for FRBs scattered by their host galaxies by reciprocity (or by time reversal of propagation).   

\Fig\ref{fig:taudm} shows  the  distribution of $\taud$  vs. \DM\ for Galactic pulsars after applying this geometrical correction.   
The cyan band in Figure~\ref{fig:taudm} is a schematic depiction of the fit to Galactic pulsars. 
  Also shown are scattering times from FRBs in the first CHIME catalog of 535 distinct FRBs and a measurement of the largest measured scattering for 
FRB~20191221A  \citep[][]{2021arXiv210708463T},
$\tau(0.6~{\rm GHz}) = 340\pm10$~ms or $\tau(1~{\rm GHz}) \simeq 44\pm 1.3$~ms. 
This latter point is included to show the wide range of values for FRB scattering.  

The abscissa in  \Fig\ref{fig:taudm} should in principle stand for the \DM\ of the relevant extragalactic scattering medium with any due redshift correction, but of course we do not know the redshifts of most FRBs.   
Using the nominal total DM values shows a long-recognized feature  \citep[e.g.][]{2019ARAA..57..417C} of FRB scattering that they are ``underscattered'' compared to Galactic pulsars.   This signifies that the scattering properties of a significant portion of  the total DMs are deficient in scattering strength.   

We further analyze FRB scattering in terms of a parameterized cloudlet model for the scattering medium. 

\begin{figure}[t!] 
   \centering
   \includegraphics[width=\linewidth]{\Otherfigures/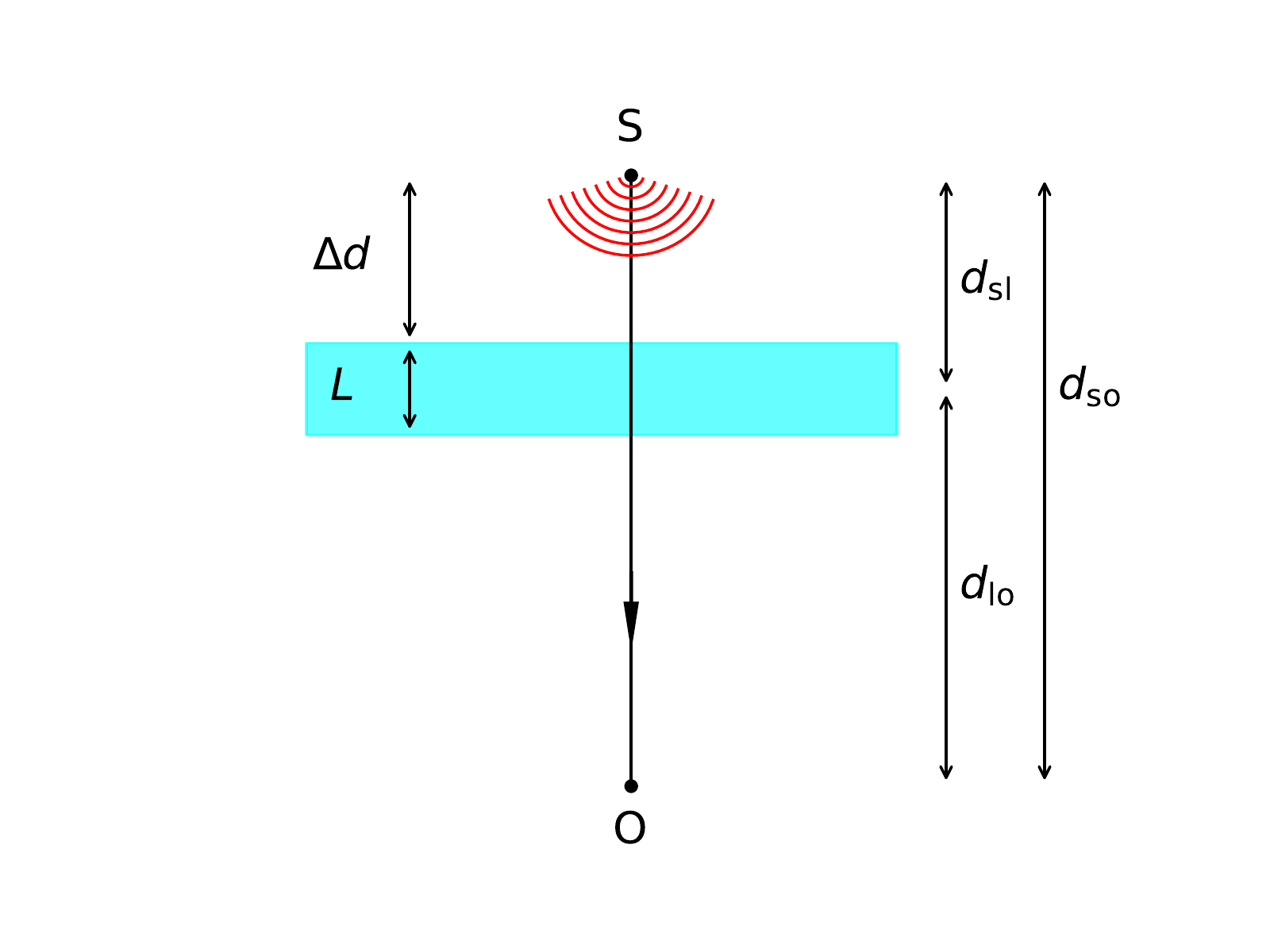} 
   \caption{
   Scattering geometry for a single layer of thickness $L$ and  offset from the source by $\Delta d$.  
   Its midpoint is a distance $\dsl$ from the source and $\dlo = \dso-\dsl$ from the observer.  This can represent a
   source  behind its host galaxy or it could represent an intervening galaxy as in  \Fig\ref{fig:fourgeometries}.
   \label{fig:geometry}
   }
\end{figure}

\subsection{ $\taud(\DM)$  for an Ionized Cloudlet Medium}
\label{sec:cloudlet}

As implied by the $\taud-\DM$ relation for Galactic pulsars,  free electrons  both  disperse and scatter pulses and bursts.   However, while all non-relativistic free electrons cause dispersion,   scattering requires small scale density fluctuations that are likely very different exist  in warm $\sim 10^4$~K plasma and  hot-phase gas ($> 10^6$~K).   Consequently, we expect the $\taud-\DM$ relation to differ greatly between interstellar media in galaxies and hot, tenuous plasma in galaxy halos and in the IGM.  This is demonstrated to be the case using existing scattering measurements. 

To model the scattering medium in any one component (e.g. the MW,  a host or intervening galaxy, or subregions within galaxies, such as HII complexes), we use a population of small clouds of ionized gas, each with internal  density fluctuations.   Following the formalism first presented in \citet[][]{cwf+91} and further developed by
\citet[][]{tc93,cl02, cws+16, 2013apj...776..125m,2020ApJ...897..124O, Ocker_2021}, cloudlets have internal electron densities $\nebar$ and fractional rms density fluctuations 
$\varepsilon = \sigma_{\nelec} / \langle \nebar \rangle \le 1$ (angular brackets denote ensemble average).  Variations between cloudlets are given by  $\zeta = \langle \nebar^2 \rangle  / \langle \nebar \rangle^2 \ge 1$. 
Cloudlets have a volume filling factor $\ff$. We assume internal fluctuations follow a power-law spectrum 
$\propto \cnsq q^{-\beta}\exp[-(2\pi q / \linner)^2]$ for wavenumbers $2\pi/\louter \le q \lesssim 2\pi/\linner$,
where $\louter$ and $\linner \ll \louter$ are the outer and inner scales, respectively.   We use a Kolmogorov spectrum  with $\beta = 11/3$ as a reference spectrum.  

The resulting  broadening time  from a layer with dispersion depth $\DMl$  is  derived in Appendix~\ref{app:cloudlet},
\be
\taud(\DMl, \nu) &=&  C_\taud  \nu^{-4} \Clinner \Ftilde \Gscatt \,\DMl^2
\nonumber \\
&\simeq& 0.48~{\rm ms} \times 
 \nu^{-4} \Clinner \Ftilde G\, \times\DM_{100}^2
\label{eq:taucloudlet2}
\ee
with  $\nu$ in GHz and  $\DM_{100} \equiv \DMl / (100 \ \DMunits)$.  \added{The quantity $C_\taud$ is a numerical constant defined in the Appendix.   The quantity $A_\tau$ depends on the inner scale $\linner$  and spectral index $\beta$ and accounts for the shape of the pulse broadening function, as described in the Appendix. It can range from $\sim 1/6$ to unity.}  Other parameters that characterize density fluctuations combine into the quantity
\be
\Ftilde = \frac{\zeta \varepsilon^2}{f (\louter^2 \linner)^{1/3}},
\ee
\added{which has units of ${\rm (pc^2\, km)^{-1/3}}$ for the outer scale in parsecs and the inner scale in kilometers.}
\explain{This text deleted: For the outer scale in parsecs and the inner scale in kilometers, $\Ftilde$ has units ${\rm (pc^2\, km)^{-1/3}}$.}
The location of the scattering layer relative to the source  strongly affects $\taud$ and determines the geometric factor, $\Gscatt$
(see next section).   \added{ In most of our analysis, the composite quantity $A_\taud \Ftilde G$ is constrained by observations though we expect $G=1$ for the lines of sight considered in this paper.}

For cosmological distances, a source at redshift $\zs$ and a scattering region in a host or intervening galaxy at
$\zl$  gives
\be
\tau(\DMl, \nu, \zl, \zs) &&
\nonumber \\
\simeq 0.48~{\rm ms} 
&\times &
\frac{\Clinner \Ftilde G(\zl, \zs ) \DM_{l, 100}^2} {\nu^4  (1+\zl)^3} .
\label{eq:taudmz_xgal}
\ee
Here $\DMl$ is in the rest frame of the scattering layer (i.e. a host or intervening galaxy or halo),
\added{which contributes to the measured DM as $\DMl / (1 + \zl)$. }

\begin{figure*}[t!] 
   \centering
   \includegraphics[width=\figwidthfour, trim = {3.9cm 1.2cm 1.3cm 1.4cm}, clip]{\Otherfigures/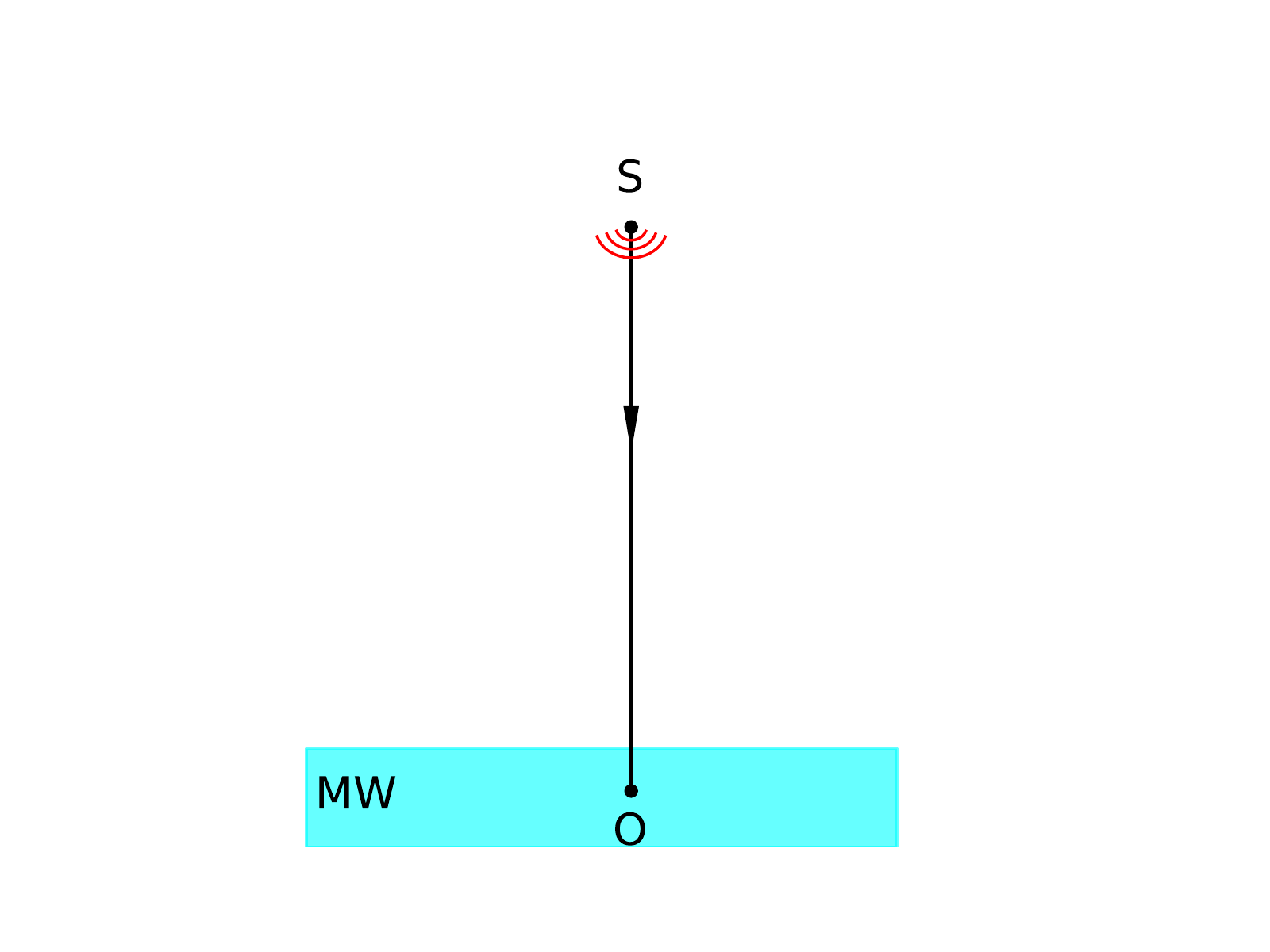}
   \hspace{-0.3in}
   \includegraphics[width=\figwidthfour, trim = {3.9cm 1.2cm 1.3cm 1.4cm}, clip]{\Otherfigures/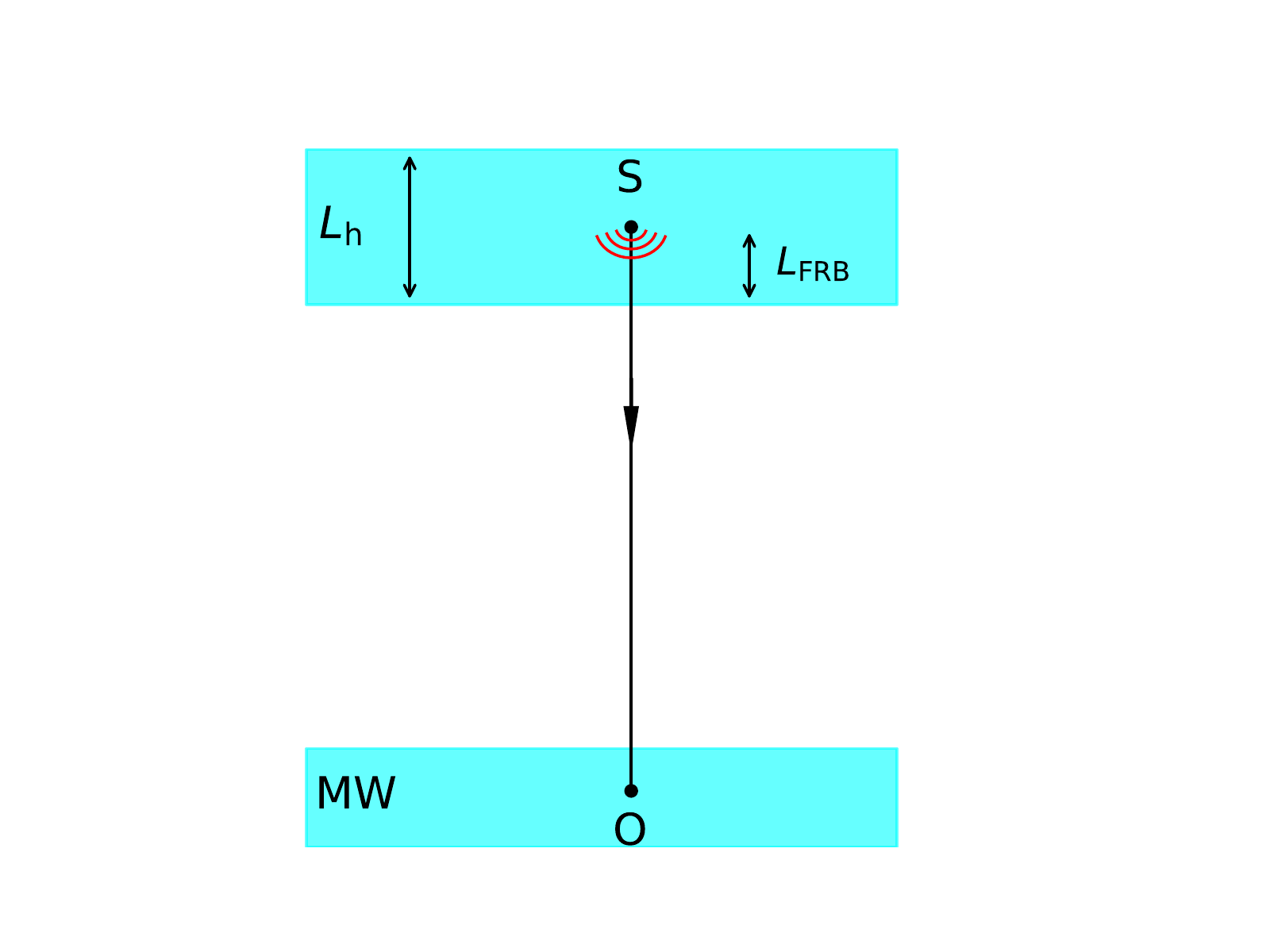}
      \hspace{-0.3in}
    \includegraphics[width=\figwidthfour, trim = {3.9cm 1.2cm 1.3cm 1.4cm}, clip]{\Otherfigures/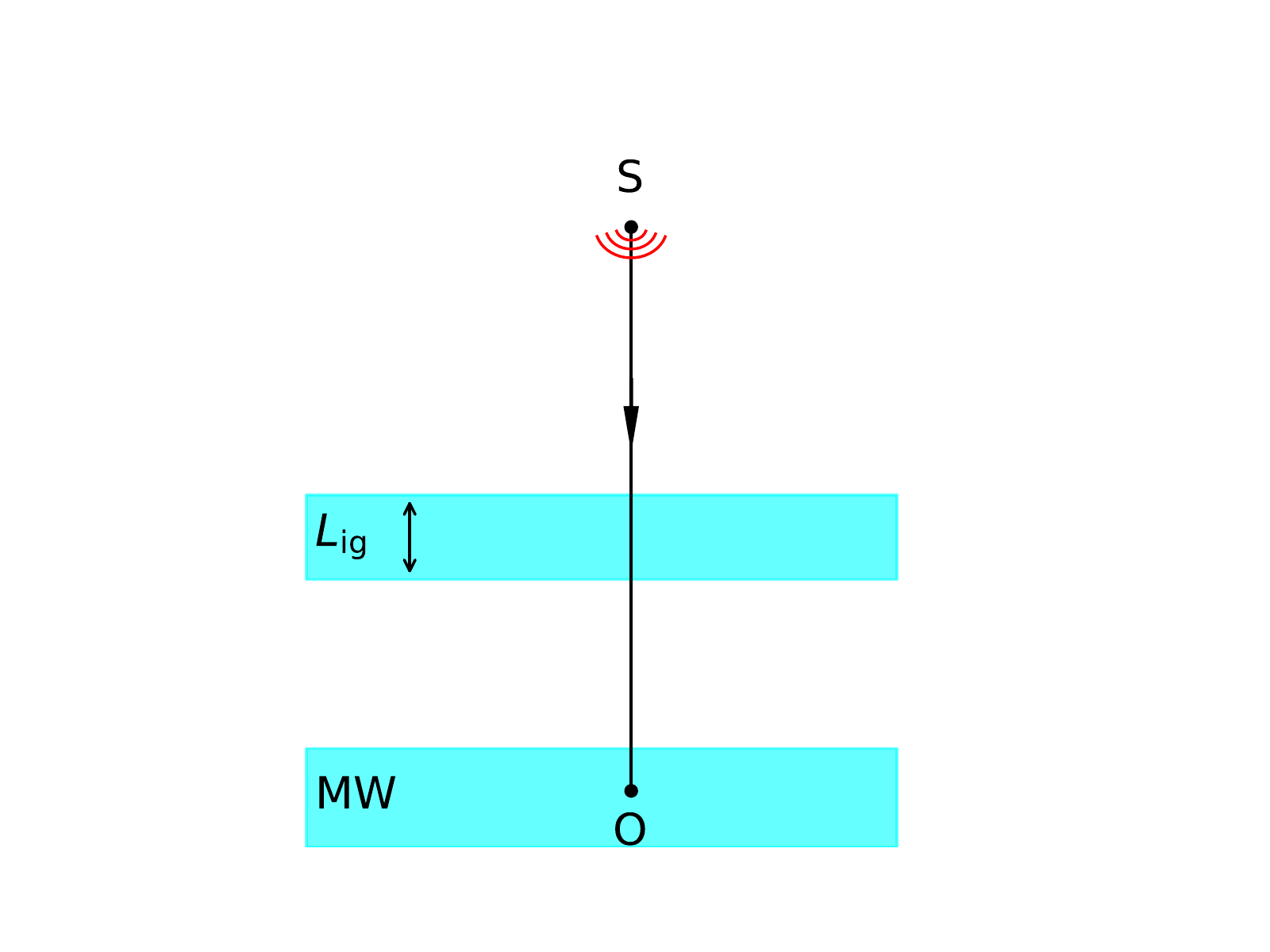}
       \hspace{-0.3in}
    \includegraphics[width=\figwidthfour, trim = {3.9cm 1.2cm 1.3cm 1.4cm}, clip]{\Otherfigures/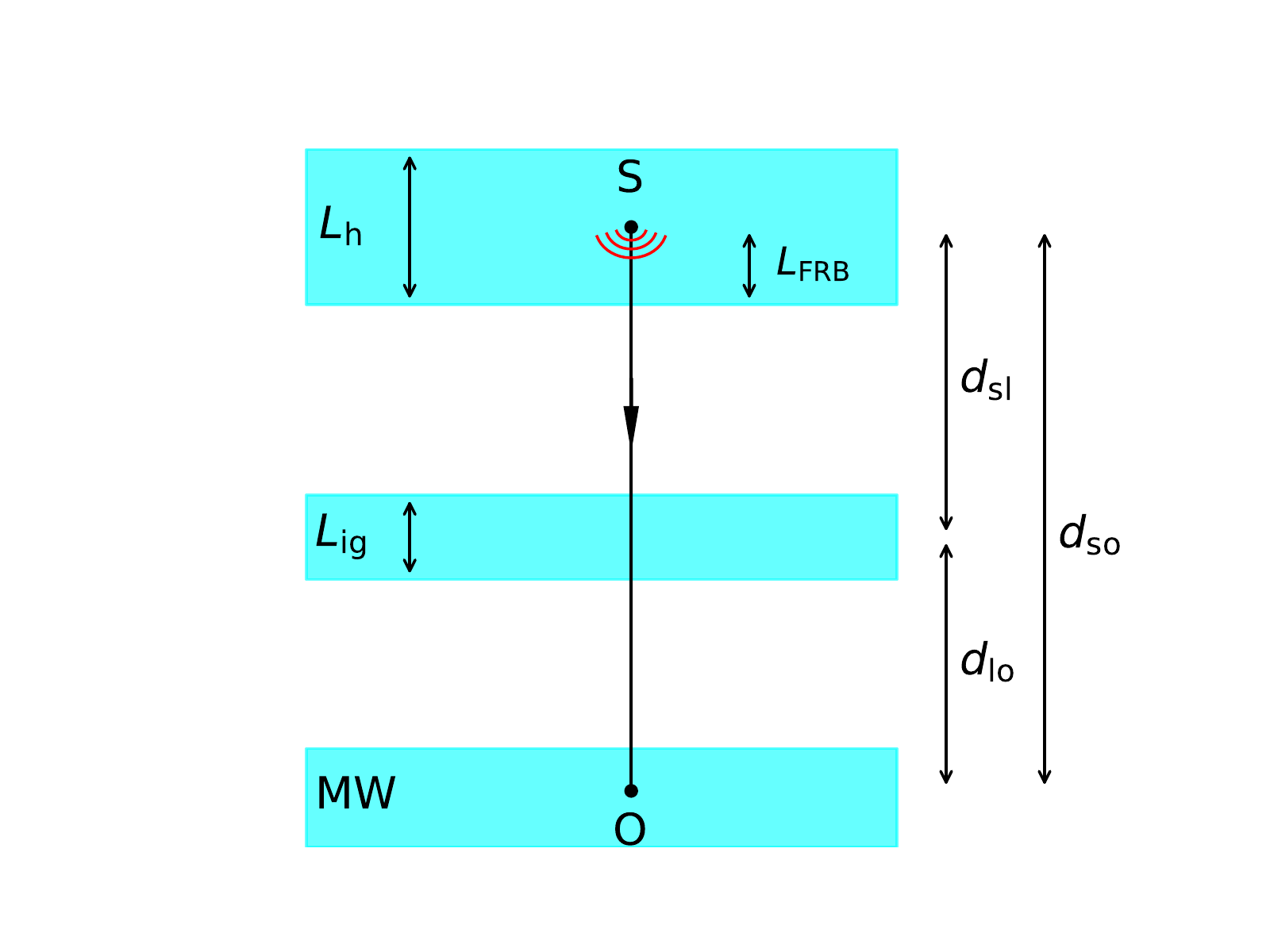}
   \caption{Scattering geometries for extragalactic sources that include the Milky Way 
   and cases without and with scattering layers in a host galaxy (h) or intervening galaxy (ig). 
   The direct line of sight is shown from source S to observer O.  
   A scattering layer in an intervening galaxy is at a distance
   $\dsl$ from the source and $\dlo = \dso-\dsl$ from the observer.
   The thickness of the host galaxy $L_{\rm h}$ is generally larger than the path length through the galaxy to the source, $L_{\rm FRB}$. 
    }
   \label{fig:fourgeometries}
\end{figure*}

\subsubsection{Scattering Geometries}
\label{sec:geometries}

We define the dimensionless geometric factor  $\Gscatt$ so that it is unity for a source embedded in the scattering medium, such as its host galaxy, and the source distance is much larger than the thickness of the scattering medium.    Generally, $\Gscatt$ is a strong function  of the \LOS\ distribution of scattering electrons and can exceed unity by many orders of magnitude. 
In Euclidean space
\be
\Gscatt = 
\frac{\int_{\rm layer} ds\, s (1 - s/d)}{\int_{\rm host} ds\, s (1 - s/d)}.
\ee
For scattering within the MW or in a distant FRB host galaxy $\Gscatt = 1$, but is $\gg 1$  for an intervening galaxy or halo.

First we derive the geometric factor $\Gscatt$   with reference to  the  geometry shown in \Fig\ref{fig:geometry}  for a statistically homogeneous (i.e. constant $\cnsq$) layer of thickness $L$  that is offset from the FRB source by $\Delta d$.  
Letting $x = L/\dso$ and $y = \Delta d / \dso$ for a source-observer distance $\dso$ and defining
$g(a,b) = \int_a^ b ds\, s(1-s)$,  the geometric factor is defined  so that $\Gscatt=1$ for
a slab representing a host galaxy ($y=0$) or the MW ($y=1-x$),
\be
\Gscatt(x,y) &=& \frac{g(y, x+y)}{g(0, x)} 
\nonumber \\
&&\!\!\!\!\!\!\!\! = \frac{1 - 2x/3 +(2y/x)(1-y-x) }{(1-2x/3) }.
\ee

Treating intervening galaxies and halos  as thin slabs ($x\ll 1$) that are  close to neither the source or observer,
we have 
$\Gscatt \simeq (2y/x)(1-y) \gg 1$, illustrating that scattering from an intermediately positioned slab yields much greater pulse broadening, all else being equal.  
 The strong dependence of $\Gscatt$ on $y/x$ suggests that some of the scatter 
 {\it at fixed dispersion measure} in the 
 cyan band   shown in \Fig\ref{fig:pulsar_tauDM} derives from different pulsars having different
concentrations of scattering regions along their \LOS.   This `Galactic variance'  yields different values of $G$ and thus $\taud$
for objects with identical values of \DM. 

\Fig\ref{fig:fourgeometries} shows four scattering  configurations involving the MW, a host galaxy, and  an intervening galaxy \added{that are likely to be encountered in FRB observations.}  Two cases apply to an FRB source that is unaffiliated with or on the near side of a host galaxy and  cases are shown with and without an intervening galaxy.  \added{While objects discussed in this paper   involve only the first two cases in the figure,
we also need to dismiss the possibility that the other two cases  apply to the current sample, as discussed below.}

\begin{figure}[t!] 
   \centering
   \includegraphics[width=\linewidth]{\NewFigures/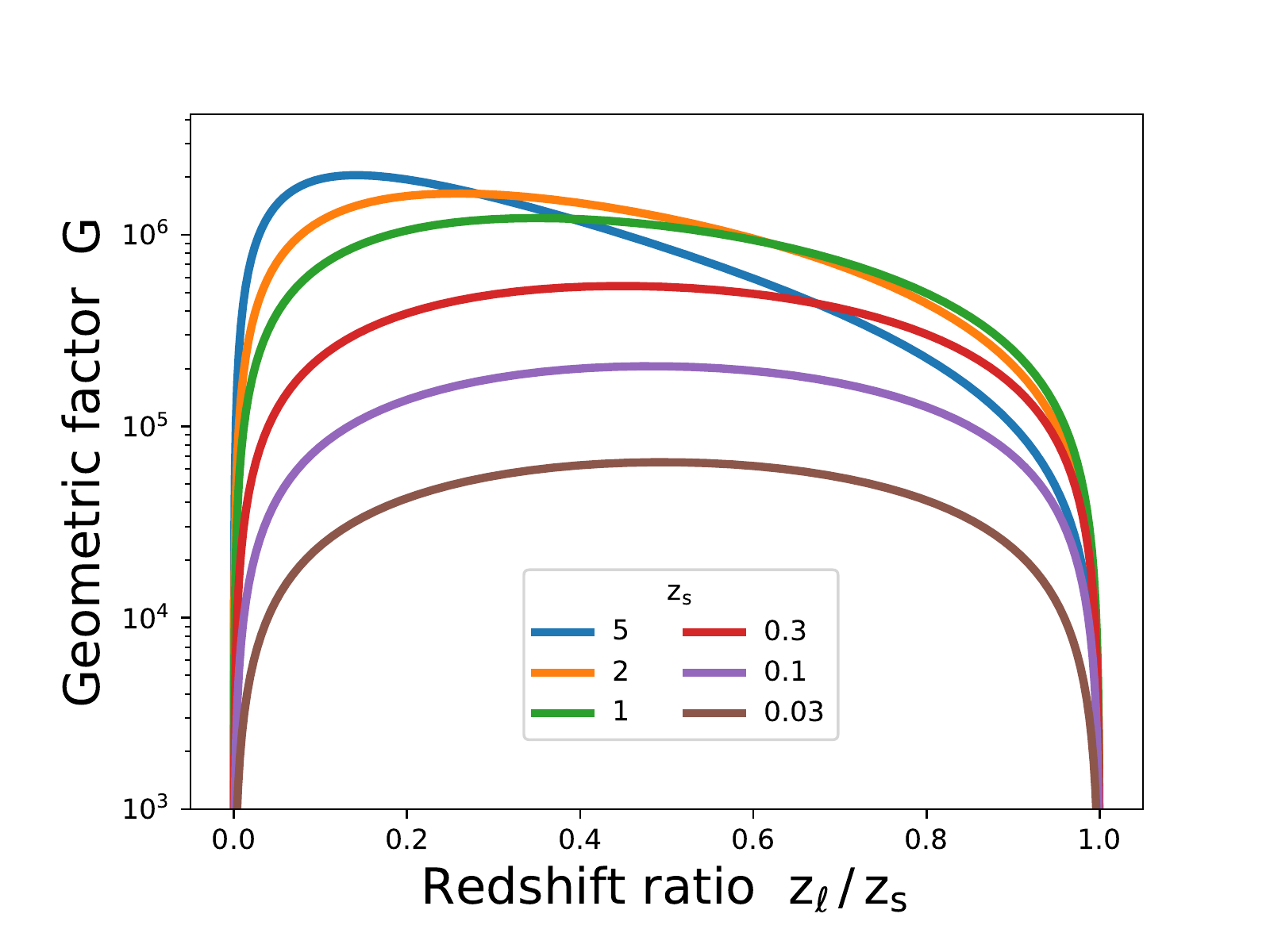}
    \caption{
Geometrical  factor $\Gscatt$ vs. redshift ratio  for a source at redshift $z_s$ and scattering layer  at redshift $z_{\rm l}$.  The thickness of the scattering layer is assumed to be 
   $L = 1$~kpc, which is indicative of a galaxy disk.  For a thicker layer, $\Gscatt \propto L^{-1}$ yields a smaller value. 
   }
   \label{fig:Gvsz}
\end{figure}

For non-negligible redshifts,   the expression 
for $\Gscatt$ is replaced by one involving  angular diameter distances, yielding     
$\Gscatt(\zl, \zs) = 2\dsl\dlo / L \dso$, where $\dsl$ and $\dlo$  are  distances from the source to the scattering layer and from layer to observer, respectively, and $\dso$ is the source distance.    
\Fig\ref{fig:Gvsz} shows $\Gscatt$ vs redshift ratio for several values of the source redshift, $z_{\rm s}$, which we  take to be the redshift of a host galaxy (though generally a source need not be associated with a galaxy). 
\added{For low redshifts $\Gscatt$ is symmetric about the midpoint where $\zl / \zs = 1/2$.   However, for large source redshifts, $\Gscatt$ maximizes at progressively smaller values of $\zl / \zs$, though at intervening redshifts $\zl$ that are still cosmological.    This effect enters into any consideration of scattering of high redshift FRBs that we defer to another paper in progress 
 \citep[][]{2022arXiv220213458O}.}

 For scattering in a host galaxy, $\dlo / \dso \to 1$ and $\dsl \to L/2$ yielding 
$\Gscatt \to 1$  as with the Euclidean expression. 
For  Gpc distances ($\dsl, \dlo$, $\dso$)  and $L= 1$~kpc,  $\Gscatt \sim {\rm Gpc / kpc} \sim 4\times 10^5$. However, unless a galaxy disk is encountered with edge-on geometry, the dispersion measure $\DMl$ may be small. 
Nonetheless, even with $\DMl = 10~\DMunits$, the scattering time for an intervening galaxy would be
$\tau \simeq 2$~s if $\Ftilde$ is similar to Galactic values.  This fact can be used to rule out whether any observed scattering occurs in an intervening galaxy instead of a host galaxy if an FRB can be detected along an \LOS\ that intersects a galaxy disk.   More likely, given the large implied scattering for nominal parameters,  FRB detections would be strongly suppressed along any such \LOS, \added{such as those in the two cases shown on the right
in Figure~\ref{fig:fourgeometries}}.

Our results indicate that FRB lines of sight that pierce an intervening galaxy {\it disk} are unlikely to be seen at frequencies $\nu \lesssim 1.5$~GHz because the scattering is much larger than the intrinsic burst width.  If only a halo is intersected,  the FRB's DM will be enhanced but the scattering will not increase significantly. 

 \subsubsection{Required Values of $\Ftilde G$ for Galactic Pulsars}

The quadratic scaling with $\DM$ in Equations~\ref{eq:taucloudlet2},\ref{eq:taudmz_xgal}     (see also Eq.~\ref{eq:taucloudlet}) contrasts with the empirical scaling  in 
Eq.~\ref{eq:hockeystick} for pulsars shown Figure~\ref{fig:taudm},  which has shallower and steeper dependences for small and large  DM, respectively.    These differences reflect the strong spatial dependence of $\Ftilde$ across the Galaxy  because large pulsar DMs necessarily probe the inner part of the Galaxy where population I activity (e.g. supernovae) is more intense than near the Sun where low-DM pulsars reside \citep[][]{cwf+91}.    The NE2001 model in fact uses values of $\linner^{1/3}\Ftilde$  that are larger by $\gtrsim 500$ in the thin disk and spiral arm components compared to the smaller value in the  thick disk sampled by low-DM pulsars.

Lines of constant $\Ftilde G$ shown in \Fig\ref{fig:taudm} for 
$\Ftilde G = 10^{-3}$ to $10^2~\FtGunits$ demonstrate that large values are needed to account for the scattering of inner Galaxy pulsars while much smaller values suffice for shorter \LOS\ to pulsars in the solar neigbhorhood. 
Pulsars at high Galactic latitudes sample the thick disk of free electrons and yield
$\Ftilde = (3\pm2)\times 10^{-3}~\FtGunits$. 
 In addition, Galactic scattering to FRB~20121102A in the anticenter direction places an upper bound   $\Ftilde \lesssim 3\times 10^{-2}~\FtGunits$ for the MW halo and scattering toward two other FRBs with lines of sight near or close to galaxy halos yield $\Ftilde \lesssim 10^{-3}~\FtGunits$ \citep[][]{Ocker_2021} for those halos. 

\subsection{Scattering in the IGM}

On both observational and theoretical grounds, the IGM's contribution to scattering is likely negligible in comparison
with contributions from the interstellar media of  galaxy disks, including the MW, host, and intervening galaxies.   Not all FRBs with large measured DMs $\gtrsim 10^3~\DMunits$  show large scattering times, which might have been expected if scattering were IGM dominated even with cosmic variance taken into account. 

The $\taud(\DM)$ relation for galaxy disks does not change qualitatively in a cosmological context once redshift dependences are included, as in \Eq\ref{eq:tauexpression} \citep[see also][]{2013apj...776..125m}.  Given that the IGM contributes \DM\ values comparable to those of galaxy disks, one might expect  scattering to also be similar.  However, the $\Ftilde G$ factor is  likely to be quite different.   Assume the product
$\zeta\varepsilon^2$ is the same, since  it measures fractional fluctuations that are of order unity
in the ISM, and consider equal contributions to the total \DM.  Ignoring redshift factors, which are close to unity for low-$z$ objects,  the ratio of scattering times from the IGM and from a galaxy's  ISM  is
\be
\frac{\strut {\taud}_{\rm\, IGM}} {\strut{\taud}_{\rm\, ISM}\strut}
\approx
\frac
{\mathstrut \left[ \ff (\louter^2 \linner)^{1/3} \right]_{\rm ISM}}
{\left[ \ff (\louter^2 \linner)^{1/3} \right]_{\rm IGM}} ,
\ee
where  $G=1$ applies to a source embedded in  the ISM of a host galaxy. 

The outer scales alone are probably very different because in standard turbulence pictures, they correspond to the scales on which energy is injected.   ISM scales are $\lesssim$kpc and IGM scales $\gtrsim$Mpc, giving
$
{\taud}_{\rm\, IGM} / {\taud}_{\rm\, ISM} 
\lesssim  ({\rm  kpc} / {\rm Mpc})^{2/3} 
\lesssim 10^{-2}$.

The filling factor and inner scale are also likely larger for the IGM, further reducing the ratio.
For example, the inner scale for the solar wind  may be linked to  the thermal proton gyroradius $r_{g,p} = v_p(T) / \Omega_{g,p}$ (where $v_p(T)$  is the RMS thermal speed and $\Omega_{g,p}$ is the gyrofrequency) or to the proton inertial length, $\ell_{i,p}  = c / \omega_{p,p}$, where $\omega_{p,p} $ is the proton plasma frequency \citep[][]{2015RSPTA.37340147G}.  These would imply  $\linner \propto T^{1/2} / B$  or $\linner \propto \nelec^{-1/2}$, respectively.
Given the higher temperature, smaller magnetic field, and smaller plasma density of the IGM compared to an ISM, the ratio ${\taud}_{\rm\, IGM} /{\taud}_{\rm\, ISM}$ might be  reduced by  another order of magnitude.

\citet[][]{lg14} argue similarly that the outer scale for the IGM must be comparable to Galactic values to allow a significant contribution to $\taud$ but they also  point out that the resultant 
turbulent heating would cause the IGM to be hotter than inferred from observations.  In the following we therefore exclude any contribution from the diffuse IGM to scatteriing. 

\begin{figure*}[b] 
   \centering
    \includegraphics[width=\figwidthtwo]{\Figures/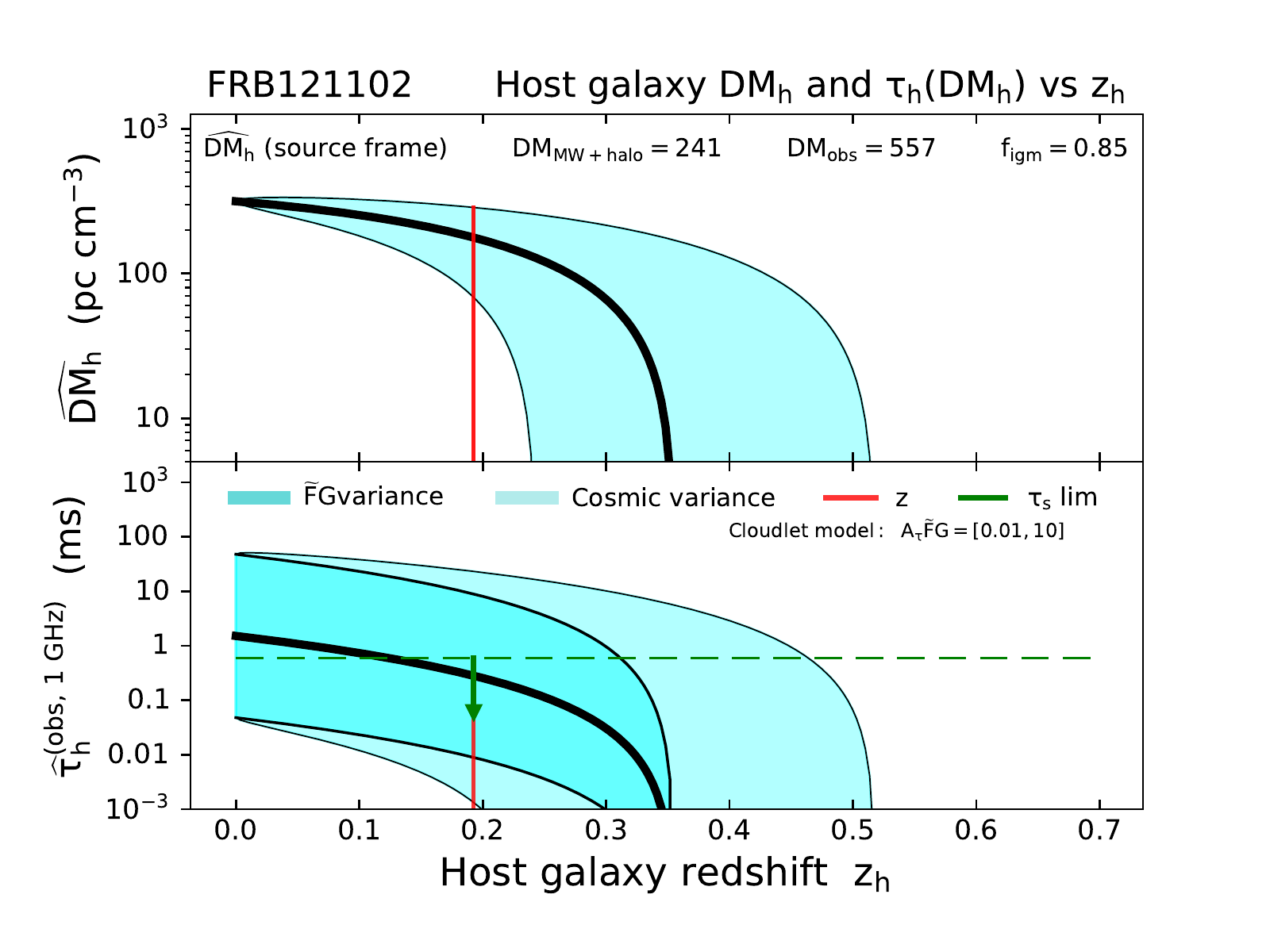}
        \includegraphics[width=\figwidthtwo]{\Figures/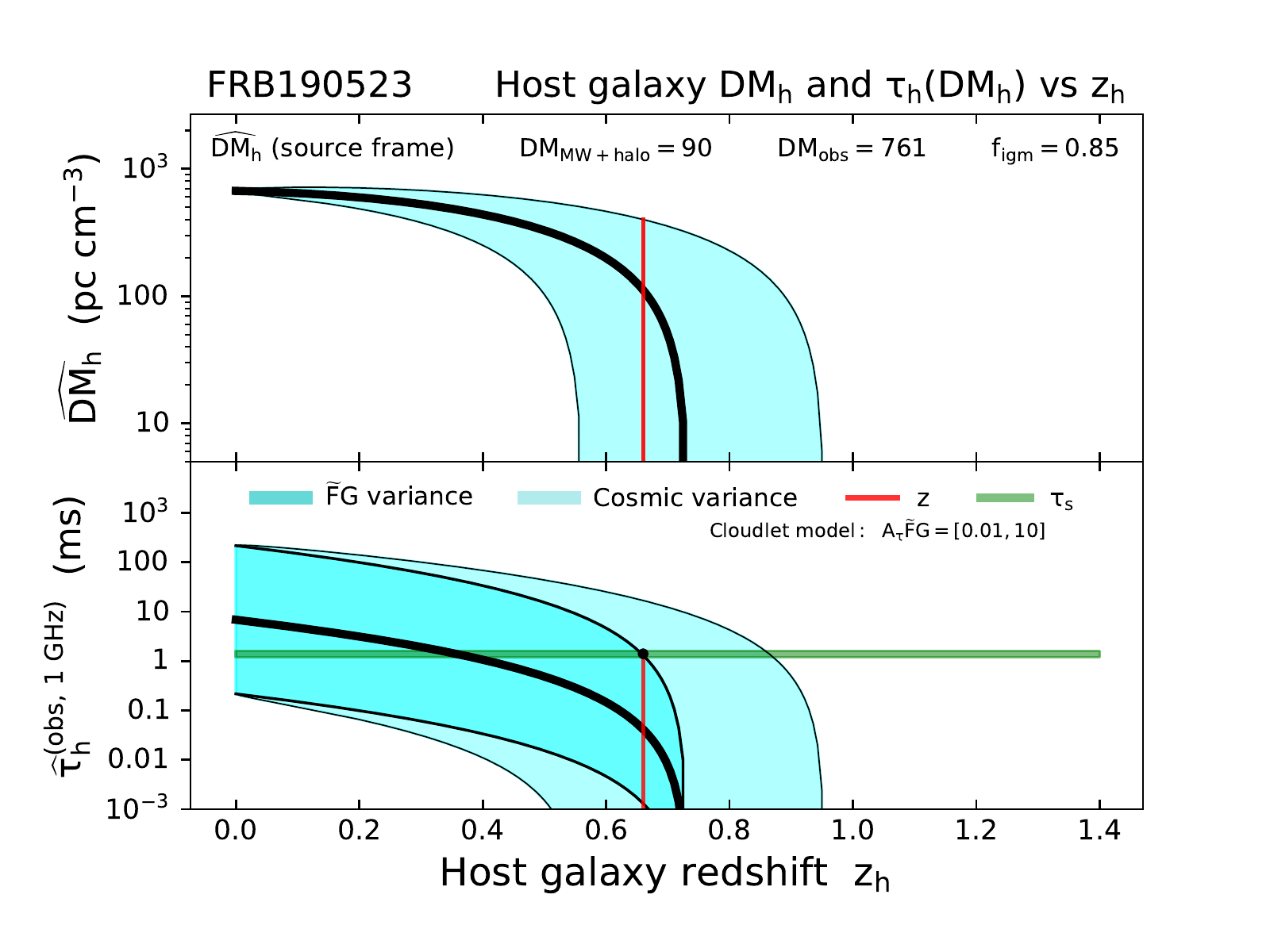}
   \caption{Analysis for FRB~20121102A and FRB~20190523A using the cloudlet model and a baryonic fraction
   $\figm = 0.85$.    Two panels are shown for each FRB:
   (Top)   Host  galaxy DM  vs. redshift using methodology discussed in the text.  $\DMhhat$ is expressed in the rest-frame of the galaxy and includes uncertainties in the IGM's contribution to the total DM shown as the light shaded band (cyan).   The vertical red line marking the measured redshift  shows the range of possible 
   $\DMh$ values; for FRB~20121102A these  are consistent with the results of \cite[][]{tbc+17}. 
   (Bottom)  Estimated scattering time in the observer's frame at 1~GHz using the cloudlet \added{model} discussed in the text.    The heavier shaded band indicates the extent of scattering times $\taud$ for cloudlet models with
   $\FtG$ in  the range $[0.1, 10]~\FtGunits$.  The lighter  shading indicates the effect on $\taud$ of the cosmic variance of the IGM's contribution to $\DMhhat$  shown in the top panel.    
   The green line marks  the upper limit (dashed for FRB~20121102A) or measured (solid for FRB ~20190523A) scattering time.   
   }
   \label{fig:twopanel_twofrbs}
\end{figure*}

\subsection{Scattering in  Galaxy Halos}

Galactic pulsars show a strong Galactic latitude dependence  for scattering times indicative of contributions from a strongly scattering thin disk \citep[][]{cl02,2017apj...835...29y} and a thick disk with a scattering scale height of one half the scale height $H_{\nelec} \sim 1.6$~kpc for the electron density  \citep[][]{2020ApJ...897..124O}.   Comparison with the scattering of AGNs, which sample the entire MW halo (unlike pulsars in or near the thick disk or pulsars in the Magellanic clouds), shows no increase in scattering over that provided by the disk components.  
This implies a modest DM contribution from the Galactic halo  along with a small value of
$\Ftilde G$.   We therefore exclude  contributions to scattering from the Galactic halo.   This may be true for the halos of other galaxies.    A specific case is  FRB~20200120E in a globular cluster near M81 that likely samples only the halos of M81 and the Milky Way along with the MW disk components.   Burst amplitude substructure is seen down to tens of microseconds (combined with shot pulses at the resolution limit of 31.25~ns)  that shows no hint of  scattering from  outside the MW disk \citep[][]{2022NatAs.tmp...43N}.

Nonetheless, given the dynamic processes involved with halo evolution \citep[e.g.][]{2020ApJ...905...60S} that might also drive turbulence and the prospects for there being substructure in halos that might also cause  radio scattering \citep[][]{2019MNRAS.483..971V}, the possibility is still open that some halos may contribute to scattering. 

\subsection{Pulse Broadening from the Milky Way}

All FRBs have been found from directions  where pulse broadening from the Milky Way is too small to detect
at   observation frequencies larger than 1~GHz \added{but it  has been measured at 0.15~GHz for FRB~190816 
\citep[][]{2021Natur.596..505P}. }  MW scattering has also been measured  in the form of intensity variations with a characteristic scintillation bandwidth $\dnud$ for several objects \citep[e.g.][]{mls+15b,2019ApJ...876L..23H,2020Natur.577..190M,2020ApJ...901L..20B}. For the low latitude FRB~20121102A ($b = -0.2^{\circ}$), the implied scattering is only about  $\taud \sim 1/2\pi\dnud \simeq 20~\mu$s at 1~GHz for its Galactic anti-center direction, in agreement with the NE2001 prediction within a factor of two \citep[][]{{Ocker_2021}}.   \added{Future observations will likely probe a wide range of scattering strengths as more bursts are found at low frequencies and low Galactic latitudes.}
 \explain{This text deleted:  Eventually MW scattering will be strongly manifested  in any  FRBs discovered in low-latitude directions toward the inner Galaxy and in low-frequency observations \citep[][]{2019ARAA..57..417C}.    }

For the remainder of the paper, we  ignore contributions to $\taud$ from  the MW disk along with those from the halo and from the IGM.   As with the DM,  we also ignore for now any scattering from intervening galaxies
and  their halos and also from any intercluster medium,  leaving only host galaxies as the main contributor to pulse broadening.
The observed scattering time $\taudobs$  is then given by \Eq\ref{eq:taudmz_xgal} using $\ \DMh$
and $\zg = \zs = \zh$. 
Future studies are likely to include FRBs with significant scattering from the MW.   For these cases, the pulse broadening is simply  the sum of the contributions from the host galaxy and the MW.

\section{Dispersion and Scattering in Host Galaxies}

\added{
In this section we present several analyses  that provide the basis for redshift estimation using both dispersion and scattering.   In the first, we show how the coupling of these two processes in host galaxies depends on redshift if redshift is treated as an independent variable.  In the second, we demonstrate that the extragalactic contribution to scattering is most economically  understood as originating in host galaxies rather than in  intervening galaxies.  If that were not the case, values for  $\AFtG$ would have to be significantly different from those encountered in the Milky Way and in host-galaxies.    The third analysis presents  joint constraints on the host-galaxy contribution to DM and the scattering parameter,
$\AFtG$, to demonstrate their covariance using examples for two FRB values.   }

\subsection{Dispersion and Scattering vs. Redshift}

\explain{The next two paragraphs replace the very short description that had been in the text.}

\added{
The interplay between dispersion, scattering, and redshift is shown in 
\Fig\ref{fig:twopanel_twofrbs} for  two cases, FRB~20121102A and FRB~20190523A. In the top panel of each frame,
$\DMh$ is plotted against redshift using Equations~\ref{eq:DMexpression} and \ref{eq:DMIGMbar}
and taking into account cosmic variance in $\DMigm$ as described in \S\ref{sec:DMigm}.   If the redshift were unknown and only the measured DM is available (along with a model for the Milky Way's contribution), a wide range of redshifts is allowed, roughly a factor of two in both cases.    The actual redshifts shown as vertical red lines indicate a somewhat narrow range for $\DMh$ for FRB~20121102A but a much wider range for
FRB~20190523A. }

\added{
The bottom panels show how the scattering time estimate depends on redshift (using Eq.~\ref{eq:tauexpression}  and \ref{eq:taucloudlet2}), again taking into account cosmic variance in $\DMigm$, but including a wide range for $\AFtG$ in the host galaxy.   This `interstellar variance' expands the range of possible scattering times .    The upper bound on $\tau$ for FRB~20121102A is compatible with the this range while the measured $\tau$ for FRB~20190523A is at the high end of the range of $\AFtG$ at
the measured redshift,  but overall  is not inconsistent with the predicted ranges when cosmic variance of $\DMigm$ is also taken into account. 
}

  \subsection{Scattering in Host vs.  Intervening Galaxies}
 
\begin{figure}[t!] 
   \centering
   \includegraphics[width=\linewidth]{\NewFigures/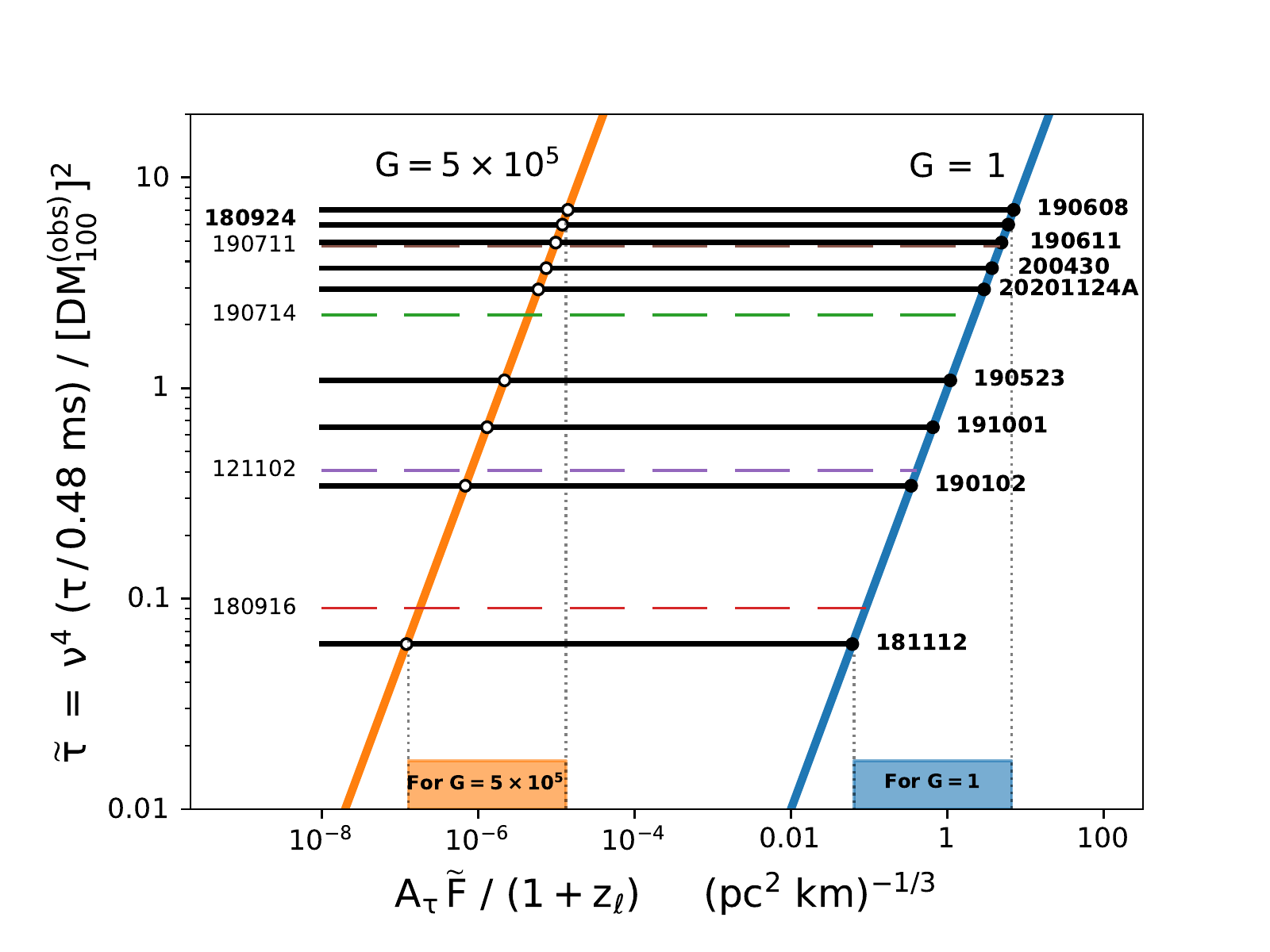}
   \caption{\footnotesize Scaled scattering time $\tautilde$ vs $\Clinner \Ftilde / (1 + \zl)$ 
   \added{(see Eq.~\ref{eq:tautilde})
   where the quantity $\DM_{100}^{\rm obs}$ is the observer-frame DM contributed by a `layer' in either a host galaxy or an intervening galaxy
   (see text). }
   Horizontal lines denote measured values (solid) or upper limits (dashed); colored bands show the 
 \added{ vertical}  uncertainties for measurements due primarily to cosmic variance of $\DMigm$.  The slanted lines
   show $\tautilde \propto G$ for $G=1$, which applies to scattering in host galaxies, and 
   $G=5\times10^5$ that is a typical value for a 1-kpc thick scattering region in an intervening galaxy midway to the source
   and at $\sim 1$~Gpc.
   The horizontal bars at the bottom of the figure indicate the range of values for the abscissa spanned by FRB measurements  for each of the two values of $G$.
   We exclude the upper limit $\tautilde\le 0.0015$ for FRB~20200120E because its line of sight is qualitatively different from those of the other FRBs, which are evidently influenced by propagation through their host galaxies. }
   \label{fig:tautilde_vs_Ftilde}
\end{figure}

\added{
Next we  compare scattering in host galaxies with that in  intervening galaxies. 
As shown in Figure~\ref{fig:Gvsz},  the geometric factor  used in Eq.~\ref{eq:taudmz_xgal} that enhances scattering is orders of magnitude larger for  an intervening galaxy compared to $\Gscatt = 1$ in a host galaxy.   
A consequence is that $\Ftilde$ needs to be proportionately smaller in intervening galaxies if they are not to cause scattering times vastly exceeding measured values.  These very small values of $\Ftilde$ would imply that intervening galaxies can produce significant  contributions to DM without corresponding scattering times like those   derived from pulsars in the Milky Way.  This in turn would require an explanation for why FRBs sample  dispersive gas in intervening galaxies with significantly different turbulence properties.   A simpler hypothesis is that extragalactic scattering occurs in host galaxies, not in any intervening galaxies in the sample we have analyzed.}

To compare host and  intervening galaxies,  we define a scaled scattering time,
\be
\tautilde 
= \frac{\nu^4}{[\DM^{\rm obs}_{100}]^{2}} \left( \frac{\tau}{0.48\, {\rm ms}}\right)
= \frac{\Clinner \Ftilde G(\zl, \zs)}{1+ \zl} ,
\label{eq:tautilde}
\ee
which involves observable quantities after the first equality and unknown quantities after the second.  The redshift of the scattering layer $\zl$ is  either that of  an  intervening galaxy or  a region in a host galaxy (with $\zl$ very slightly smaller than $\zh$ so that $\dsl = L/2$).   
\added{The DM contributed by the layer $\DM^{\rm obs}_{100}$ is expressed in the observer's frame in units of 100~$\DMunits$.}
\Fig\ref{fig:tautilde_vs_Ftilde} shows $\tautilde$ vs $\Clinner \Ftilde G(\zl, \zs) / (1+ \zl)$ for two values of the geometric factor,  $G=1$ for host galaxies and $G = 5\times 10^5$ that is typical for an intervening galaxy at
a redshift that maximizes $G$ (c.f. \Fig\ref{fig:Gvsz}).  In both cases we have used the inferred $\DMh$ as the dispersion measure  contributed by the layer expressed in the observers frame. 
Measurements and upper limits on $\tautilde$ yield
a range for the abscissa of $\Clinner\Ftilde / (1+\zh) \sim 0.018$~to~7.9$~\FtGunits$ if scattering occurs in host galaxies
with $G=1$ (blue band along horizontal axis).   However, if  scattering were to occur in intervening galaxies, the values  would be
\added{smaller by a factor $(5\times 10^5)^{-1}$ or $\Clinner\Ftilde / (1+\zh) \sim 3.6\times10^{-8}$~to~$1.6\times10^{-5}~\FtGunits$
(yellow band along horizontal axis).  
These values are significantly smaller than those that apply to the ISM of the MW, which range from about $10^{-3}$ to 10~$\FtGunits$. }
We conclude that scattering of FRBs with known redshifts occur in host galaxies with interstellar media similar to those in the Milky Way as gauged by $\Ftilde$.   Based on this, in \S\ref{sec:zhat} we adopt a flat prior for $\Clinner\Ftilde$ over the range  0.01 to 10~$\FtGunits$.  This is consistent with values in Figure~2 of \citet[][]{Ocker_2021} based on measurements of Galactic pulsars.\footnote{While Figure~2 of \citet[][]{Ocker_2021} shows values of $\Ftilde$ extending outside the range we adopt, the bulk of the measurements are in that range.  Note also that the values in that paper assume $\Clinner=1$, so  they can alternatively be interpreted as the range for $\Clinner \Ftilde$.}

\begin{figure*}[htbp] 
   \centering
   \includegraphics[width=\figwidthtwo]{\Figures/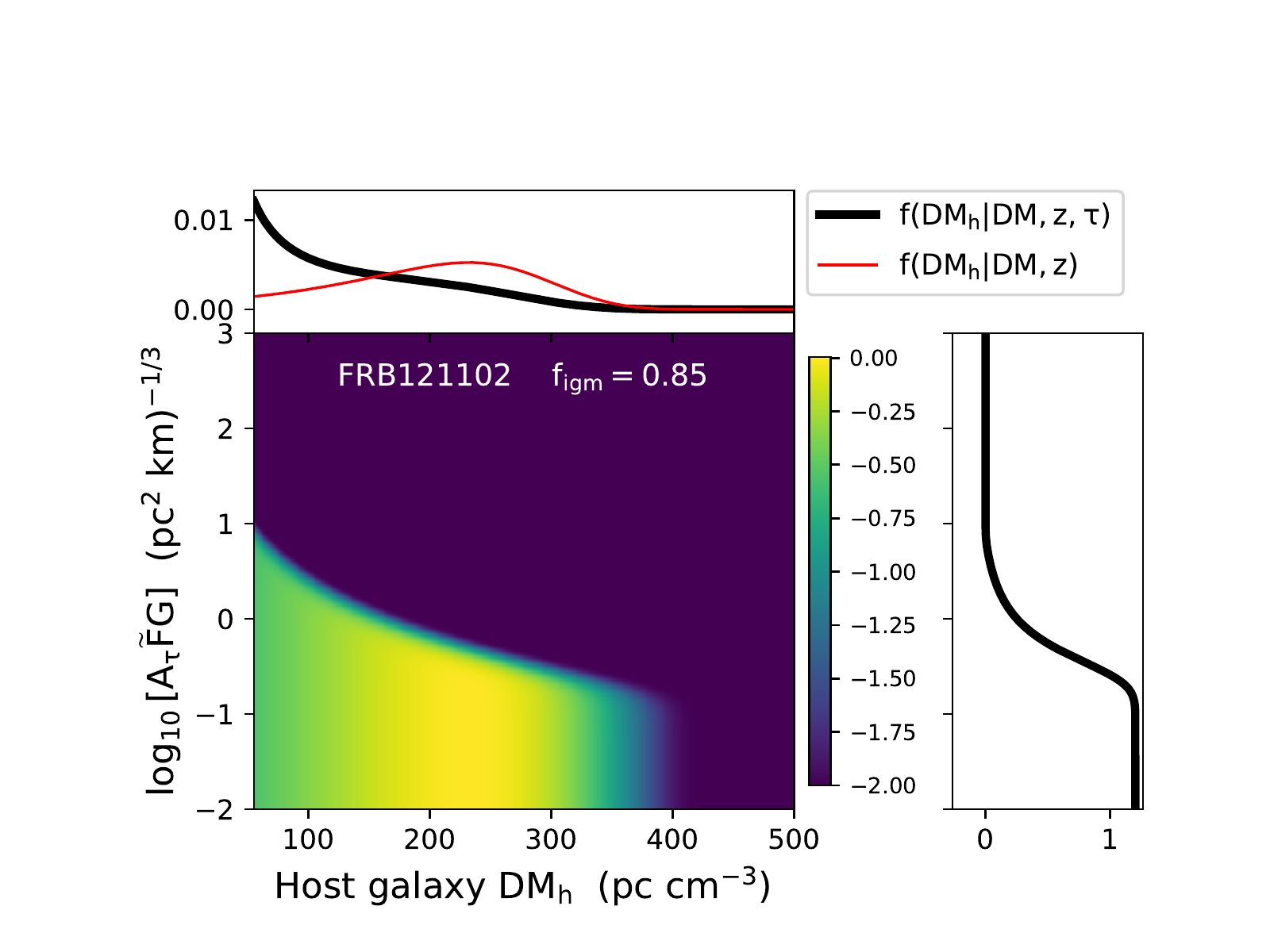}
   \includegraphics[width=\figwidthtwo]{\Figures/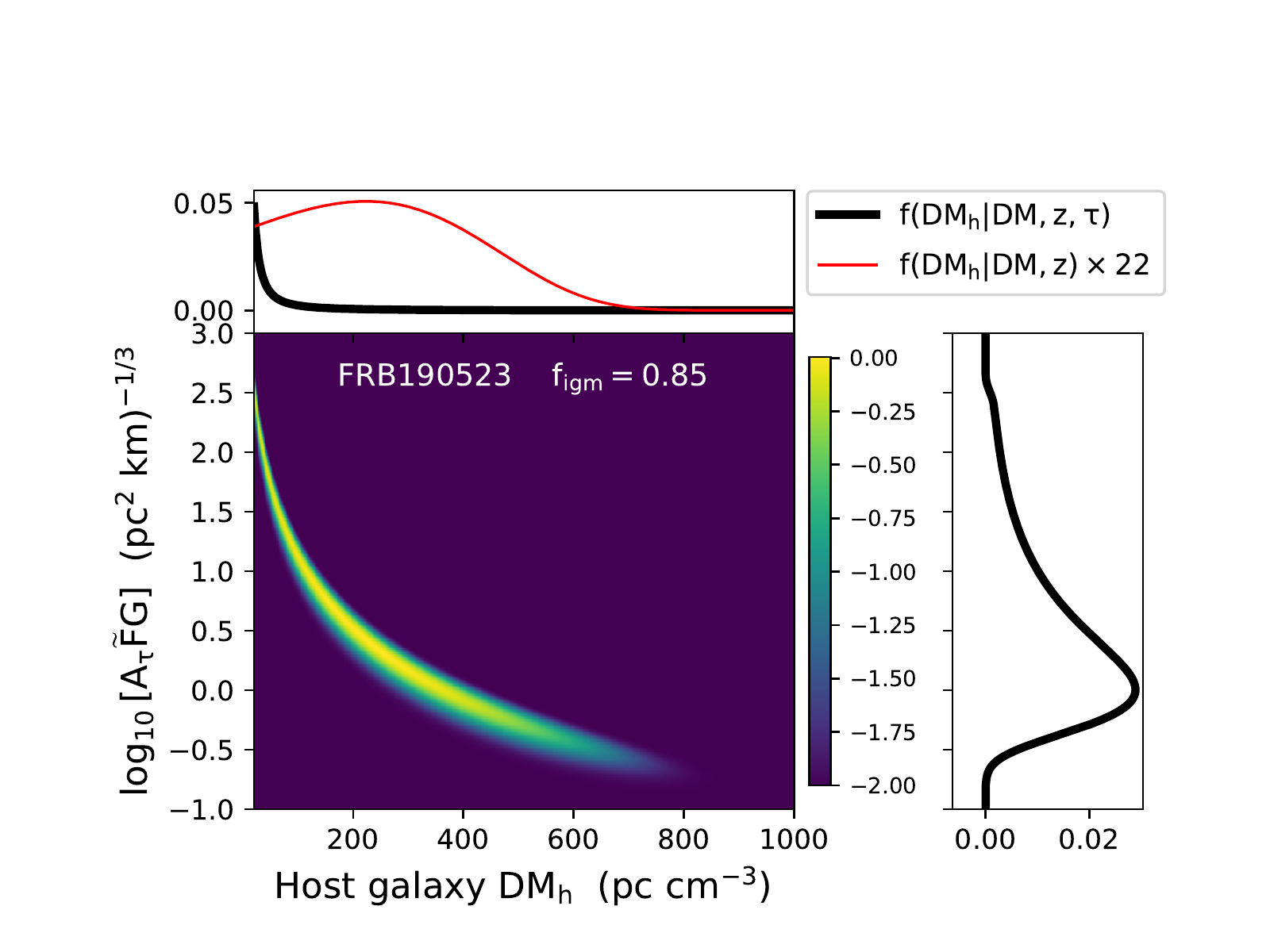}
   \caption{Posterior PDFs for $\DMh$ and $\FtG$ for two FRBs.
   The large panel shows probability density vs. $\AFtG$ and $\DMh$ assuming flat priors for each quantity, with $\DMh$ constrained within
   50 to 500~$\DMunits$ or 50 to 1000~$\DMunits$ for FRB~20121102A and FRB~20190523A, respectively.   The ranges for $\log_{10}\AFtG$ are
   $-2$ to 3 for FRB~20121102A and $-1$ to 3 for FRB~20190523A.   The black curves in the top and side panels are marginalized, one-dimensional PDFs.  The red curves in the top panels are the posterior PDFs for $\DMh$ derived from the DM-inventory analysis of \S\ref{sec:postDMh}. 
   }
   \label{fig:post2D_a}
\end{figure*}

\subsection{Posterior PDFs for $\DMh$ and $\FtG$}

The PDF of $\tau$ given the redshift and host galaxy $\DMh$ is
$\delta(\tau - \tauhat)$ with $\tauhat =   C_\tau \nu^{-4} \AFtG \, \DMh^2 / (1+z)^3$. 
If the redshift  is known to high precision 
but the measured broadening has an error distribution
$\pdftauobs(\tau - \tauobs, \sigtau)$, where $\tauobs$ is a nominal value and $\sigtau$ is the uncertainty, we calculate  the likelihood 
function for $x \equiv \DMh$ and $\phi =  \AFtG$, the two parameters we use to  characterize the interstellar medium of a host galaxy,  
 \be
\like(x, \phi \, \vert \, \DM,  \zh, \tau)  &=&
\nonumber \\
\pdftau(\tauhat \, \vert \, \DMh, \zh)  &=& \pdftauobs(\tauhat-\tauobs) .
\ee
Combined with the MW-marginalized PDF for $\DMh$ in \Eq\ref{eq:postDMh}, 
and assuming an uninformative flat prior for  $\phi$, the posterior PDF  is
\be
&&\pdfDMhFtG(x, \phi  \vert \DM,  \zh, \tau)  
\propto
 \\ 
&&\quad\quad {\pdfh(x \vert \DM,  \zh) f_\tau(\tau- \Clinner C_\tau \nu^{-4} \phi \, x^2 / (1+z)^3)} .
 \nonumber
\label{eq:postDMhFtG}
\ee

Figure~\ref{fig:post2D_a}  gives two examples of the joint posterior PDFs for $x$ and $\phi$
using $\figm = \figmused$ for FRB~20121102A,  an object with only an upper limit on scattering, and FRB~20190523A, which has a significant measurement of the scattering time. 
The flat prior for $\DMh$ extends from 50~$\DMunits$ in both cases  up to different maximum values, 500~$\DMunits$ and 1000~$\DMunits$, 
respectively. 

For FRB~20121102A,  the marginalized PDF for $\AFtG$  in the side panel includes a tail to 
values of $\AFtG$ larger than unity  that are still consistent with the upper limit on $\tau$ because the corresponding values of $\DMh$ are very small.  These small values are strongly disfavored by Balmer line measurements that indicate $\DMh \sim 55$ to 380~$\DMunits$ (as indicated in \Fig\ref{fig:DMh_PDF_CDF_page3}).   The marginalized PDF for $\DMh$ in the upper panel (black curve) indicates these values along with a wider range extending to $\sim 400~\DMunits$.   The same frame shows in red the posterior PDF for $\DMh$ resulting from the DM inventory also shown in \Fig\ref{fig:DMh_PDF_CDF_page3}.    

FRB~20190523A, by comparison,  shows a joint PDF with a shape determined by the relationship $\phi = \AFtG  \propto \tau \DMh^{-2}$.   
\added{Without other constraints,   the line of sight through the host galaxy can encounter ionized gas with  values
for $\DMh$ and $\AFtG$ anywhere along the curved ridge of high probability density.    Formally,  the broad prior in $\DMh$ allows a solution with 
small $\DMh$ and a corresponding value for $\AFtG$ much larger (by one or two orders of magnitude) than is encountered in the Milky Way.   While it is conceivable there could be  such regions along the \LOS\ to an FRB, their required properties run counter to those in the ISM of the Milky Way and other galaxies.  
\hfil\break\indent
A simpler conclusion is that the actual range of values for $\DMh$ and thus also for $\AFtG$ are significantly smaller than the plotted ranges in Figure~\ref{fig:post2D_a}.  In particular,  the blue horizontal band in Figure~\ref{fig:tautilde_vs_Ftilde} corresponds
to values for $\AFtG$ in the interval $[0.02, 6]~\FtGunits$ and in the next section we will use a slightly larger range,
$[0.01, 10]~\FtGunits$ as one of the flat priors used in redshift estimation. }

\section{Redshift Estimators}
\label{sec:zhat}

Only a small fraction of the current sample of FRBs has been reliably localized to host galaxies with redshifts, and while efforts are underway to provide routine high precision localizations of a large number of FRBs with e.g., CHIME outriggers \citep[][]{2021AJ....161...81L, 2022AJ....163...65C}, DSA-2000 \citep{2019BAAS...51g.255H}, and other facilities, it will take some time for such efforts to come to fruition. 

Meanwhile,   we assess a method that uses  scattering times $\tau$ along with dispersion measures to constrain FRB redshifts.   The gist of the method is that a host galaxy requires a large-enough $\DMh$ to provide the scattering time given a plausible value for 
$\AFtG$.     The resulting constraints on $\DMh$ in turn yield a plausible range for $\DMigm$ and thus redshift $\zh$. 
For the current sample of objects, this approach also provides a test for the actuality of FRB-galaxy associations.   
\added{In particular, if a candidate host galaxy is at a redshift that implies a small $\DMh$ (because the IGM dominates the DM budget) but the FRB has a large amount of scattering, there are two possibilities.   There may be an intervening galaxy that scatters the FRB with a relatively small contribution to DM owing to the geometrical effects discussed in \S\ref{sec:geometries}.  Alternatively, the association may be incorrect with the `intervening' galaxy in the first instance being the actual host galaxy.}

\subsection{DM-based Redshifts}
\label{sec:dm2z}

DM-based redshift estimation follows from previous work {\citep[e.g.][]{2020Natur.581..391M}} and expressions in    \S~\ref{sec:DMigm}.
First we express the  DM based redshift $\hat z_{\rm DM}$ in terms of an assumed value for $\DMh$, either an
{\it a apriori} value  or one based on Balmer-line measurements of a host galaxy  to determine an emission measure  
$\EM$ from which  a galaxy-wide estimate for $\DMh$ is estimated.   This implies a point estimate for the IGM's contribution (c.f. \Eq\ref{eq:DMexpression}),
\be
\DMigmhat = \DM - \DMMWhat - \DMhhat / (1 + z)
\ee
that yields a redshift by inverting the function $\rtilde_1(z)$ defined in \Eq\ref{eq:DMIGMbar},
\be
\zdm = \rtilde_1^{\,-1} (\DMigmhat / \nezero \dH) \simeq  \DMigmhat / \nezero \dH, 
\ee
where the approximate equality is  for small redshifts. 

\begin{figure*}[t] 
   \centering
   \includegraphics[width=\figwidththree]{\Figures/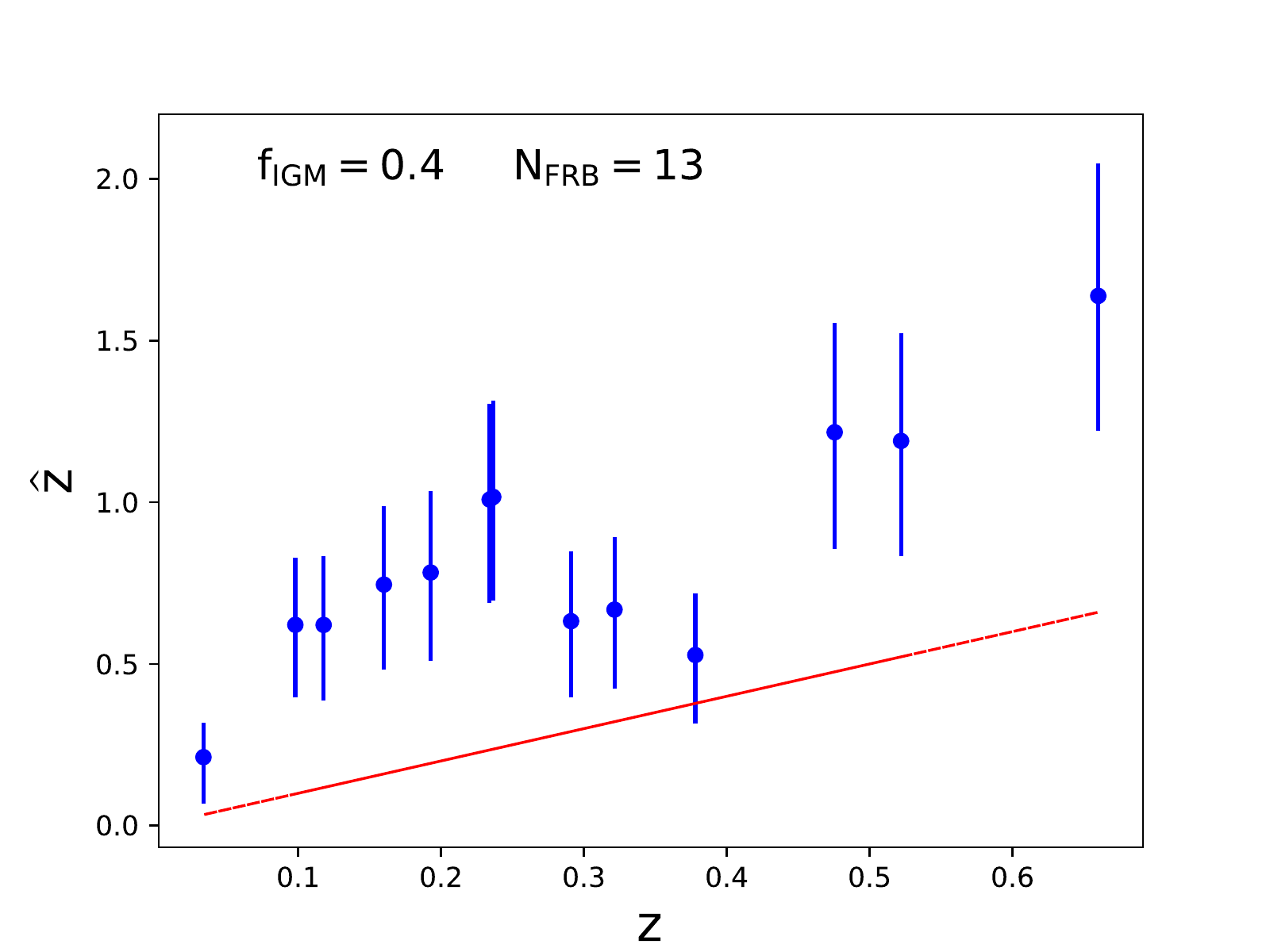}
   \includegraphics[width=\figwidththree]{\Figures/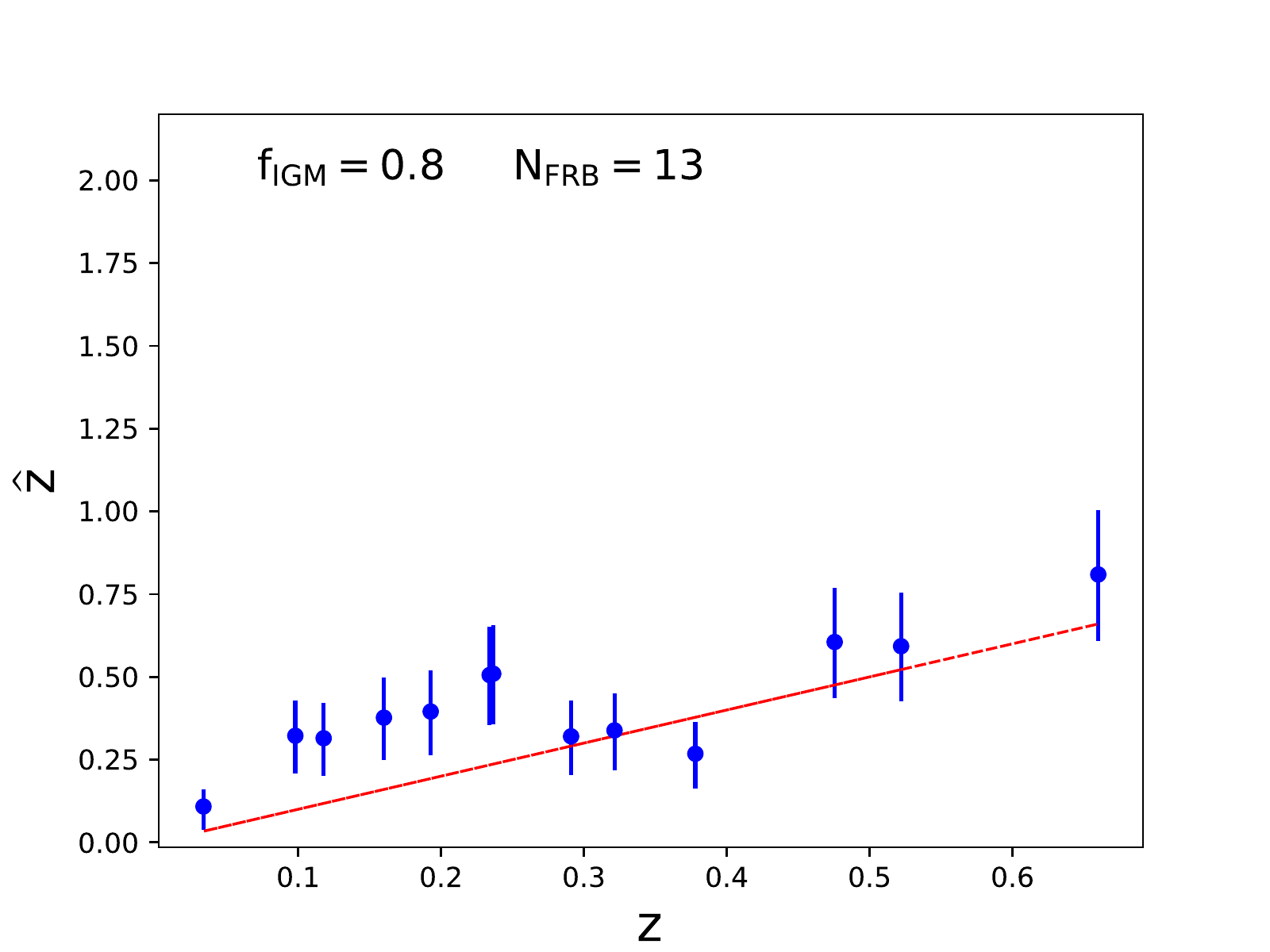}
      \includegraphics[width=\figwidththree]{\Figures/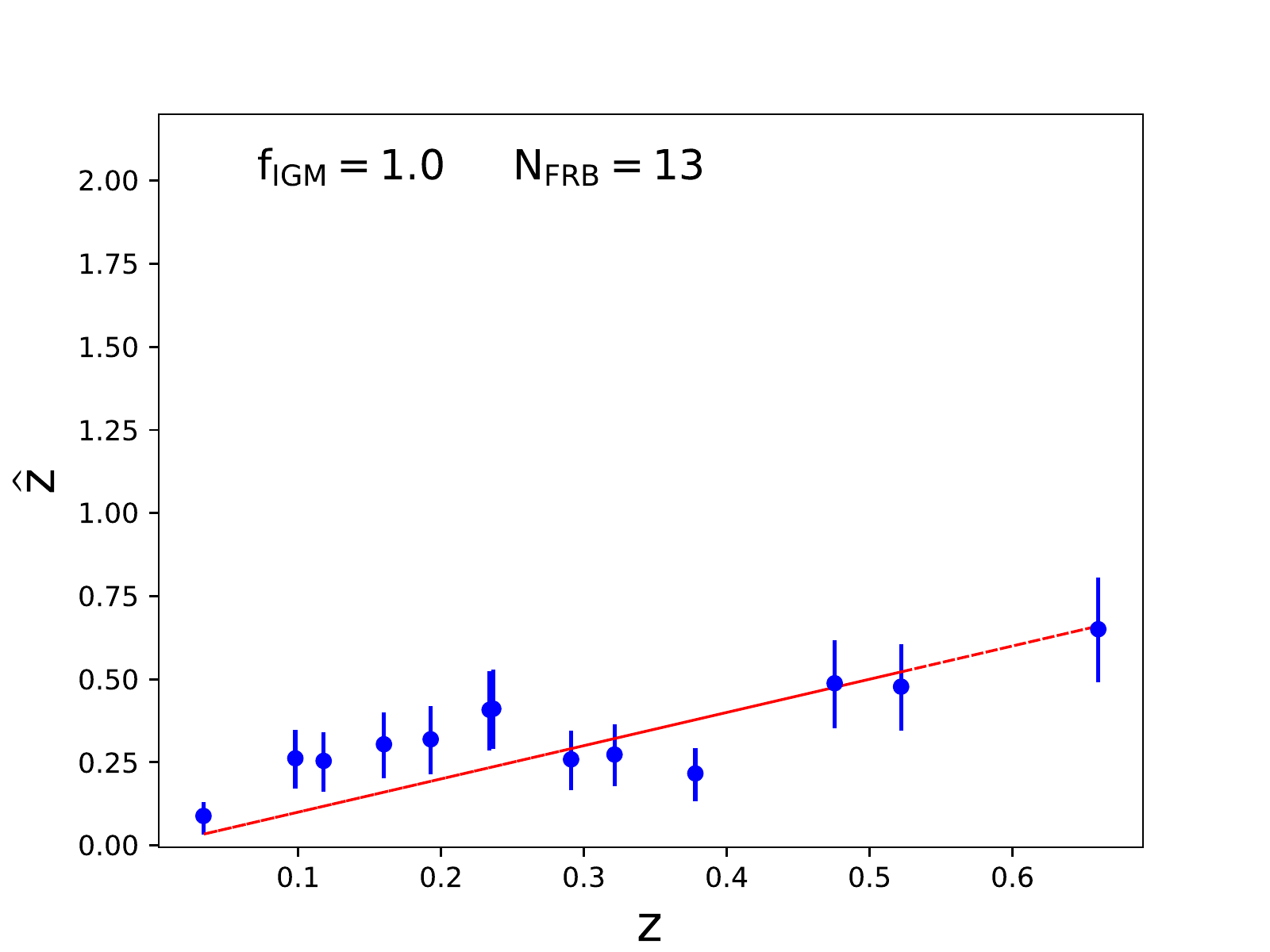}
   \caption{Redshift estimates using only the DM inventory vs. true redshift for 13 FRB where redshifts are available. 
            The three frames are for $\figm = 0.4$, 0.8, and 1.0 from left to right.        
            Vertical  bars represent the 68\% credible regions for $\zhat$ from the posterior PDFs. 
            The red lines indicate $\zhat = z$. 
            FRB~20200120E has been excluded.            
}
   \label{fig:zhat_vs_z_dm_only}
\end{figure*}

\begin{figure}[b] 
   \centering
   \includegraphics[width=\linewidth]{\Figures/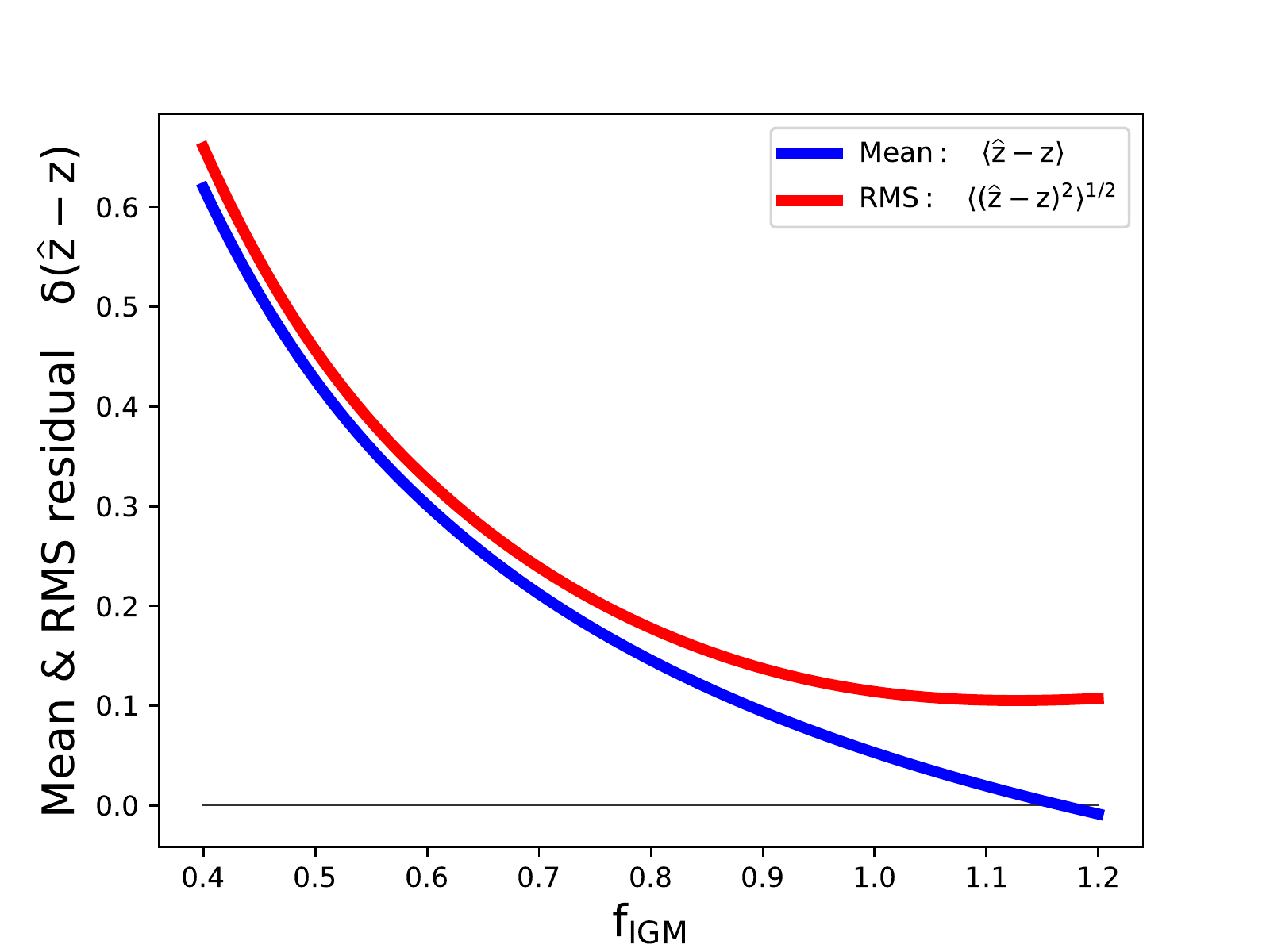}
   \caption{Mean and RMS redshift residual vs. baryonic fraction $\figm$ for the DM-only redshift estimator. 
}
   \label{fig:mean_rms_dz_dm_only}
\end{figure}

More useful is   the posterior PDF for redshift based on a likelihood function
\be
\like(\DMh, z \vert \DM; \figm)  &=&
 \\
\delta(\DM - \DMMW - \DMigm(z) &-& \DMh (1+z) ).
\ee
 Using a flat, unconstraining prior $f_{\zh}(\zh)$ for the host galaxy's redshift and integrating over  prior PDFs for $u=\DMMW, v=\DMh$ and $w=\DMigm$ yields a posterior redshift PDF
\be
&&\pdfz(\zh \vert \DM; \figm) 
\nonumber \\
&&\quad
 \propto 
     f_{\zh}(\zh) \iiint du\, dv\, dw\, \pdfmw(u) \pdfh(v) 
 \nonumber \\
  && \quad\quad\quad
  \times \  \pdfigm(w, \zh; \figm) \delta(\DM - v / (1+\zh)  - u - w)
\nonumber \\
&&\quad
 \propto 
     f_{\zh}(\zh) \iint  dv\, dw\,    \pdfh(v) \pdfigm(w, \zh; \figm)
     \nonumber \\
     && \quad\quad\quad
     \pdfmw(\DM - v / (1+\zh) - w)
\label{eq:postzdm}
\ee
In the following we hold $\DMh$  fixed at a nominal value  to compare our results  with the common practice of setting $\DMh = 50~\DMunits$.  Other fixed values can also be used and there is some tradeoff between $\figm$ and a choice for $\DMh$.  
\explain{Deleted these two sentences because they were out of place and don't add to the paper:
The goal in this paper is to demonstrate methods for improving redshift estimates that include scattering measurements.    Not all FRBs show measurable scattering so future work can apply \Eq\ref{eq:postzdm} using a prior for $\DMh$ based on a larger sample of FRBs. }

\explain{Next sentence moved here from end of subsection.}
However, foreshadowing later results, it is unrealistic to assume a constant value for $\DMh$ given the wide variety of galaxies  found to harbor FRB sources, as well as their different locations in those galaxies and the orientations of those galaxies relative to the line of sight.

\Fig\ref{fig:zhat_vs_z_dm_only} shows the DM-based redshift estimator plotted against true redshift  for 13 objects.   This sample excludes  FRB~20201120E because its association with M81 makes it too close to the Milky Way to be characterized with a redshift, which  is negative.   The three panels  for baryonic fractions $\figm = 0.4, 0.8$, and 1 demonstrate the much larger bias and scatter for $\figm = 0.4$ (left panel) compared to the two larger values used for the center and right panels.   

\added{The larger  bias and scatter for $\figm = 0.4$ arises because a larger redshift is needed to provide the IGM contribution to DM, on average, when $\figm$ is smaller and the cosmic variance of $\DMigm$ is correspondingly larger.   This may be seen from Eq.~\ref{eq:DMIGMbar}
which gives $\DMigmbar \propto \figm \rtilde_1(z) \propto \figm z$ (for $z\ll 1$), which implies that
$\zhat \propto \figm^{-1} \times \text{(required $\DMigm$)} $.     The RMS $\DMigmsig \propto \sqrt{\figm z}$ translates into an error on $\zhat $ that then scales as $\sigma_{\zhat} \propto \sqrt{\zhat / \figm}$, which is also larger for smaller $\figm$.}

As measures of the goodness of fit,  we show in \Fig\ref{fig:mean_rms_dz_dm_only}  the mean residual
$\delta z = \langle \zhat-z \rangle$, which measures the estimation bias,    and the RMS residual $\sigma_{\delta\zhat} = \langle (\delta z)^2 \rangle^{1/2}$ vs. $\figm$.   We have used values for $\figm$ that exceed unity here (and in further analyses below)  to include the possibility that FRBs reside in regions of atypically high baryon fraction 
\citep[e.g.][]{2019ApJ...886..135P}.  Angular brackets denote a weighted average using weights equal to the reciprocal of the variance of $\zhat$ for each FRB (determined from the 68\% probability region centered on the median of the posterior CDF).     The figure shows these to be monotonically decreasing with larger $\figm$. If a larger fixed value of $\DMh$ were used instead of 50~$\DMunits$,  the bias would be reduced for the FRBs with $z \lesssim 0.25$ but would increase for larger redshifts.  

\subsection{DM and $\tau$-based Redshifts}
\label{sec:dmtau2z}

Scattering can further constrain redshifts  if it is measurable and sufficiently large to require a substantial host-galaxy $\DMh$.  For a scattering time $\tauh$  attributed to a host galaxy at redshift $\zh$,   the  host-galaxy contribution to the DM (in the host frame) is
\be
\DMhhat(\tauh) 
&=& \left[ \frac{(1+\zh)^3 \nu^4 \tau(\nu)}{ C_\taud \AFtG} \right]^{1/2}
\nonumber \\
&&
\!\!\!\!\!\!\!\!\!\!\!\!\!\!\!\!\!\!\!\!\!\!\!\!
\simeq
144~\DMunits \left[ \frac{(1+\zh)^3 \nu^4 \tau_{\rm ms}(\nu)}{\AFtG} \right]^{1/2},
\label{eq:dmhhattau}
\ee
for $\nu$ in GHz, $\tau$ in ms, and $\Ftilde$ in ${\rm (pc^2\, km)^{-1/3}}$
in the approximate equality.
This in turn yields  a scattering-based point estimate for $\DMigm$,
\be
\DMigmhat(\tauh, \zh) = &&
\nonumber \\
\DM - \DMMWhat &-& \DMhhat(\tauh) / (1 + \zh),
\ee
from which a  DM-$\tau$ based redshift is estimated by inverting the dimensionless quantity $\rtilde_1(z)$ (defined in  Eq.~\ref{eq:DMIGMbar}),
\be
\zdmtau = \rtilde_1^{-1} (\DMigmhat(\tauh, \zh)/ \nezero \dH).
\ee

\newcommand{\priorDMhFtG} {f_{\DMh, \AFtG}}

To obtain the posterior redshift PDF we use the likelihood function 
\be
&&\like(x, \phi, z \vert \DM, \tau) 
\nonumber\\
&&\quad\quad\quad\quad = \pdftau(\tau \vert \DMh, z)  = \pdftauobs(\tauhat-\tau) ,
\ee
where, as before, $\pdftauobs(\delta\tau)$ is the measurement error PDF for the scattering time, $x\equiv\DMh$,  and $\phi \equiv \AFtG$ . 
We marginalize over the joint PDF of $x$ and $\phi$ and use 
a prior $f_{\zh}(\zh)$ for the host galaxy's redshift,
\be
&&\pdfz(\zh \vert \DM,  \tau) 
\nonumber \\
&&\quad
 \propto 
     f_{\zh}(\zh) \iint dx \,d\phi\, \priorDMhFtG(x,\phi) \, 
 \nonumber \\
  && \quad\quad\quad
  \times \ \pdftauobs(\tau - \Clinner C_\tau \nu^{-4} \phi \, x^2 / (1+z)^3) .
\label{eq:postz}
\ee
Errors in the estimates $\zdm$ and $\zdmtau$ are  due  to the usual uncertainties in the MW contribution 
to \DM, the measurement error in $\tauh$, and the astrophysical variance in $\Ftilde$ but mostly from cosmic variance in
$\DMigm$ and uncertainty in $\figm$.

\begin{figure*}[t!] 
   \centering
    \includegraphics[width=\linewidth]{\NewFigures/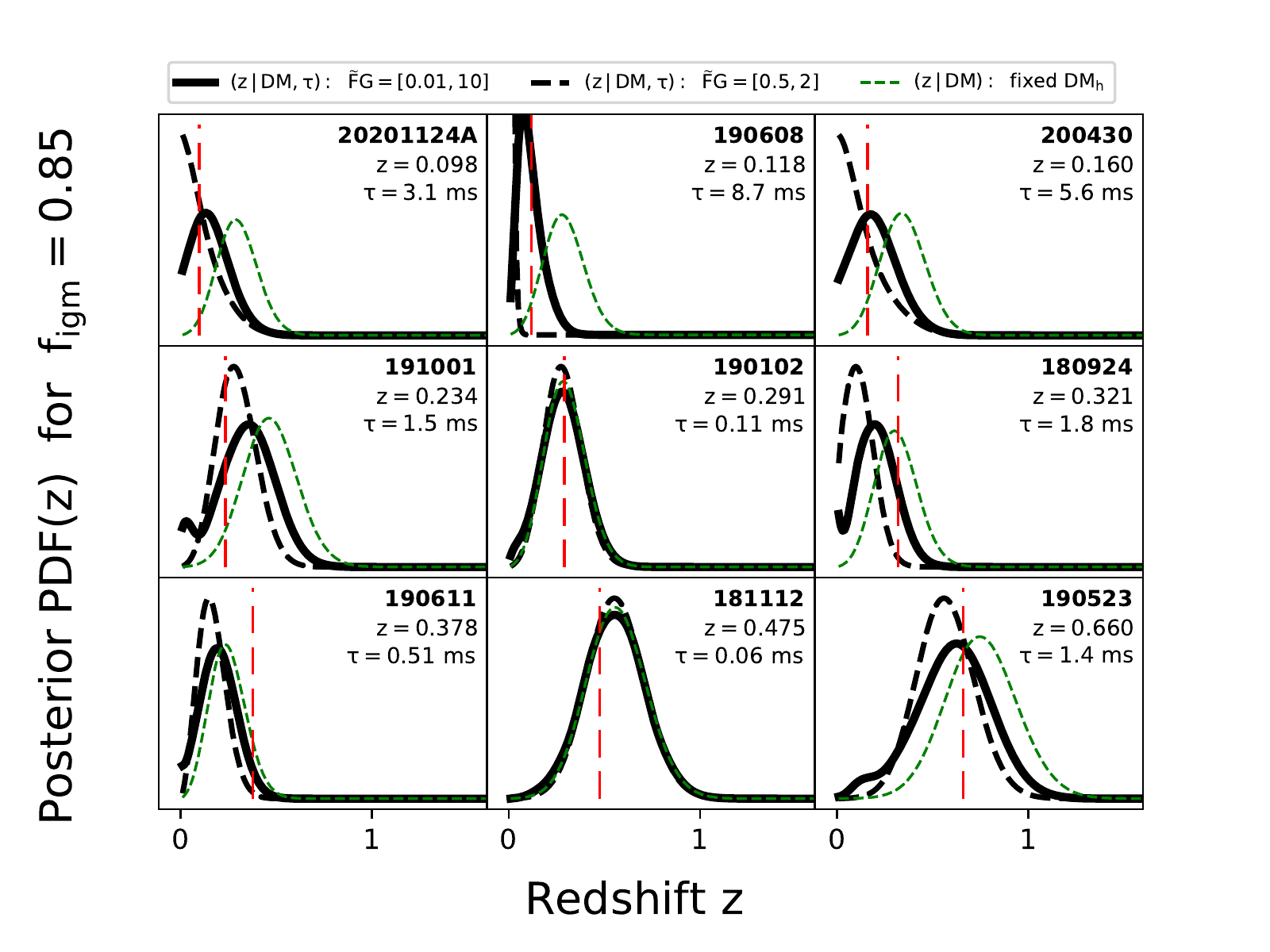}
   \caption{\footnotesize   Posterior redshift PDFs for three different redshift estimators applied to nine FRBs with both redshift and scattering measurements and using an
   ionized baryonic fraction $\figm = \figmused$.  Two (solid and dashed black lines) use  the measured DM along with the scattering time $\tau$ but with different ranges for $\AFtG$ .  The third  (green dashed line) uses only the measured DM.   In each panel, the red dashed vertical line indicates the measured redshift of the associated host galaxy.  In each panel the redshift and scattering time at 1~GHz are given.}
   \label{fig:pdfz_9pack}
\end{figure*}

\explain{This paragraph replaces previous text   to give more motivation and details about the shapes of the priors used.}
\added{
We evaluate \Eq\ref{eq:postz} using a  flat, uninformative redshift prior $f_{\zh}(\zh)$ and a flat PDF for
$x\equiv\DMh$  over a range $\DMh = [20, 1600]~\DMunits$.   For $\phi \equiv  \AFtG$, we use a flat PDF
that is sampled at logarithmic intervals over two ranges to provide two different  priors for $\phi$:  a broad range 
$[\phi_{\rm min}, \phi_{\rm max} = [0.01, 10]\ \FtGunits$ and a  narrow range, $[0.5, 2]\ \FtGunits$.   We take this approach to illustrate the effects of alternative priors for the current limited sample of nine objects with scattering measurements and redshifts.  The broad range  is consistent with most of the pulsar and FRB measurements.   In the future when more redshifts and scattering times are available, we will explore usage of an alternative prior for $\AFtG$, such as a log-normal distribution. }
 
\Fig\ref{fig:pdfz_9pack} shows the posterior PDFs for the three estimators applied to  the nine FRBs with available redshifts (again excluding the nearby FRB~20200120E) and scattering times and using $\figm = \figmused$, a choice that is discussed below.   
Scattering is constraining on the redshift if $\zdmtau$ is substantially smaller than $\zdm$ or equivalently 
if $\DMhhat(\tauh)$ is substantially larger than an {\it a priori} chosen value.       The scattering-based redshift estimator is 
likely more accurate for cases where the resulting change in $\DMigmhat$ from a scattering-based estimate of $\DMh$  is larger than one standard deviation from cosmic fluctuations, or $\DMhhat(\tauh) > (1+\zh)\DMigmsig(\zh)$.      

Applying this constraint  using  \Eq\ref{eq:DMIGMbar}, 
\ref{eq:DMIGMsig} and \ref{eq:dmhhattau}  and approximating $\rtilde_1(\zh) \sim \zh$ for redshifts $\zh \lesssim 1$, the criterion for when   scattering influences redshift estimates is
\be
\tau_{\rm 1\, GHz} \gtrsim 2~{\rm ms} \times \AFtG (\figm / 0.85)\, \zh .
\label{eq:tau_constraining}
\ee
This expression is consistent with the posterior  PDFs shown in  \Fig\ref{fig:pdfz_9pack} for FRBs with different redshifts and scattering times.    Those with small scattering times yield nearly identical PDFs for the DM-only and DM+scattering estimators while those with scattering times greater than about one millisecond are clearly influenced by scattering.   This unsurprising result simply underscores the consistency of the  method. 

Table~\ref{tab:FtG_and_zhatscatt} gives redshift estimates for the  objects in
Table~\ref{tab:FRBproperties}.  The columns are the FRB name,  measured DM, redshift,
median host-galaxy $\DMh$, the scattering quantity  $\AFtG$, and median redshift estimates and credible ranges using  narrow and wide ranges for $\AFtG$.   
\added{
Estimates for $\AFtG$ are made by inversion of 
Eq.~\ref{eq:taudmz_xgal} (again with $\nu$ in GHz and $\tau$ in ms)
}
\be
\AFtG &=& 2.1~\FtGunits 
\nonumber \\
&& \quad\quad
 \times  \frac{\nu^4 (1+\zh)^3 \tau(\nu)} {(\DMh / 100~\DMunits)^2}.
 \label{eq:AFtG}
\ee
\added{
The five FRBs with scattering upper limits yield upper limits on $\AFtG$. 
The upper limit on  $\AFtG$ for FRB~20200120E is very small, consistent with the absence  of scattering from either  the halo of M81 or the halo  of the Milky Way.   The values and upper limits on $\AFtG$ are all consistent with the adopted prior that is flat between $0.01$ and 10~$\FtGunits$. }

For the four cases with $\tau \ge 3.1$~ms (at 1 GHz), the true redshift is below that of the DM-based estimator and more consistent with the scattering based estimator using the broader range of
$\AFtG$.   For small scattering times, $\tau \lesssim 0.1$~ms, the three estimators give the same result because the measured scattering does not require a large $\DMh$ for either of the ranges for $\AFtG$.  The intermediate cases  FRB~20191001A and FRB~20180924A with $\tau = 1.5$~ms and 1.8~ms, respectively,
are mixed, with the former object being more consistent with the scattering-based redshift and the latter slightly more consistent with the DM-only estimator.      The outlier in this sample of nine is FRB~20190611B where the true redshift is larger than the mode, mean, or median of any of the estimators but is not improbable for the DM-only estimator or the scattering estimator with a broad range of $\AFtG$.     \citet[][]{2020Natur.581..391M} noted that the association of the FRB with the galaxy at $z= 0.378$ is tentative and the redshift estimations here may reflect that possibility. 
 
\subsection{Constraints on the Baryon Fraction $\figm$}

\begin{figure*}[htbp] 
   \centering
   \includegraphics[width=\figwidthtwo]{\Figures/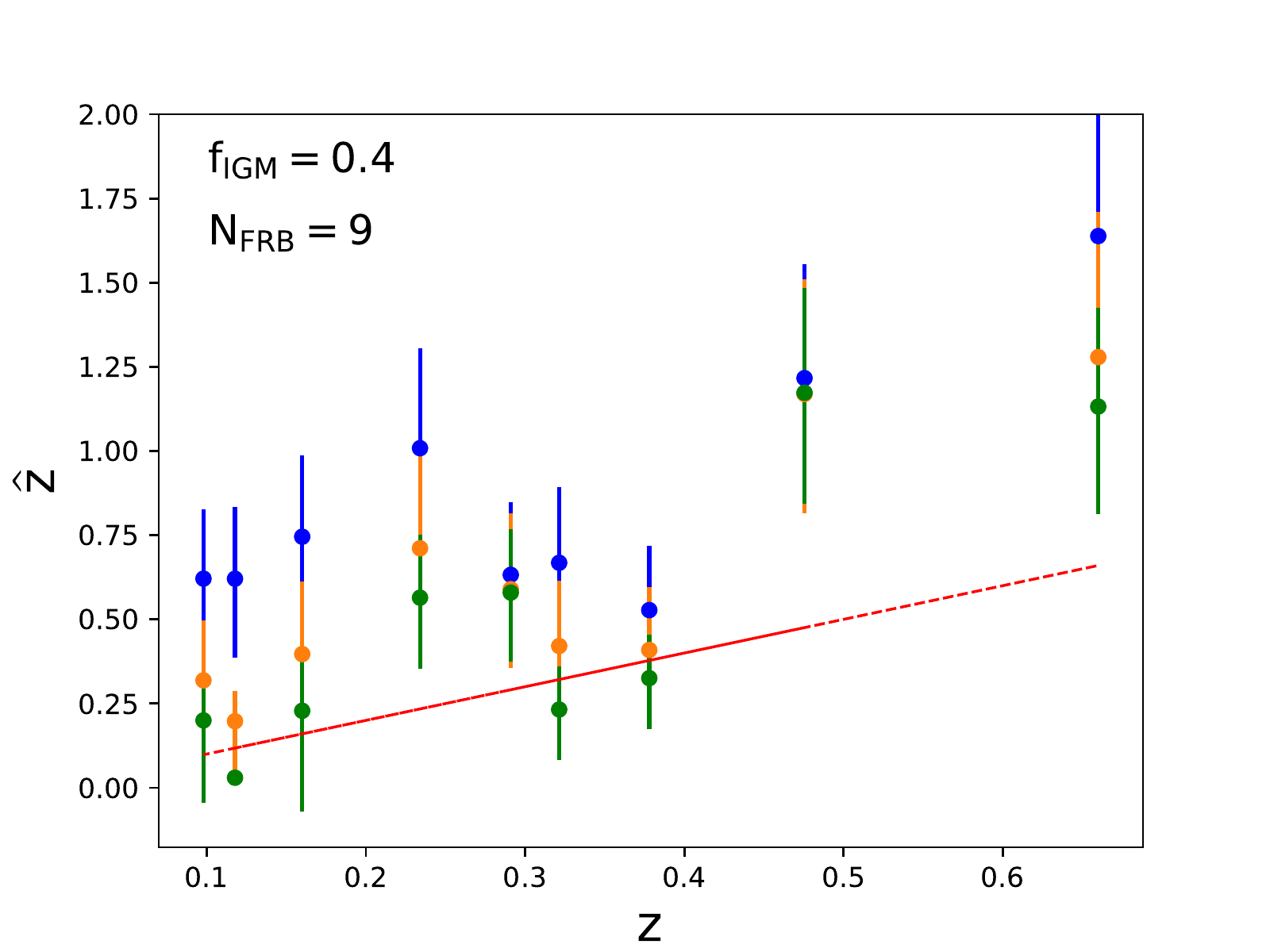}
   \includegraphics[width=\figwidthtwo]{\Figures/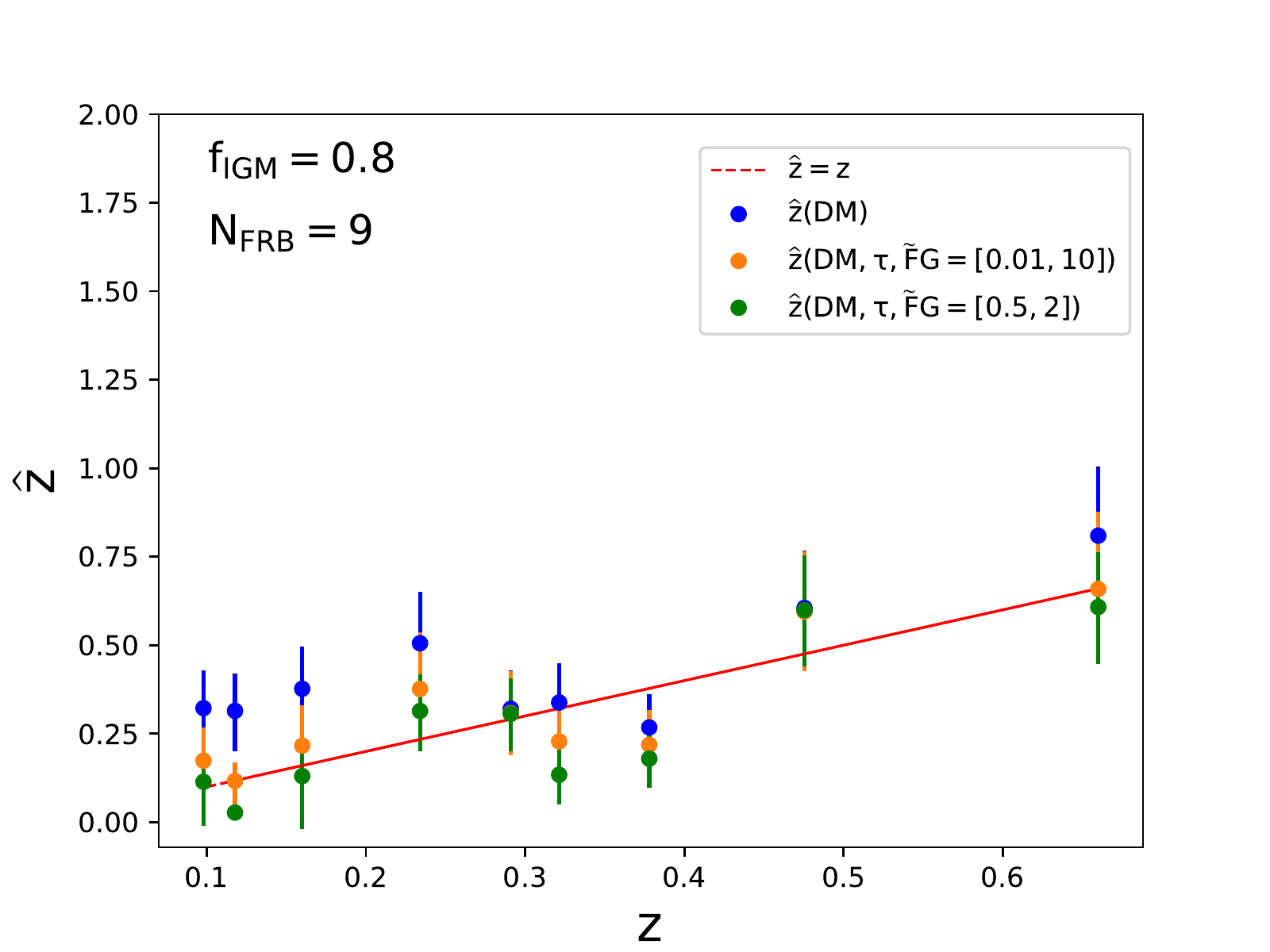}
   \caption{Redshift estimates vs. true redshift for nine FRB cases where  redshifts and scattering measurements are both available. 
    Left: $\figm = 0.4$; Right:   $\figm = 0.8$.
     Vertical bars represent the 68\% credible region centered on the median value derived from the posterior PDFs.
     The red lines show  $\zhat = z$.  The legend applies to both frames.
    }
   \label{fig:zhat_vs_z}
\end{figure*}

\Fig\ref{fig:zhat_vs_z} shows $\zhat$ plotted against $z$ using the three different redshift estimators  with values of 0.4 and 0.8 for  $\figm$ that allow comparison with two of the panels in Figure~\ref{fig:zhat_vs_z_dm_only}.   The plotted points are the median values of the posterior PDFs.   
Note that here the posterior PDF for $\DMh$ is calculated only for the nine objects with $\taud$ measurements compared to 13 objects use in Figures~\ref{fig:zhat_vs_z_dm_only} and \ref{fig:mean_rms_dz_dm_only}. The other two estimators incorporate scattering  using the two different ranges for $\AFtG$ described above   The larger value $\figm = 0.8$ yields much greater consistency between $\zhat$ and $z$ than the smaller value.  

\begin{figure*}[t!] 
   \centering
   \includegraphics[width=\figwidthtwo]{\NewFigures/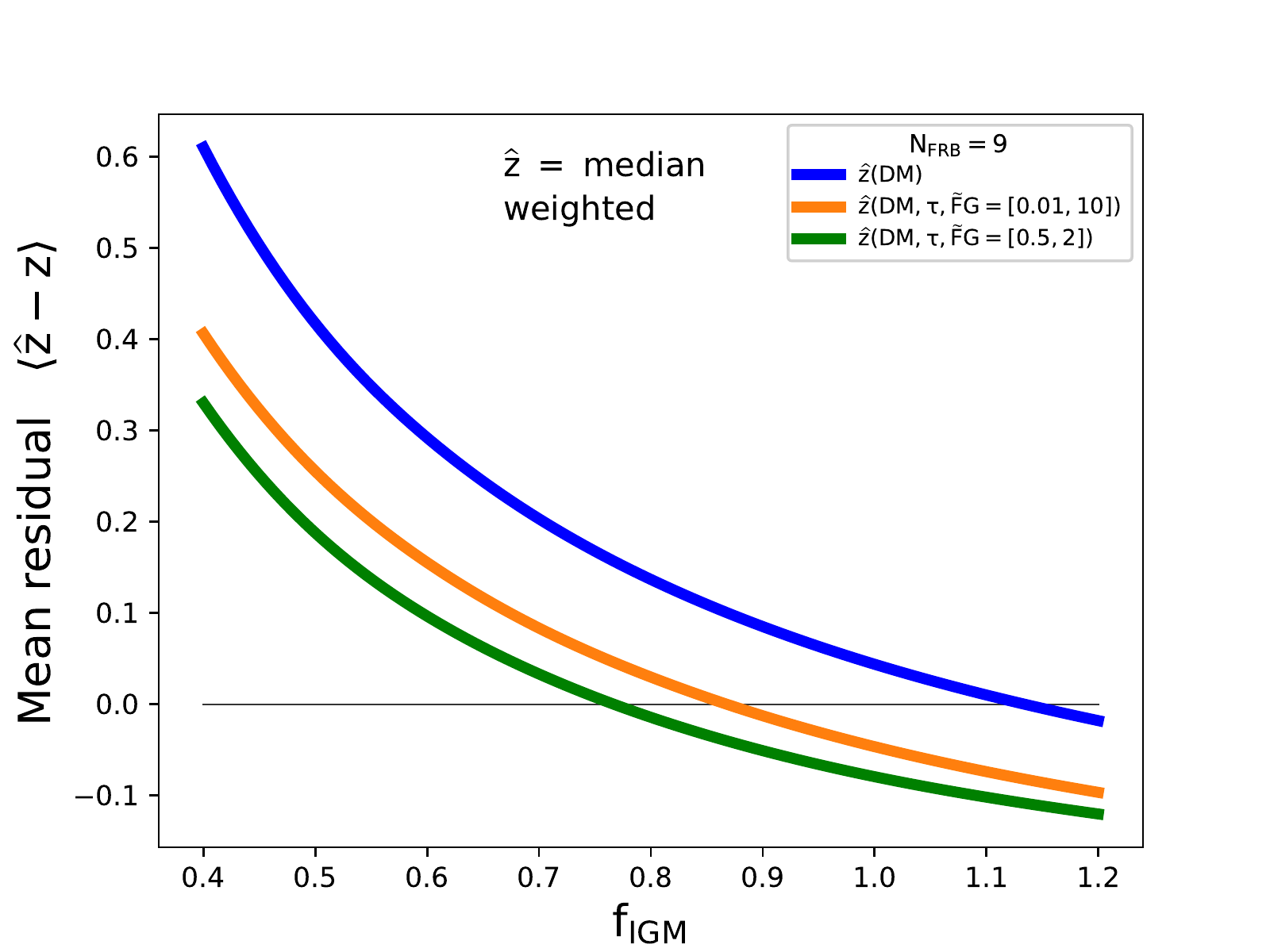}
   \includegraphics[width=\figwidthtwo]{\NewFigures/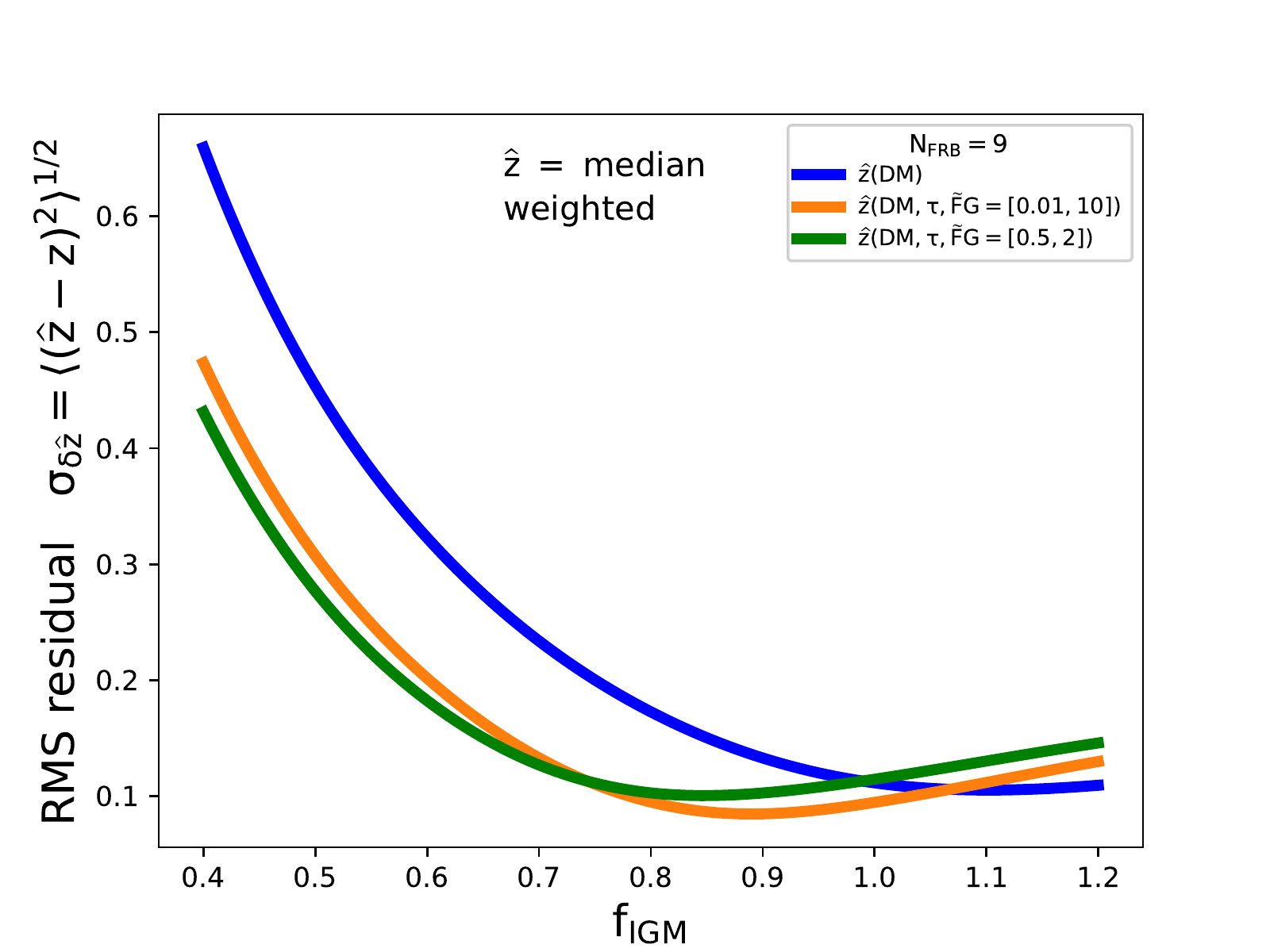}
   \caption{\footnotesize  Mean (top) and RMS (bottom) residual difference between estimated and true redshifts estimates vs. $\figm$.  Redshift estimates are median values calculated from posterior PDFs for $\widehat z$.    
The three curves for each set correspond to $\widehat z$ estimators using DM only and two using  the combined DM and $\tau$ estimator with a wide or narrow ange of $\Ftilde G$, as indicated in the legend.   }
   \label{fig:mean_and_rms_zhat_vs_z}
\end{figure*}

To identify the plausible range for $\figm$, we show the mean and RMS residual
$\langle \delta z\rangle = \langle \zhat-z \rangle$ and $\sigma_{\delta\zhat} = \langle (\delta z)^2 \rangle^{1/2}$ for the three estimators  in the left and right-hand panels of \Fig\ref{fig:mean_and_rms_zhat_vs_z}.   
The figure shows $\langle \delta z \rangle$  to be monotonically decreasing with larger $\figm$ for all three estimators
and $\sigma_{\delta\zhat} $ decreasing up to $\figm \sim \figmused \pm \figmusederr$.  
It is notable that  all three of the redshift estimators are better for  larger values of  the baryonic fraction, 
$\figm\gtrsim 0.8$, showing less bias and less scatter about the measured redshifts. 

The DM-based estimator remains positive for  values of $\figm \lesssim 1$ and is therefore biased.   The wide-range $\AFtG$ scattering estimator crosses zero at $\figm \sim 0.85$ and the narrow-range estimator crosses at $\figm \sim 0.75$.    The RMS residual curve for the DM-only estimator decreases monotonically and is slightly below that of the narrow-$\AFtG$ estimator at $\figm = 1$ but is larger than both scattering-based estimators for $\figm \lesssim 0.85$.  The two scattering-based estimators bottom out at $\figm \simeq 0.8$ to 0.9.   Considering both the bias and the minimum RMS residual,  a value $\figm \simeq 0.8$ to 0.9 appears to give the best match.   For these values the bias of the wide-$\AFtG$ scattering estimator is $\vert \langle \delta\zhat\rangle \vert \lesssim 0.02$ and the RMS redshift error is $\sigma_{\delta \zhat} \sim 0.1$.    

\iffrbFAST
	{\bf Will likely exclude this paragraph; the figure alluded to is not included here but it shows dramatically how the dependence 		on $\figm$ varies}
	Results are shown with and without inclusion of FRB~190520, the source with the very large implied host-galaxy contribution to 		DM.   When included,   the best estimator uses scattering with a narrow range of
	$\Ftilde G = [0.5, 2.0]$ and yields $\sigma_{\widehat z}\sim  0.12$ that is substantially smaller than the RMS residuals for the 		other two methods.   When excluded, all three estimators improve substantially 
	but the one using the wider range of $\AFtG$ is the best of the three.
\fi


\begin{deluxetable*}{l c r    c    C  r r r   C r r r  C r r r }[t]
\tabletypesize{\footnotesize}
\tablecaption{ FRB Host Galaxy Parameters and Redshift Estimates \label{tab:FtG_and_zhatscatt}}
\tablehead{
\colhead{FRB} &
\colhead{\DM} & 
\colhead{$z_{\rm h}$} &
\multicolumn{1}{c}{$\DMh$} &
\multicolumn{1}{c}{$\AFtG$} &
\multicolumn{3}{c}{$\zhat(\DM, \tau, {\rm narrow})$} && 
\multicolumn{3}{c}{$\zhat(\DM, \tau, {\rm wide})$} 
\\
\cline{6-8}
\cline{10-12}
\colhead{} & 
 \colhead{($\DMunits$)} &
 \colhead{} & 
\multicolumn{1}{c}{($\DMunits$)}  &  
 \multicolumn{1}{c}{$(\FtGunits)$}  & 
\\
\colhead{(1)} &
\colhead{(2)} &
\colhead{(3)} &
 \multicolumn{1}{c}{(4)}  &
 \multicolumn{1}{c}{(5)}  & 
 \colhead{(6)} &
  \colhead{(7)} &
  \colhead{(8)} &&
  \colhead{(9)} &
  \colhead{(10)} &
  \colhead{(11)}  
 }
\tablecolumns{18}
\startdata
20121102A  &    557    &   0.193   &   215   &  $<0.46$  &  $\cdots$  &  $\cdots$  &  $\cdots$  &&  $\cdots$  &  $\cdots$  &  $\cdots$  \\ 
20180916B  &    349    &   0.034   &   82   &  $<0.092$  &  $\cdots$  &  $\cdots$  &  $\cdots$  &&  $\cdots$  &  $\cdots$  &  $\cdots$  \\ 
20180924A  &    361    &   0.321   &   99   &  $8.7$  &   0.128   &   $-$0.066   &   $+$0.080    &&   0.216   &   $-$0.097   &   $+$0.106    \\ 
20181112A  &    589    &   0.475   &   206   &  $0.094$  &   0.566   &   $-$0.147   &   $+$0.151    &&   0.561   &   $-$0.159   &   $+$0.159    \\ 
20190102B  &    364    &   0.291   &   100   &  $0.48$  &   0.290   &   $-$0.094   &   $+$0.101    &&   0.292   &   $-$0.111   &   $+$0.113    \\ 
20190523A  &    761    &   0.660   &   261   &  $2.0$  &   0.576   &   $-$0.147   &   $+$0.153    &&   0.623   &   $-$0.206   &   $+$0.190    \\ 
20190608B  &    339    &   0.118   &   190   &  $7.0$  &   0.027   &   $-$0.005   &   $+$0.009    &&   0.111   &   $-$0.049   &   $+$0.075    \\ 
20190611B  &    321    &   0.378   &   58   &  $8.3$  &   0.170   &   $-$0.067   &   $+$0.079    &&   0.207   &   $-$0.092   &   $+$0.099    \\ 
20190711A  &    593    &   0.522   &   171   &  $<8.0$  &  $\cdots$  &  $\cdots$  &  $\cdots$  &&  $\cdots$  &  $\cdots$  &  $\cdots$  \\ 
20190714A  &    504    &   0.236   &   289   &  $<2.5$  &  $\cdots$  &  $\cdots$  &  $\cdots$  &&  $\cdots$  &  $\cdots$  &  $\cdots$  \\ 
20191001A  &    507    &   0.234   &   287   &  $0.72$  &   0.298   &   $-$0.099   &   $+$0.109    &&   0.356   &   $-$0.152   &   $+$0.143    \\ 
20200430A  &    380    &   0.160   &   217   &  $3.9$  &   0.124   &   $-$0.075   &   $+$0.141    &&   0.205   &   $-$0.108   &   $+$0.128    \\ 
20200120E  &    88    &  $\cdots$  &   13   &  $<0.014$  &  $\cdots$  &  $\cdots$  &  $\cdots$  &&  $\cdots$  &  $\cdots$  &  $\cdots$  \\ 
20201124A  &    414    &   0.098   &   172   &  $3.9$  &   0.109   &   $-$0.064   &   $+$0.117    &&   0.165   &   $-$0.088   &   $+$0.109    \\ 
 \enddata
\end{deluxetable*}

\subsection{Discussion of Individual FRBs with Measured Scattering Times}
\label{sec:taumeas}

In previous sections, scattering has been attributed to host galaxies,  yielding  a range for $\AFtG$ of $\sim [0.1, 9]~\DMunits$ (Table~\ref{tab:FtG_and_zhatscatt}), from which estimates for the host galaxy \DM\ contribution and redshift were made. 
We now discuss individually each FRB for which there are both scattering and redshift measurements.
Quoted scattering times are referenced to 1~GHz.   

{\bf FRB~20180924A} ($\tau = 1.78 \pm0.08 $~ms, $z = 0.321$):   \citet[][hereafter H20]{2020ApJ...903..152H} designate the galaxy association as high probability (A class);  the FRB is only slightly offset from the  galaxy center.  Scattering is constraining on the redshift  for smaller values of the baryon fraction, $\figm \lesssim 0.6$ but the DM-only estimate matches the measured redshift for $\figm \gtrsim 0.7$. For these larger $\figm$,   $\AFtG \simeq 8.7$ is needed to better match the redshift.  

{\bf FRB~20181112A} ($\tau = 0.06 \pm 0.003 $~ms, $z = 0.475$):  Also designated a high probability association  by H20.  The scattering is too small to be constraining, in accordance with the criterion in
Eq.~\ref{eq:tau_constraining}.  The measured redshift is slightly less than the median $\zhat$ for all three estimators. 

{\bf FRB~20190102B} ($\tau = 0.11 \pm 0.008$~ms, $z=0.291$):  Another high-probability association (H20).  Scattering is too small to be constraining.   The $\zhat$ estimators favor $\figm \gtrsim 0.6$.


{\bf FRB~20190523A} ($\tau = 1.4 \pm 0.2$~ms, $z=0.66$):  The FRB is offset from the galaxy center by 27~kpc and there is a 7\% probability of a chance association, yielding a C classification by H20.  Scattering is constraining and requires $\AFtG \sim 2$ in order to match the measured redshift for  $\figm = 0.85$.   The DM-only estimator requires $\figm \gtrsim 0.6$.  

Given the large offset from the galaxy center,  it is possible that the required value of $\AFtG$ receives a significant contribution from a geometric factor, $G>1$.   This  could arise from a contribution to DM from a galaxy halo or disk with a very small value of $\Ftilde$  but with a large geometric boost.   An alternative is   that the candidate galaxy association is incorrect.  Given that the galaxy has the largest redshift in the sample, a mis-association would require reassessment of the empirical $\DM(z)$ statistics.

{\bf FRB~20190608B} ($\tau = 8.7 \pm 0.5 $~ms, $z=0.118$): An A class association with the FRB coincident in projection with a spiral arm \citep[H20; see also][]{2020Natur.581..391M,2021ApJ...917...75M}.
Scattering is strongly constraining on redshift and requires smaller values of $\AFtG$ in the $[0.1, 10]$ range.   The DM-only estimate is a poor match for all values of $\figm$. 

\explain{The discussion of FRB~20190611B is rewritten and the conclusion about the galaxy association is tempered from the original submission.    Some of the numbers quoted in the original submission were incorrect.}
{\bf FRB~20190611B} ($\tau = 0.51 \pm 0.06$~ms, $z = 0.378$):  The FRB-galaxy association was called `tentative' by \citep[][]{2020Natur.581..391M} but designated as   class A by H20 in spite of a significant offset $\sim 11 \pm 4$~kpc from the galaxy center compared to an $i$-band radial size of $\sim 2$~kpc. 

The galaxy's redshift  is within the credible region for $\zhat$ only for relatively small values of
$\figm \lesssim 0.7$ for the DM-only estimator and for $\figm \lesssim 0.6$ and $\lesssim 0.5$ using the 
DM-$\tau$ estimators with wide and narrow ranges of $\AFtG$, respectively. 
This implies that a smaller than normal IGM contribution to DM is needed to match the DM inventory and allow the host-galaxy $\DMh$ to be large enough to account for the measured scattering for the two ranges of $\AFtG$ considered in the analysis. 

From Table~\ref{tab:FRBinventory},  the median $\DMigm$ using the 
the log-normal model of \S\ref{sec:DMigm} is $320~\DMunits$, nearly equal to the measured $\DM = 321~\DMunits$ without any consideration of contributions from the Milky Way or host galaxy.   When those contributions are included,  the measured DM is estimated to have  a total extragalactic contribution, 
$\DMxgal \simeq \DMigm + \DMh / (1+\zh) = 211 \pm 17~\DMunits$ and an implied IGM contribution,
$\DMigm  = \DMxgal - \DMh / (1+\zh) \simeq 179~\DMunits$, with an uncertainty of about 40~$\DMunits$ (where we have used the geometric mean of the asymmetric confidence interval values for $\DMh$ in quadrature with the uncertainty in the MW contribution).    

This IGM value is  $\sim 141~\DMunits$ below the median IGM  from the log-normal model of \S\ref{sec:DMigm}, or about 1.5 times the 68\% confidence range, $\sigma_- = 97~\DMunits$, to smaller $\DMigm$ values (column 7 of Table~\ref{tab:FRBinventory}).   While this is not overly improbable in a 9-object sample,  the necessarily smaller IGM contribution implies that the line-of-sight to this FRB needs further study. 


The measured scattering requires the second largest value of $\AFtG$ given in Table~\ref{tab:FtG_and_zhatscatt} (column 3), which is based on  $\figm = 0.85$, a value that is   consistent with the entire set of objects.  Using Eq.~\ref{eq:AFtG} along with the median inferred value for $\DMh = 58~\DMunits$, we require $\AFtG \simeq 8.3$ (Table~\ref{tab:FtG_and_zhatscatt}). 

The summary for this source is that the DM inventory requires a lower-than average contribution from the IGM for the redshift of the proposed galaxy association. However, it is not so extreme that  the association is necessarily incorrect.  However an incorrect association is certainly a possibility.

%
%
%

{\bf FRB~20191001A} ($\tau = 1.5\pm0.1$,  $ z = 0.234$):  An A-class association (H20) with a 2:1 offset relative to the $i$-band radial size but with the FRB overlaying a spiral arm \citep[][]{2021ApJ...917...75M}.  The redshift is overestimated by the DM-only estimator but is consistent with either of the scattering-based estimators. 

{\bf FRB~20200430A} ($\tau = 5.6\pm2.8$~ms, $z = 0.160$):  An A-class association (H20).   Scattering is constraining on the redshift. The DM-only redshift estimator is disfavored compared to the DM-$\tau$ estimators, especially for smaller values of $\figm \lesssim 0.6$.       But even for $\figm \ge 0.9$, the scattering constraint gives a better estimate. 

{\bf FRB~20201124A} ($\tau = 3.13 \pm  1.7$~ms, $z=0.098$): The lowest redshift galaxy in the sample,  J0508+2603, has a stellar mass comparable to the two  candidate galaxies for FRB~20190523A and FRB~20191001A \citep[][H20]{2021arXiv210609710R}  with an extended source spatially coincident with the FRB and associated with star formation activity.   The scattering time is large enough to constrain  the redshift with both $DM-\tau$  estimates yielding credible values  superior to the DM-only result for any value of $\figm$.  The two $\DM-\tau$ estimates are equally good for $\figm \gtrsim 0.7$.     


In summary,   combined measurements of the scattering time $\tau$ and DM yield better estimates for redshift than DM-only estimates in the majority of the nine FRBs for which such measurements exist along with redshifts. The exceptions are when the scattering time is too small to constrain the host-galaxy DM using plausible values of the fluctuation-geometry parameter $\AFtG$.     

\subsection{FRBs with Scattering Upper Limits}
\label{sec:tauul}

Four FRBs in Table~\ref{tab:FRBproperties}   have A-class galaxy associations (H20).    Three of these (FRBs~20121102A, 20180916B, and 20190711A) have scattering upper limits  too large to be  constraining on the host-galaxy \DM\ or on the redshift.  
The fourth case, FRB~20200120E, has a very low upper limit that is informative about scattering in the halos of the MW and M81, as previously mentioned.    A detailed interpretation of that case is deferred to another paper (in preparation). 

FRB~20121102A  warrants additional discussion because Balmer line measurements place ancilllary constraints on $\DMh$
\citep[][]{tbc+17}.  FRB~20121102A  is in a dwarf, star-forming galaxy at  redshift $\zh = 0.193$ and produces bursts with a total
$\DM \approx 570~\DMunits$  contributed to roughly equally by the MW, the IGM, and its dwarf host galaxy \citep[][]{tbc+17,bta+17,2017ApJ...844...95K}.    Intensity scintillations  imply a small Galactic contribution to temporal broadening ($\sim 20~\mu s$ at 1 GHz)   and an upper bound on extragalactic scattering is $\taud(1~{\rm GHz}) < 0.6$~ms. 
The dependences of  $\DMh$  and extragalactic scattering on redshift and other quantities shown in Figure~\ref{fig:twopanel_twofrbs} (left panel) indicate that the upper bound on $\taud$ is consistent with the plausible range of $\DMh$ (c.f. \Fig\ref{fig:DMh_PDF_CDF_page3}) combined with possible
 values for $\AFtG$ that match the ranges for $\AFtG$  implied by FRBs with measured $\taud$ and with the expectations based on Galactic pulsars.  

\section{Summary and Conclusions} 

We have used dispersion and scattering measurements on FRBs with candidate host galaxy associations and their redshifts to characterize scattering.  We have shown 
in \S\ref{sec:dmtau2z} that the combined $\DM-\tau$ redshift estimator more accurately predicts the redshift than a $\DM$-only based estimate.   Overall the results are consistent with our assumption that scattering of FRBs is dominated by galaxy disks, including that of the Milky Way, but is not significant from galaxy halos or the IGM.     

We have  derived an expression (\Eq\ref{eq:dmhhattau}) for the host-galaxy dispersion measure, 
$\DMh \propto (1+z)^{3/2}  (\taud / \AFtG)^{1/2}$ in terms of the scattering time $\tau$ and turbulence fluctuation parameter 
$\Ftilde$. This is useful by itself for providing an order of magnitude estimate where the (considerable) uncertainty derives primarily from that for $\Ftilde$.    When combined with  a model for the IGM's contribution  to the DM inventory, the expression for $\DMh$ also provides the basis for a scattering (and dispersion) based redshift estimate.    The scattering time is constraining on the redshift if it satisfies the inequality given in \Eq\ref{eq:tau_constraining} and a specific range of values for the
composite scattering quantity $\AFtG$ is known or assumed.     Similarly if independent constraints on the host galaxy's dispersion measure, $\DMh$, is known from 
(e.g.) emission line measurements, then $\AFtG$ can be estimated using \Eq\ref{eq:AFtG}.

 Imperfections in this method are also tied to the question of whether some of the candidate FRB-galaxy associations are genuine. 
Two of the  FRBs (FRB~20190523A and FRB~20190611B) are offset significantly from  their proposed galaxy associations.  If the associations are real significant contributions to DM and to scattering must come from the outskirts of the galaxies or from their halos. That is in contrast to the MW and raises the second caveat that the  large offsets may cast doubt on the reality of those associations. 

As discussed in \S\ref{sec:taumeas}, the redshift of 0.378 for FRB~20190611B is larger than all three of the propagation-based redshift estimates, whereas more typically  the DM-based redshift  overestimates the redshift.    This suggests that the host galaxy might be closer than $z=0.378$ and would necessarily be much dimmer to avoid identification in  the images from 
\citet[][]{2020Natur.581..391M} and   \citet[][]{2020ApJ...903..152H}.   

FRB~20190523A has the largest redshift (0.66) in the sample that is not inconsistent with the DM or scattering based redshift estimates.   It  requires a large extragalactic contribution  $\DMxgal \sim 740~\DMunits$  of which about 20\%  is from the host galaxy at the stated redshift (c.f. Table~\ref{tab:FRBinventory}.   However, as noted above, the galaxy  association \citep[][]{2019Natur.572..352R}
is in the C class defined by H20, so conclusions about the source of the large DM may be premature at this point. 

We conclude that FRB studies can benefit from redshift estimation that incorporates scattering measurements.  This is true especially now when optically-determined redshifts are few in number.  But we expect it will also help in the future even when  more precise redshifts are measured for some but certainly not all FRBs.  For this to be the case, scattering measurements need to be robustly differentiated from frequency-time structure in FRBs that differs from that produced by scattering in either a host galaxy or the Milky Way.

\begin{acknowledgements}
J.M.C., S.K.O., and S.C. acknowledge support from the National Science Foundation (NSF AAG-1815242) and are members of the NANOGrav Physics Frontiers Center, which is supported by the NSF award PHY-2020265.
\end{acknowledgements}

\appendix
\section{Pulse Broadening for the Cloudlet Model}
\label{app:cloudlet}

We describe electron density  fluctuations inside a turbulent cloud  with a power-law spectrum 
$\propto \cnsq q^{-\beta}\exp[-(2\pi q / \linner)^2]$ for wavenumbers $2\pi/\louter \le q \lesssim 2\pi/\linner$,
where $\louter$ and $\linner \ll \louter$ are the outer and inner scales, respectively.   We use a Kolmogorov spectrum  with $\beta = 11/3$ as a reference spectrum.    This builds upon the model originally presented in  \citet[][]{cwf+91}.

The integral of the spectrum gives the variance of the electron density inside a cloud, 
$\sigma_{\nelec}^2 = (\varepsilon \nebar)^2 = \CSM^{-1} \cnsq \louter^{\beta-3}$ 
with $\CSM = (\beta-3) / 2(2\pi)^{4-\beta}$.  
We then relate the volume-averaged electron density $\nelec = f \nebar$ to the internal density $\nebar$ using the filling factor, $\nelec = f \nebar$,  and express  the volume-averaged $\cnsq$  for the Kolmogorov case ($\beta = 11/3$) as
\be
\cnsq = \CSM F \nelec^2
\label{eq:cnsq}
\ee
where $\CSM = [3(2\pi)^{1/3}]^{-1}$ and the parameter $F = \zeta \varepsilon^2 / f \louter^{2/3}$
characterizes the fluctuation properties of the medium.  \added{Different components of the NE2001 model have different values of $F$}. 

With these definitions, we relate the scattering time $\taud$ to the \DM\ of a medium as follows.  
In Euclidean space, the {\it mean} scattering time is given by the line of sight integral   
\be
 \langle \tau \rangle = \frac{1}{2c}  \int_0^d ds\, s(1-s/d) \eta(s),
\label{eq:tau}
\ee
where $\eta(s)$ is the mean-square scattering angle per unit distance, which may vary along the line of sight, and is given at a single location by \citep[][]{cr98}
\be
\eta(s) 
=    
\frac{\Gamma(3-\beta/2)} {4-\beta} 
\left(\frac{2\pi}{\linner} \right)^{4-\beta}  \lambda^4 \relec^2  \cnsq(s)
= 
\Gamma(7/6)   \relec^2  \lambda^4   \Ftilde  \nelec^2
\label{eq:eta}
\ee
based on \Eq\ref{eq:cnsq} and using $\beta=11/3$ for the second equality.  
The   parameter  $\Ftilde$ in \Eq\ref{eq:eta} is 
\be
\Ftilde \equiv F\linner^{-1/3}  = \frac{\zeta\varepsilon^2}{f (\louter^2 \linner)^{1/3}}.
\label{eq:ftilde}
\ee
\deleted{where $F$ is a parameter used in the NE2001 model. }
For a layer of thickness $L$ with constant density and constant $F$, the dispersion measure is
$\DMl = \nelec L$ and  the mean scattering time becomes
\be
\langle \tau \rangle 
= C_\taud \nu^{-4} \Ftilde  \DMl^2 (1 - 2L/3d) \Gscatt 
\ee
where 
$C_\taud =  \Gamma(7/6) c^3 \relec^2 / 4$ is a constant that depends only weakly on any modification from
the $\beta = 11/3$ spectrum. 

The last consideration is how scattering times estimated from pulse or burst shapes are related to the mean
scattering time $\langle \tau \rangle$.    The scattered shape for an emitted narrow impulse is the 
pulse broadening function (PBF), often assumed to be a one-sided exponential,
$p(t) = \tau_e^{-1} \exp(-t / \tau_e) \Theta(t)$ where $\Theta$ is the Heaviside function.    The scattering time
is typically estimated by comparing the measured pulse shape with  the convolution of $p(t)$ with an assumed intrinsic pulse shape  to determine a best-fit value.  In this case, 
the mean scattering time is identical to the $e^{-1}$ time.    However, scattering from a power-law wavenumber spectrum can show PBFs with much longer tails than an exponential \cite[e.g.][]{lr99}, yielding 
$\langle \tau \rangle > \tau_e$ by an amount that depends on details of the wavenumber 
spectrum (the inner scale,  the spectral index $\beta$, and the amplitude $\cnsq$). 
To account for  how empirical estimates for the scattering time are related more closely  to the $e^{-1}$ time than to the mean, we define a factor $\Clinner \equiv \tau_e / \langle \tau \rangle \le 1$.   Assuming also that the scattering region is thin, $L/d \ll 1$, we  write the measured $\tau$ as (dropping the `e' subscript) 
\be
\tau \simeq \Clinner C_\taud \nu^{-4} \Ftilde  \DMl^2  \Gscatt.
\label{eq:taucloudlet}
\ee
%
\added{
The factor $\Clinner$ depends on the ratio, $\linner / \ld$,  of the inner scale  to the diffraction scale, where the latter is  related to the characteristic  scattering angle and thus to the width of the PBF.   When  the ratio is small,
$\linner / \ld \ll 0.1$, $\Clinner \sim 1/6$ for a Kolmogorov spectrum with $\beta = 11/3$  but increases to 
$\Clinner \sim 0.7$ for $\linner / \ld = 1$.    For a fixed inner scale, $\linner/\ld$ is larger for stronger scattering 
and $\Clinner \to 1$ (unpublished notes by JMC).   For FRBs   that show significant scattering as pulse broadening,  it is likely that $\Clinner$ is close to unity. 
Because of its model dependence, we simply include  $\Clinner$ in Equation~\ref{eq:taucloudlet} as one of the factors in the lumped quantity, $\Clinner\FtG$ and recognize that it is model dependent.  }





\listofchanges
\end{document}